\documentclass[aps,twocolumn,showpacs,preprintnumbers,prb,superscriptaddress]{revtex4-2}
\usepackage{graphicx}  
\usepackage{dcolumn}   
\usepackage{tikz}
\usepackage{bm}        
\usepackage{amssymb}   
\usepackage{framed}
\usepackage{amsmath}
\usepackage{appendix}
\usepackage{hhline}
\definecolor{bluegreen}{rgb}{0,0.2,0.8}
\usepackage{subfigure,amsmath,verbatim,moreverb}
\usepackage{tabularx}
\usepackage{adjustbox}
\usepackage{lipsum}
\usepackage{longtable}
\usepackage{booktabs}
\usepackage{latexsym}
\usepackage{dcolumn}
\usepackage{amsmath}
\usepackage{epsf}
\usepackage{float}
\usepackage[breaklinks=true,colorlinks,citecolor=blue,linkcolor=blue,urlcolor=blue]{hyperref}
\usepackage[normalem]{ulem} 
\usepackage{multirow}
\usepackage{siunitx}
\usepackage{longtable}
\usepackage{booktabs}
\usepackage{threeparttable}
\usepackage[T1]{fontenc}

\usepackage{color}

\usepackage{etoolbox}
\AtBeginEnvironment{align}{\setcounter{subeqn}{0}}
\newcounter{subeqn} %

\begin{document}

\title{
Meta-GGA dielectric-dependent and range-separated screened hybrid functional for reliable prediction of material properties
}
\author{Subrata Jana}
\email{Corresponding author: subrata.niser@gmail.com}
\affiliation{Institute of Physics, Faculty of Physics, Astronomy and Informatics, Nicolaus Copernicus University in Toru\'n,
ul. Grudzi\k{a}dzka 5, 87-100 Toru\'n, Poland}
\author{Abhishek Bhattacharjee}
\affiliation{School of Physical Sciences, National Institute of Science Education and Research, An OCC of Homi Bhabha National Institute, Bhubaneswar 752050, India}
\author{Suman Mahakal}
\affiliation{Department of Physics, Vidyasagar Metropolitan College, Kolkata, India}
\author{Szymon \'Smiga}
\affiliation{Institute of Physics, Faculty of Physics, Astronomy and Informatics, Nicolaus Copernicus University in Toru\'n,
ul. Grudzi\k{a}dzka 5, 87-100 Toru\'n, Poland}
\author{Prasanjit Samal}
\affiliation{School of Physical Sciences, National Institute of Science Education and Research, An OCC of Homi Bhabha National Institute, Bhubaneswar 752050, India}

\begin{abstract}

We propose a range-separated hybrid exchange-correlation functional to calculate solid-state material properties. The functional mixes Hartree-Fock exchange with the semilocal exchange of the meta-generalized gradient approximation (meta-GGA) and the fraction of Hartree-Fock exchange is determined from the dielectric function. First-principles calculations and comparison with other meta-GGA approximations show that the functional leads to reasonably good performance for the band gap and optical properties. We also show that the present functional also successfully resolves the well-known ``band gap problem'' of narrow gap Cu$-$based semiconductors, such as Cu$_3$SbSe$_4$ and Cu$_3$AsSe$_4$
, where, in general, a considerably large band inversion energy leads to a ``false'' negative or metallic band gap for all other methods. Furthermore, reasonable accuracy for the occupied $d-$ bands and transition energies is also obtained for bulk solids. Thus, overall, our results demonstrate the predictive power of range-separated meta-GGA hybrid functionals for quantum materials simulations.

%

\end{abstract}

\maketitle

\section{Introduction}

Over the past decades, Kohn-Sham (KS) density functional theory (DFT)~\cite{kohn1965self,hohenberg1964inhomogeneous} has become the workhorse for electronic-structure calculations on all kinds of electronic systems: atoms, molecules, surfaces, bulk, nano-structure, liquids, etc. In principles, DFT includes exactly all the many-body effects beyond the Hartree approximation through the so-called exchange-correlation (XC) functional \cite{burke2012perspective,engel2013density,Jones2015,CoheMoriYang2012,HasnRefsProb2011,kummel2008orbital,dftsharing2022}. 
Since the inception of the KS framework, a plethora of approximate XCs has been proposed ~\cite{perdew2001jacob,scuseriaREVIEW05,della2016kinetic,perdew2008density,perdew2005prescription,LehtolaSX18} satisfying various constraints, among which most popular are the Perdew, Burke, and Ernzerhof (PBE)~\cite{perdewPRL96} and Heyd, Scuseria, and Ernzerhof (HSE06)~\cite{heyd2003hybrid,krukau2006influence,heyd2004efficient,jana2020screened,jana2020improved} for having proper balance of efficiency and accuracy.
The PBE XC has its roots in inhomogeneous electron gas (HEG) approximation, which poses a serious challenge to the accuracy of the PBE band gap in solids. While the HSE functional improves upon this limitation employing exact non-local exchange, it still falls short of achieving the expected accuracy for materials with large band gaps \cite{perdew2017understanding,tran2007band,Borlido2020,patra2019efficient,fabien2018assessment,jana2018assessing,tran2017importance,fabien2019semilocal}.
The many-body perturbation-based methods $GW$ are considered state-of-the-art for the reliable band gap prediction for semiconductors and insulators. However, the $GW$ methods are expensive, and the results of the widely used single-shot $GW$ (also known as $G_0W_0$) largely depend on the initial KS orbitals. In this respect, the hybrid functionals with system-dependent screening parameters have become 
the cheap alternative of $G_0W_0$ within KS-regime for studying excited state properties (such as gaps and optical transitions~\cite{paier2008dielectric,wing2019comparing,stadele1999exact,petersilka1996excitation,kim2002excitonic,terentjev2018gradient,sharma2011bootstrap,rigamonti2015estimating,van2002ultranonlocality,cavo2020accurate,jana2020improved,OhadWingGant2022,WingHabeJonah2019,AshwinDahvydLeeor2019}) of quantum materials.

The self-interaction error (SIE) \cite{PhysRevB.23.5048}, which is mainly responsible for the band gap underestimation, can be reduced by using a fraction of the Hartree-Fock (HF) exchange. The exchange potential behaves differently for finite and dielectric mediums such as solids. In a finite system, when an electron is 
displaced, the coulomb potential it experiences falls as $-1/r$ asymptotically, while for solids, the behavior becomes $-\epsilon^{-1}(r,r')/r$, where $\epsilon(r,r')$ is the screening function or dielectric function. Thus, considering the analogy to the static version of the Coulomb hole plus screened-exchange (COH-SEX)~\cite{AndrKressHinu2014}, a system-dependent hybrid functional should capture the screening effect of the material. Based on these concepts, system-dependent hybrid functionals, also popularly known as dielectric-dependent hybrid (DDH) have been proposed \cite{ShimNaka2014,SkonGovoGall2014,BrawVorosGovo2016,SkonGovoGall2016,WeiGiaRigPas2018,CuiWangZhang2018,MichPeteThom2020}. However, for simplicity and to make the implementation of such methods easier one may replace $\epsilon(r,r')$ by the static version of the dielectric function $\epsilon_\infty$, also known as the macroscopic dielectric constant (optical or high-frequency part). Thus, DDH functionals can be regarded as simplified versions of the $GW$ method, and its time-dependent version as the analog to the Bethe-Salpeter equation (BSE) for spectral properties~\cite{wing2019comparing}. One may also note that, unlike DDHs with unscreened HF exchange, the range-separated versions~\cite{BrawVorosGovo2016,SkonGovoGall2016,WeiGiaRigPas2018,Jana2023Simple} can be constructed to have the correct asymptotic behavior at long-range separation.

Typically, the DD range-separated hybrids (RS) which are based on the Coulomb attenuation method (CAM) perform considerably better than screened hybrid HSE06 for excited state properties of semiconductors and insulators~\cite{WeiGiaRigPas2018,Bischoff2019Nonempirical,Wang2022Accurate,Jana2023Simple,Yang2023Range,Miceli2018Nonempirical,Yang2022OneShot}. They are also excellent for the absorption spectra~\cite{tal2020accurate,Jana2023Simple,Jin2024Self}, photoelectron spectra of aqueous solutions~\cite{Gaiduk2016Photoelectron}, and quantum information related applications~\cite{Jin2022Vibrationally}. So far, all
the proposed RS DDHs are limited to the semilocal GGA framework \cite{SkonGovoGall2014,ZhengGovoniGalli2019,WeiGiaRigPas2018}.
However, over the past few years, functionals of the next rung of Jacob's ladder \cite{perdew2001jacob}, the meta-GGAs, have made significant progress~\cite{della2016kinetic,sun2015strongly,tao2016accurate,furness2020accurate,Neupane2021Opening,patra2019relevance,JanaACSC,Lebeda2024Balancing} providing new physical insights for material simulations~\cite{Borlido2020,tran2017importance,GhoshJanaRauch2022}.
Thus, developing RS DDH functionals could greatly benefit from the enhanced accuracy of semilocal functionals and exchange hole models at the meta-GGA level.

This work fills this gap by proposing an RS DDH functional based on a meta-GGA exchange hole~\cite{Jana2022Solid}. Moreover, we establish its usability and effectiveness by applying it to band gaps, optical properties for solids, and its applications for topologically challenging materials where other functionals are not so effective.

\section{Theory}\label{theory}
\subsection{Generic form \label{sec:theo:A}}

A generic form of range-separated hybrid (RSH) functional can be obtained by a RS of the Coulomb interaction into short- and long-range (abbreviated below as SR and LR) utilizing the following CAM~\cite{YANAI200451,Skone2016Nonempirical,kronik2018dielectric} decomposition of Coulomb operator
\begin{eqnarray}
\frac{1}{r}&=&\underbrace{\frac{\alpha\,\text{erf}(\mu r)+\beta(1-\text{erf}(\mu r))}{r}}_{\text{HF exchange}} \nonumber\\
& & +\underbrace{\frac{1-[\alpha\,\text{erf}(\mu r)+\beta(1-\text{erf}(\mu r))]}{r}}_{\text{DFT exchange}},
\label{eq1}
\end{eqnarray}
where, as indicated, the first and second terms are for the exact HF exchange and DFT semilocal exchange, respectively.
From Eq.~(\ref{eq1}), one can see that $\alpha$ and $\beta$ are the fractions of exact HF exchange at long and short ranges, respectively, and $\mu$ is the screening parameter required to interchange between the HF or and DFT exchange. Then, using the above range separation, a general form of DDH exchange-correlation (XC) potential can be written as~\cite{Skone2016Nonempirical}
\begin{eqnarray}
v_{xc}^{\text{DDH}}({\bf{r}},{\bf{r}}')&=&\alpha v_{x}^{\text{LR-HF}}({\bf{r}},{\bf{r}}',\mu)+\beta v_{x}^{\text{SR-HF}}({\bf{r}},{\bf{r}}',\mu)\nonumber\\
&+&(1-\alpha) v_{x}^{\text{LR-DFT}}({\bf{r}},\mu)+(1-\beta) v_{x}^{\text{SR-DFT}}({\bf{r}},\mu)\nonumber\\
&+&v_c^{\text{DFT}}({\bf{r}}),
\label{eq2}
\end{eqnarray}
or, alternatively, as \cite{Liu2019Assessing}
\begin{eqnarray}
v_{xc}^{\text{DDH}}({\bf{r}},{\bf{r}}')&=&[\beta-(\beta-\alpha)\text{erf}(\mu|{\bf{r}}-{\bf{r}}'|)] v_{x}^{\text{HF}}({\bf{r}},{\bf{r}}')\nonumber\\
&+&(1-\alpha) v_{x}^{\text{LR-DFT}}({\bf{r}},\mu)+(1-\beta) v_{x}^{\text{SR-DFT}}({\bf{r}},\mu)\nonumber\\
&+&v_c^{\text{DFT}}({\bf{r}}),
\label{eq3}
\end{eqnarray}
where $v_x^{\text{(LR/SR)-HF}}$ is the nonlocal HF exchange potential, $v_x^{\text{(LR/SR)-DFT}}$ is a semilocal DFT exchange potential, and $v_c^{\text{DFT}}$ is a semilocal DFT correlation potential.
As discussed, for instance, in Refs.~\cite{Marques2011density,SkonGovoGall2014}, the fraction $\alpha$ of long-range HF in the above equations can be chosen as a function of the inverse of the macroscopic dielectric function, $\alpha=\alpha[\epsilon_\infty^{-1}]$, leading to DDH.

A correspondence between the exchange potential of DDHs and the self-energy ($\Sigma$) of the static COH-SEX (Coulomb
hole and screened exchange) approximation of $GW$ can be made \cite{CuiWangZhang2018}:
%
\begin{eqnarray}
\Sigma_{\text{COH}}({\bf{r}},{\bf{r}}',\omega=0)\delta({\bf{r}}-{\bf{r}}')
&\approx&(1-\alpha[\epsilon_\infty^{-1}]) v_{x}^{\text{LR-DFT}}({\bf{r}},\mu)+\nonumber\\
& & +(1-\beta) v_{x}^{\text{SR-DFT}}({\bf{r}},\mu)
\label{eq-coh}
\end{eqnarray}
 and 
 \begin{eqnarray}
\Sigma_{\text{SEX}}({\bf{r}},{\bf{r}}',\omega=0)
&\approx&\alpha[\epsilon_\infty^{-1}] v_{x}^{\text{LR-HF}}({\bf{r}},{\bf{r}}',\mu)\nonumber\\
&&+\beta v_{x}^{\text{SR-HF}}({\bf{r}},{\bf{r}}',\mu)
\end{eqnarray}
So far, the range-separated DDHs have been constructed using mostly the GGA PBE exchange approximation for the semilocal Coulomb hole, and among the different variants of DDHs, we can mention the two following, that have different values of $\beta$~\cite{Chen2018Nonempirical}:

(i) RS-DDH with $\beta=0.25$ and $\alpha[\epsilon_\infty^{-1}]=\epsilon_\infty^{-1}$ (proposed in Ref.~\cite{Skone2016Nonempirical})
\begin{eqnarray}
v_{xc}({\bf{r}},{\bf{r}}')&=&\left[\frac{1}{4}-\left(\frac{1}{4}-\epsilon_\infty^{-1}\right)\text{erf}(\mu|{\bf{r}}-{\bf{r}}'|)\right] v_{x}^{\text{HF}}({\bf{r}},{\bf{r}}')\nonumber\\
& &+(1-\epsilon_\infty^{-1}) v_{x}^{\text{LR-DFT}}({\bf{r}},\mu)+\frac{3}{4}v_{x}^{\text{SR-DFT}}({\bf{r}},\mu)\nonumber\\
&  & +v_c^{\text{DFT}}({\bf{r}})
\label{eq4}
\end{eqnarray}

(ii) DD-RSH-CAM with $\beta=1$ and $\alpha[\epsilon_\infty^{-1}]=\epsilon_\infty^{-1}$ (proposed in Ref.~\cite{WeiGiaRigPas2018})
\begin{eqnarray}
v_{xc}({\bf{r}},{\bf{r}}')&=&[1-(1-\epsilon_\infty^{-1})\text{erf}(\mu|{\bf{r}}-{\bf{r}}'|)] v_{x}^{\text{HF}}({\bf{r}},{\bf{r}}')\nonumber\\
&&+(1-\epsilon_\infty^{-1}) v_{x}^{\text{LR-DFT}}({\bf{r}},\mu)+v_c^{\text{DFT}}({\bf{r}})
\label{eq5}
\end{eqnarray}
%

We note that the RS-DDH and DD-RSH-CAM potentials have both the correct limit $\epsilon^{-1}_{\infty}/|{\bf{r}}-{\bf{r}}'|$ at long range. However, at short-range, the RS-DDH exchange potential tends to $\beta v_{x}^{\text{HF}}({\bf{r}},{\bf{r}}')$ instead of being unscreened [i.e., $v_{x}^{\text{HF}}({\bf{r}},{\bf{r}}')$] as it should, which is the case with the DD-RSH-CAM potential. Thus, only when $\beta=1$ is the short-range limit of the Fock exchange is it correctly unscreened. Also, note from Ref.~\cite{WeiGiaRigPas2018} that DD-RSH-CAM performs much better for several solid properties compared to RS-DDH. 

Constructing such a range-separated method requires using the exchange or Coulomb hole model from the DFT exchange. To date, only the reverse-engineered GGA exchange hole is used to build range-separated DDHs. Although GGA-based DDHs show superior performance, there is still room for improvement in diverse material properties.


\subsection{meta-GGA DDH variant}
We start our construction by approximating Eq.~(\ref{eq-coh}) by
\begin{eqnarray}
\Sigma_{\text{COH}}({\bf{r}},{\bf{r}}',\omega=0)\delta({\bf{r}}-{\bf{r}}') &\approx& (1-\alpha[\epsilon_\infty^{-1}]) v_{x}^{LR-meta}({\bf{r}},\mu)\nonumber\\
&+&(1-\beta) v_{x}^{SR-meta}({\bf{r}},\mu)which
\label{eq8}
\end{eqnarray}
where the $LR-meta$ and $SR-meta$ denote the SR and LR part of semilocal meta-GGA functional, respectively, built using the exchange meta-GGA hole model.

At this point, we remark that the reversed-engineered meta-GGA exchange holes have been investigated only recently. To date, the Tao-Perdew-Staroverov-Scuseria (TPSS)~\cite{Tao2017Semilocal} and MGGAC exchange holes~\cite{Jana2022Solid} have been introduced in the literature using the reversed-engineered technique. We also recall that other forms of exchange holes have been utilized to construct meta-GGA range-separated hybrids~\cite{jana2019screened,janameta2018}. In this work, we employ the reverse-engineered MGGAC exchange hole variant to approximate the LR and SR parts in Eq. (\ref{eq8}) as follows:
\begin{eqnarray}
v_{x}^{SR-meta}({\bf{r}},\mu)&=&v_{x}^{\omega\text{MGGAC}}[\alpha^{iso}({\bf{r}})]\\
v_{x}^{LR-meta}({\bf{r}},\mu)&=&v_{x}^{MGGAC}[\alpha^{iso}({\bf{r}})]-v_{x}^{\omega\text{MGGAC}}[\alpha^{iso}({\bf{r}})]\nonumber \;.
\label{eq6}
\end{eqnarray}
Details of the MGGAC and $\omega$MGGAC exchange and potential energy functional can be found in Appendix~\ref{mggac-details}.

Finally, it is combined with $r$MGGAC correlation functional developed in ref.~\cite{jana2021szs}
\begin{equation}
v_{c}({\bf{r}})=v_{c}^{r\text{MGGAC}}[s({\bf{r}}),\alpha^{iso}({\bf{r}})]~.
\label{eq7}
\end{equation}
to construct the semilocal part of the full functional. The well-known ingredients are defined as follows: Pauli enhancements factor or iso-orbital indicator, $\alpha^{iso}({\bf{r}})=(\tau({\bf{r}})-\tau^{VW}({\bf{r}}))/\tau^{unif}({\bf{r}})$ and reduce gradient, $s({\bf{r}})=|\nabla n({\bf{r}})|/(2(3\pi^2)^{1/3}n^{4/3})$ with 
\begin{eqnarray}
\tau({\bf{r}})&=&\frac{1}{2}\sum_{occ}|\nabla\phi_{KS}({\bf{r}})|^2\nonumber\\
\tau^{unif}({\bf{r}})&=&\frac{3}{10}(3\pi^2)^{2/3} n({\bf{r}})^{5/3}\nonumber\\
\tau^{VW}({\bf{r}})&=& \frac{5}{3} \tau^{unif}~ s^{2}(\bf{r})~. \nonumber \\
\end{eqnarray}
Here, $n$ being the electron density, $n({\bf{r}})=\sum_{occ} |\phi_{KS}({\bf{r}})|^2$ and $\phi_{KS}({\bf{r}})$'s are the KS orbitals.
We recall that the MGGAC exchange-hole construction is quite simple because of the involvement of the only meta-GGA ingredient $\alpha^{iso}({\bf{r}})$, which is further utilized in those work to construct the meta-GGA-range-seperated DDH. Further, the modern meta-GGA-like (r)MGGAC (and SCAN-based methods~\cite{sun2015strongly}) has many new physical constraints compared to TPSS-like meta-GGAs, and it is also encouraging to use these modern meta-GGAs further to construct range-separated DDHs.

\subsection{Choices of $\beta$, $\mu$, and $\alpha[\epsilon_\infty^{-1}]$ }
\label{parameters}
Finally, the proposed range-separated DD meta-GGA functional is given using Eq.~(\ref{eq3}) with $\beta=1$ as,
\begin{eqnarray}
v_{xc}^{\text{DD-MGGAC-CAM}}({\bf{r}},{\bf{r}}')&=&[1-(1-\alpha[\epsilon_\infty^{-1}])\text{erf}(\mu|{\bf{r}}-{\bf{r}}'|)]\nonumber\\
& &v_{x}^{\text{HF}}({\bf{r}},{\bf{r}}')+(1-\alpha[\epsilon_\infty^{-1}])\nonumber\\
& &(v_{x}^{MGGAC}({\bf{r}})-v_{x}^{\omega\text{MGGAC}}(({\bf{r}},\mu)))\nonumber\\
&+&v_c^{\text{$r$MGGAC}}({\bf{r}})~.
\label{eq-meta-form}
\end{eqnarray}
Note that here $\omega$ used in the superscript of $v_{x}^{\omega\text{MGGAC}}({\bf{r}},\mu)$ is actually the range-seperated parameter denoted in ref.~\cite{Jana2022Solid}. For convenience, we will use $\omega=\mu$ to denote the same range-separated parameter throughout this paper without changing the functional naming. The choice of $\beta=1$ was already discussed in Sec.\ref{sec:theo:A}, and hence we disregard any other values of this parameter within the range $0<\beta<1$. Importantly, for the $\mu$ of the meta-GGA-DDH, we consider $\mu=\mu^{fit}_{eff}$ as given by
Eq.(19) of Ref.~\cite{Jana2023Simple}, which has the following form:
\begin{equation}
 \mu_{eff}^{fit}= \frac{a_1}{\langle r_s \rangle} + \frac{a_2 \langle r_s \rangle}{1 + a_3 \langle r_s \rangle^2}~,
 \label{eq-theo-secb-19}
\end{equation}
with $a_1= 1.91718$, $a_2= -0.02817$, $a_3=0.14954$, and
\begin{equation}
 \langle r_s \rangle=\frac{1}{\Omega_{cell}}\int_{cell} \Big(\frac{3}{4\pi(n_{\uparrow}({\bf{r'}})+n_{\downarrow}({\bf{r'}}))}\Big)^{1/3}~d^3r'~,
 \label{eq-theo-secb-13}
\end{equation}
where $\Omega_{cell}$ is the unit cell volume and $n_{\sigma}({\bf{r}})~(with~\sigma=\uparrow~or~\downarrow)$ is the spin-density. One may note that this form provides well-balanced predictions for the band gaps of bulk systems recently studied across various quantum materials~\cite{Ghosh2024Accurate,Rani2025Thermoelectric,Jana2025Nonempirical}.

Regarding the choice of $\alpha[\epsilon_\infty^{-1}]$, for DD-MGGAC-CAM, it is important to further optimize $\alpha[\epsilon_{\infty}^{-1}]$ from its conventional value $\epsilon_{\infty}^{-1}$ to improve the band gap performances.  The admixing parameter $\alpha[\epsilon_{\infty}^{-1}]=\epsilon_{\infty}^{-1}$, which includes a fraction of long-range Fock exchange to DD-MGGAC-CAM
inevitably worsens the agreement of the band gap with the experiment of both narrow and wide gap materials by opening the gap too much.

We note that the meta-GGA's semilocal form already includes some non-locality effects because of its orbital dependence and generalized KS (gKS) implementation. As noted from earlier studies~\cite{patra2019relevance,jana2021szs,patra2021efficient,Tran2021bandgap}, for semiconductors, the gKS band gaps of $r$MGGAC are already good enough. This is because meta-GGA functional incorporate (ultra)nonlocal features of the gKS potential~\cite{Aschebrock2019Ultranonlocality,Lebeda2022First,Lebeda2023Right,Aschebrock2023Exact,Lebeda2024Balancing}. The origin of the (ultra)nonlocality effect and its connection to the derivative discontinuity ($\Delta_x$) of the KS potential can be understood from Ref.~\cite{Aschebrock2019Ultranonlocality}, where it is defined as,
\begin{equation}
\Delta_x^{meta-GGA}\propto \frac{\partial e_x}{\partial\tau}~.
\end{equation}
where $e_x$ denotes the exchange energy density. Therefore, the band gap for meta-GGA functionals can be approximated as $E_g=E_g^{meta-GGA}+\Delta_x$ with $\Delta_x>0$. For LDA and GGA level approximations, $\Delta_x\approx 0$. The key quantity $\frac{\partial e_x}{\partial\tau}$, which appears in $r$MGGAC functional, already introduces a sizable discontinuity in the functional because of $\frac{\partial e_x}{\partial\tau}>0$~\cite{patra2019relevance,jana2021szs}. Hence, the dielectric-dependent version of $(r)$MGGAC includes more derivative discontinuity through its non-local form. For this reason, a better set of $\alpha$ values is required for the present meta-GGA DDH construction to enhance its performance.

\begin{figure}
  \centering
 \includegraphics[scale=0.4]{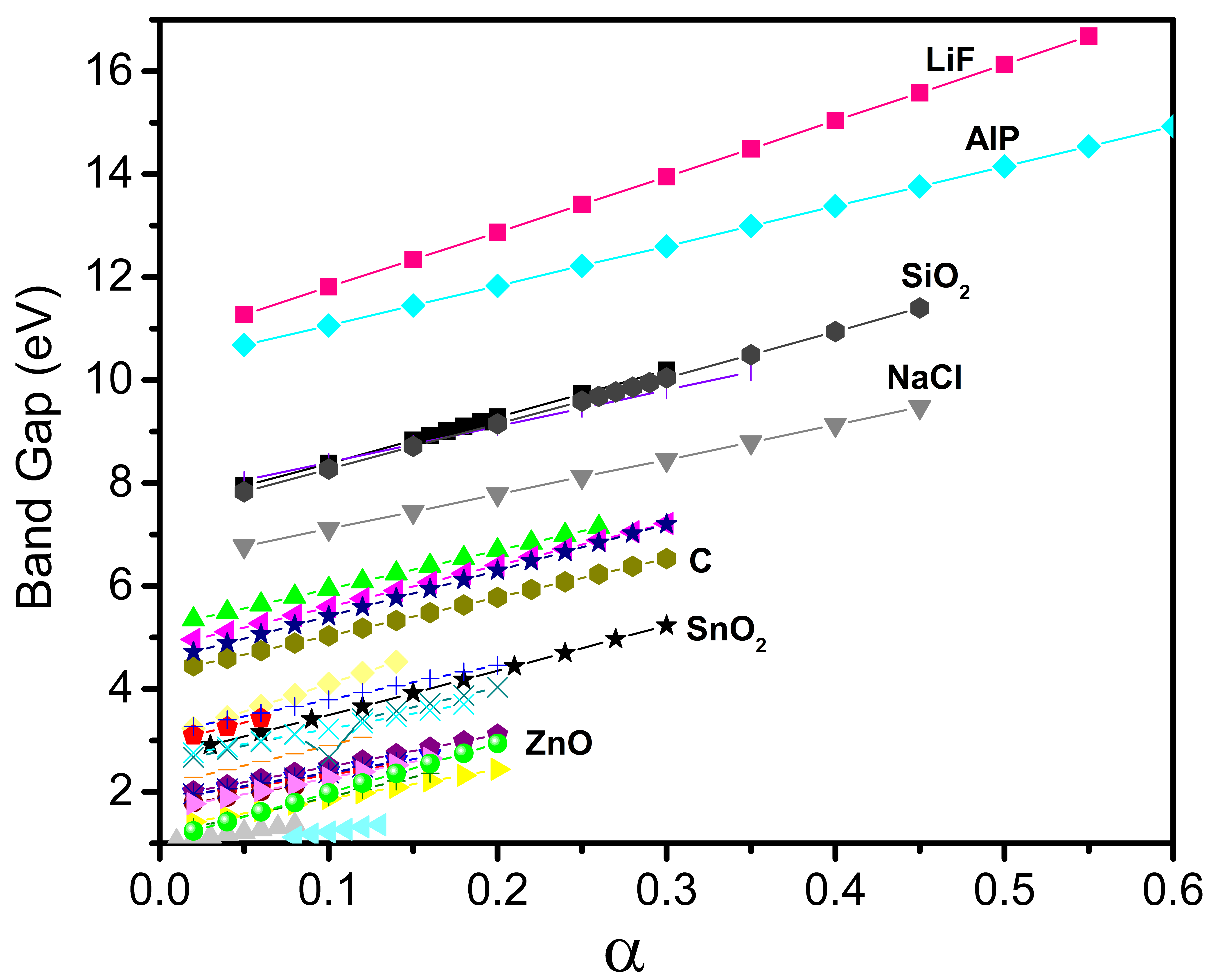}
  \caption{Shown is the linear dependence of band gaps $E_g$ with $\alpha$ with a fixed $\beta$ and $\mu$ for DD-MGGAC-CAM.}
  \label{alpha_eg}
\end{figure}

In Fig.~\ref{alpha_eg}, we plot the band gap variation for different values of $\alpha$ for the new functional form. Assuming that $E_g$ depends on $\alpha$ linearly (thus in consequence also to $\epsilon_\infty^{-1}$) as shown in Fig.~\ref{alpha_eg} (for a fixed $\beta$ and $\mu$ values) we consider a linear fitting of $\alpha$ as
\begin{equation}
\alpha[\epsilon_{\infty}^{-1}]=a\epsilon_{\infty}^{-1}+b~,
\label{eqrel}
\end{equation}
with
\begin{eqnarray}
E_g[\alpha^{exact}]&=&E_g^{Experiment}~.
\end{eqnarray}
By using the $\epsilon_{\infty}^{-1}$ of DD-MGGAC-CAM, we obtain the fitting parameters as $a = 0.757491891$ and $b = -0.009738131$. In the optimization procedure, we exclude those solids (GaAs, Ge, and InP) for which the band gap values have already been strongly overestimated at the semilocal $r$MGGAC level.
The new parametrization given by Eq. (\ref{eqrel}) leads to a large improvement of the band gap values for the present meta-GGA DDH construction with respect to the initial relation.

\begin{table*}[!ht]
\begin{center}
\caption{\label{tab1} Meta-GGA functionals according to Eq.~\ref{eq2}.}
\begin{tabular}{lcccccccccccccccccccccccc}
\hline\hline
Variant	&&	Functional	&&	$\alpha[\epsilon_\infty^{-1}]$&& $\beta$&&$\mu$&&Ref.			\\
\hline
Semilocal&&$r$MGGAC&&0&&0&&$-$&&\cite{jana2021szs}\\
RS-hybrid&&$\omega r$MGGAC&&0&&0.10&&0.11&&\cite{Jana2022Solid}\\
RS-DD-hybrid&&DD-MGGAC-CAM&&from fitting&&1.0&&$\mu^{fit}_{eff}$&&This work\\
\hline\hline
     \end{tabular}
\end{center}
\end{table*}

\section{Results and discussions}

\subsection{Technical Details}
A set of $32$ solids is considered from Ref.~\cite{WeiGiaRigPas2018} to assess the performance of the constructed functional. The computations are performed using the plane-wave pseudopotential code Vienna Ab-initio Simulation Package (VASP)~\cite{vasp1,vasp2,vasp3,vasp4}, version 6.4.2. We use $16\times 16\times 16$ $\Gamma-$centered $\text{k}-$point mesh to sample the first Brillouin zone except Al$_2$O$_3$ ($8\times 8\times 8$), AlN ($16\times 16\times 8$), Cu$_2$O ($8\times 8\times 8$), GaN ($16\times 16\times 8$), In$_2$O$_3$ ($6\times 6\times 6$), MoS$_2$ ($16\times 16\times 3$), SiO$_2$ ($10\times 10\times 8$), SnO$_2$ ($10\times 10\times 14$), TiO$_2$ ($10\times 10\times 16$), and ZnO ($16\times 16\times 8$). 
An energy cutoff of 550 eV is used in all our calculations. For meta-GGA calculations, non-spherical contributions are added using LASPH, and a more dense grid is used (using the ADDGRID tag).
The optical and exciton properties are calculated with $8\times 8\times 8$ $\text{k}-$point mesh, and symmetry is completely switched off using ISYM=-1, which includes in total 512 irreducible points. Regarding pseudopotentials, the VASP-recommended pseudopotentials are used, with deeper $3d$ states considered the valance state for Ga, Ge, and In. All calculations are performed with experimental lattice constants as given in ref.~\cite{WeiGiaRigPas2018}. For the analysis of the results, the following statistical quantities are used: ($i$) mean signed error (MSE=$\frac{1}{N}\sum_{N}$(Calculated-Experimental)), ($ii$) mean absolute error (MAE=$\frac{1}{N}\sum_{N}$(|Calculated-Experimental|)), ($iii$) mean absolute percentage error (MAPE=$\frac{1}{N}\sum_{N}$(|Calculated-Experimental|/|Experimental|$\times$100)). For a complete comparison, in this paper, we also included the performances of $r$MGGAC~\cite{jana2021szs} and $\omega r$MGGAC~\cite{Jana2022Solid}, which are the different variants of semilocal and hybrids meta-GGA XC functionals (see the Table~\ref{tab1}).

\subsection{Dielectric constants and gKS energy gaps}

\subsubsection{High-frequency or optical part of the dielectric constants}

The performances of the $r$MGGAC, $\omega r$MGGAC, and DD-MGGAC-CAM are assessed for the dielectric constants $\epsilon_\infty$ of the $32$ solid systems using the self-consistent procedure prescribed in Ref.~\cite{Ghosh2024Accurate}. Table~\ref{dielectric} reports the orientationally-averaged $\epsilon_\infty$ i.e., $\epsilon_\infty=\frac{\epsilon^{xx}_\infty+\epsilon^{yy}_\infty+\epsilon^{zz}_\infty}{3}$. Here, the $\epsilon_\infty$ is calculated using only RPA level theory, i.e., by neglecting $f_{xc}$~\cite{WeiGiaRigPas2018} with density functional perturbation theory (DFPT)~\cite{Baroni2001phonons}. We note, however, that the effect of $f_{xc}$ can be neglected as its effect in the polarizability calculation is quite small~\cite{paier2008dielectric}. For further details are equations given in ref.~\cite{Gajdo2006linear}.

We observe close performances within different methods when comparing the results of semilocal and hybrids. However, compared with the experimental results, values of the dielectric constants show underestimation. Although for semiconductors with $sp-$ bonding and insulators, meta-GGA-based methods perform close to the experiments, the agreement is not so satisfactory in the presence of semicore $d$ electrons such as GaAs, GaN, GaP, Ge, and InP. However, the calculated dielectric constants are satisfactory for oxides such as $d-$oxides, except SnO$_2$. Considering the MAPE of DD-MGGAC-CAM (9.2\%), it performs in the same line as DD-PBEH (13.0\%) and RS-DDH (9.6\%) as reported in Ref.~\cite{WeiGiaRigPas2018}. For DDH, the main quantity entered in the functional is $\epsilon_\infty^{-1}$. We observe the discrepancy of MAE of $\epsilon_\infty^{-1}$ diminishes when compared within different methods based on GGA (as reported in Ref.~\cite{WeiGiaRigPas2018}) with that of the meta-GGA-DDHs considered in this work. One may note that the high-frequency dielectric constant is calculated using the perturbative method by studying the response of the valance electrons to the applied electric field. This can be influenced strongly by the self-interaction correction of the XC approximations, which is strongly noticeable in the performance of different meta-GGAs. Although improving over PBE (see ref~\cite{WeiGiaRigPas2018}), discrepancies from meta-GGAs remain when compared with experimental values.

\begin{figure}
  \centering
 \includegraphics[width=7 cm,height=15 cm]{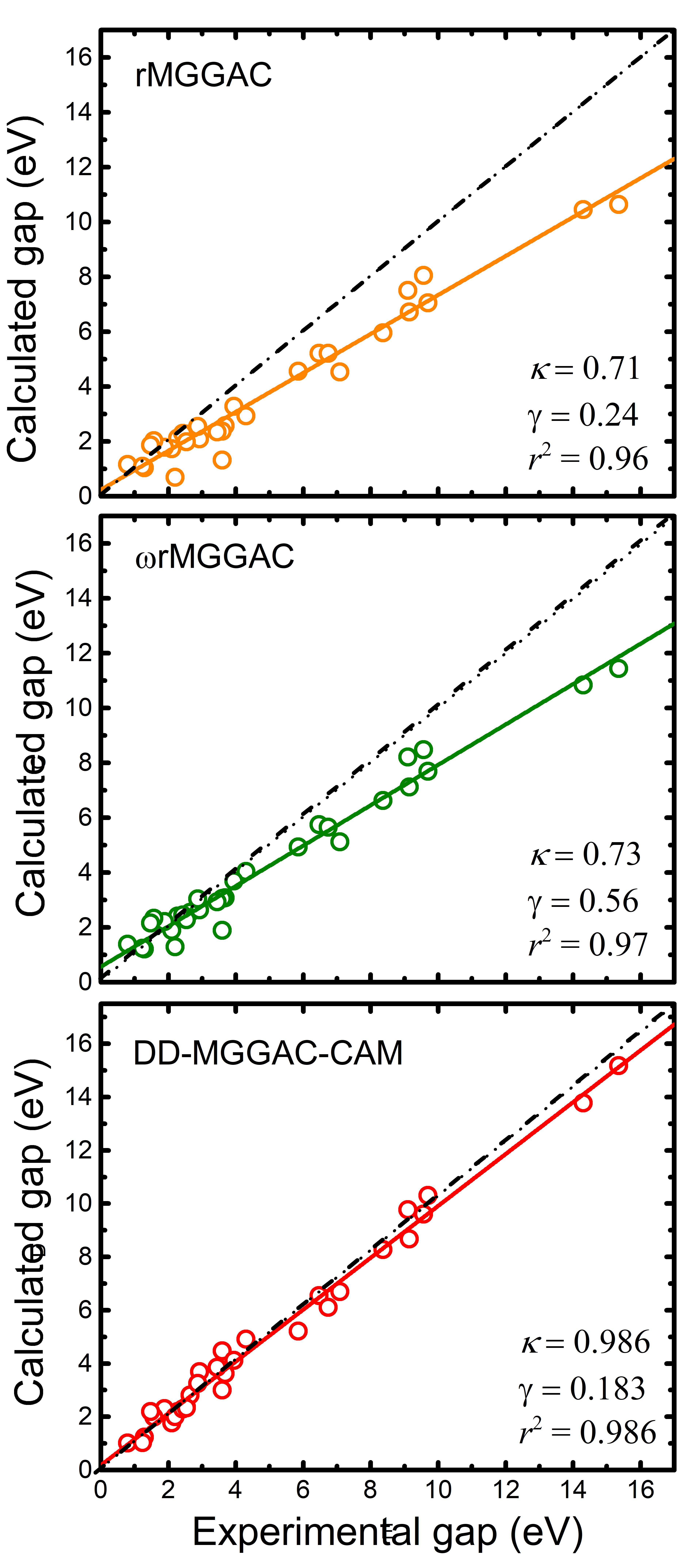}
  \caption{Shown is the calculated versus experimental band gaps for the solids of Table~\ref{band-gap}. The linear regression parameters are calculated according to $E_g^{calclated}=\kappa E_g^{experimental}+\gamma$, where $\kappa$ is the slope, and $\gamma$ (in eV) is the intercept. We also calculate the correlation coefficient $r^2$.}
  \label{band-plot}
\end{figure}

\begin{table*}[!ht]
\begin{center}
\caption{\label{dielectric} Macroscopic or optical or high-frequency dielectric constants of 32 semiconductors and insulators as calculated using different methods. For reference experimental values, see Table IV of Ref.~\cite{WeiGiaRigPas2018} and all references therein. See the main text for the details of the calculation procedures.}
\begin{tabular}{lcccccccccccccccccccccccc}
\hline\hline
Solids	&&	$r$MGGAC	&&	$\omega r$MGGAC	&&	DD-MGGAC-CAM&& Expt.			\\
\hline
Al$_2$O$_3$	&	&	2.97	&	&	2.92	&	&	2.87	&	&	3.10	\\
AlAs	&	&	7.26	&	&	7.13	&	&	7.23	&	&	8.16	\\
AlN	&	&	4.05	&	&	4.16	&	&	3.96	&	&	4.18	\\
AlP	&	&	6.69	&	&	6.55	&	&	6.84	&	&	7.54	\\
Ar	&	&	1.65	&	&	1.83	&	&	1.64	&	&	1.66	\\
BN	&	&	4.44	&	&	4.41	&	&	4.34	&	&	4.50	\\
BP	&	&	8.37	&	&	7.89	&	&	8.26	&	&	8.60	\\
C	&	&	5.75	&	&	5.33	&	&	5.63	&	&	5.70	\\
CaO	&	&	2.95	&	&	3.17	&	&	2.98	&	&	3.30	\\
CdS	&	&	4.76	&	&	4.65	&	&	4.65	&	&	5.20	\\
CdSe	&	&	5.34	&	&	5.14	&	&	5.05	&	&	5.80	\\
Cu$_2$O	&	&	8.46	&	&	6.47	&	&	6.13	&	&	6.46	\\
GaAs	&	&	9.15	&	&	8.59	&	&	9.61	&	&	10.58	\\
GaN	&	&	4.98	&	&	4.76	&	&	4.80	&	&	5.30	\\
GaP	&	&	7.98	&	&	7.64	&	&	8.06	&	&	9.10	\\
Ge	&	&	12.60	&	&	11.29	&	&	13.66	&	&	15.90	\\
In$_2$O$_3$	&	&	3.73	&	&	3.51	&	&	3.65	&	&	4.00	\\
InP	&	&	7.60	&	&	7.56	&	&	7.83	&	&	9.60	\\
LiCl	&	&	2.71	&	&	2.95	&	&	2.68	&	&	2.70	\\
LiF	&	&	1.78	&	&	2.13	&	&	1.79	&	&	1.90	\\
MgO	&	&	2.86	&	&	3.15	&	&	2.72	&	&	3.00	\\
MoS$_2$	&	&	11.46	&	&	11.40	&	&	11.50	&	&	17.80	\\
NaCl	&	&	2.27	&	&	2.25	&	&	2.21	&	&	2.30	\\
NiO	&	&	4.21	&	&	4.80	&	&	4.09	&	&	5.70	\\
Si	&	&	10.49	&	&	9.89	&	&	10.69	&	&	12.00	\\
SiC	&	&	6.23	&	&	6.26	&	&	6.20	&	&	6.52	\\
SiO$_2$	&	&	2.19	&	&	2.30	&	&	2.23	&	&	2.36	\\
SnO$_2$	&	&	3.56	&	&	3.44	&	&	3.37	&	&	4.05	\\
TiO$_2$	&	&	7.14	&	&	6.08	&	&	6.35	&	&	6.34	\\
ZnO	&	&	4.09	&	&	3.68	&	&	3.49	&	&	3.74	\\
ZnS	&	&	4.93	&	&	4.79	&	&	4.77	&	&	5.13	\\
ZnSe	&	&	5.63	&	&	5.44	&	&	5.48	&	&	5.90	\\
	&	&		&	&		&	&		&	&		\\
 \hline
MSE	&	&	-0.62	&	&	-0.83	&	&	-0.73	&	&		\\
MAE	&	&	0.82	&	&	0.88	&	&	0.73	&	&		\\
MAPE (\%)	&	&	9.82	&	&	10.39	&	&	9.22	&	&		\\
MAE ($\epsilon_\infty^{-1}$)&&0.018&&0.020&&0.019\\
\hline\hline
     \end{tabular}
\end{center}
\end{table*}
%

\begin{table*}[!ht]
\begin{center}
\caption{\label{band-gap} Band gaps (in eV) of the 32 semiconductors and insulators as obtained from different methods. The SOC corrections are included for AlAs (0.11), GaAs
(0.12), Ge (0.10), CdSe (0.14), and ZnSe(0.14), according to the Ref.~\cite{WeiGiaRigPas2018}. The experimental values are corrected for zero-phonon renormalization (ZPR). See Table V of Ref.~\cite{WeiGiaRigPas2018} for the details of the experimental values. Calculations of DD-MGGAC-CAM are performed with $\alpha[\epsilon_\infty^{-1}]=a\epsilon_\infty^{-1}+b$, where $\epsilon_\infty$ values are given in Table~\ref{dielectric}, $a=0.757491891$ and $b=-0.009738131$.
}
\begin{tabular}{lcccccccccccccccccccccccc}
\hline\hline
Solids	&&	$r$MGGAC	&&	$\omega r$MGGAC	&&	DD-MGGAC-CAM	&	&Expt	\\
\hline
Al$_2$O$_3$	&	&	7.51	&	&	8.22	&	&	9.77
	&	&	9.10	\\
AlAs	&	&	2.12	&	&	2.41	&	&	2.20 &	&	2.28	\\
AlN	&	&	5.21	&	&	5.75	&	&	6.55	&	&	6.47	\\
AlP	&	&	2.27	&	&	2.45	&	&	2.37
	&	&	2.54	\\
Ar	&	&	10.45	&	&	10.84	&	&	13.78
	&	&	14.30	\\
BN	&	&	5.21	&	&	5.65	&	&	6.11
	&	&	6.74	\\
BP	&	&	1.74	&	&	1.89	&	&	1.77	&	&	2.10	\\
C	&	&	4.56	&	&	4.94	&	&	5.22
	&	&	5.85	\\
CaO	&	&	4.53	&	&	5.12	&	&	6.7
	&	&	7.09	\\
CdS	&	&	2.22	&	&	2.55	&	&	2.81
	&	&	2.64	\\
CdSe	&	&	1.79	&	&	2.22	&	&	2.30	&	&	1.88	\\
Cu$_2$O	&	&	0.7	&	&	1.3	&	&	2.00	&	&	2.20	\\
GaAs	&	&	2.03	&	&	2.34	&	&	1.99	&	&	1.57	\\
GaN	&	&	2.58	&	&	3.08	&	&	3.63	&	&	3.68	\\
GaP	&	&	2.29	&	&	2.45	&	&	2.29	&	&	2.43	\\
Ge	&	&	1.16	&	&	1.39	&	&	1.02	&	&	0.79	\\
In$_2$O$_3$	&	&	2.09	&	&	2.64	&	&	3.69
	&	&	2.93	\\
InP	&	&	1.87	&	&	2.16	&	&	2.19
	&	&	1.47	\\
LiCl	&	&	8.06	&	&	8.48	&	&	9.61
	&	&	9.57	\\
LiF	&	&	10.64	&	&	11.44	&	&	15.18
	&	&	15.35	\\
MgO	&	&	5.96	&	&	6.63	&	&	8.27
	&	&	8.36	\\
MoS$_2$	&	&	1.04	&	&	1.22	&	&	1.24
	&	&	1.29	\\
NaCl	&	&	6.72	&	&	7.12	&	&	8.68
	&	&	9.14	\\
NiO	&	&	2.93	&	&	4.04	&	&	4.91
	&	&	4.30	\\
Si	&	&	1.09	&	&	1.24	&	&	1.03
 &	&	1.23	\\
SiC	&	&	1.99	&	&	2.28	&	&	2.34
	&	&	2.53	\\
SiO$_2$	&	&	7.06	&	&	7.7	&	&	10.31
	&	&	9.70	\\
SnO$_2$	&	&	2.37	&	&	3.06	&	&	4.48
	&	&	3.60	\\
TiO$_2$	&	&	2.35	&	&	2.94	&	&	3.86
	&	&	3.45	\\
ZnO	&	&	1.32	&	&	1.9	&	&	3.01
	&	&	3.60	\\
ZnS	&	&	3.28	&	&	3.69	&	&	4.12
 &	&	3.94	\\
ZnSe	&	&	2.55	&	&	3.04	&	&	3.25
	&	&	2.87	\\
\hline
MSE (eV)	&	&	-1.17	&	&	-0.71	&	&	0.05	&	&		\\
MAE (eV)	&	&	1.24	&	&	0.88	&	&	0.36&	&		\\
MAPE (\%)	&	&	25.19	&	&	18.65	&	&	11.43	&	&		\\
\hline\hline
     \end{tabular}
\end{center}
\end{table*}

\subsubsection{gKS gaps}

We perform the fundamental band gap calculations using the gKS method, and the results are reported in Table~\ref{band-gap}. The effects of spin-orbit coupling (SOC) on the calculated band gaps are also included whenever possible for a fair comparison with experimental results~\cite{WeiGiaRigPas2018}. The experimental band gap values are also corrected for the zero-phonon renormalization (ZPR), taken from ref.~\cite{WeiGiaRigPas2018}. One may note that ZPR significantly affects the band gap in the case of polar and ionic solids.

Turning to band gap performances using different methods
we observe a significant improvement of the gKS band gap for DD-MGGAC-CAM compared to the semilocal $r$MGGAC and screened hybrid $\omega r$MGGAC which is especially visible for insulators. One can note that for latter hybrid functionals the band gaps are systematically
underestimated for these species. Although $\omega r$MGGAC shows improvement over its semilocal form, the insulator band gaps are still away from experimental results. Nevertheless, for semiconductors with $sp-$bonding, DD-MGGAC-CAM performs close to the experiments; the agreement is less satisfactory for GaAs, Ge, and InP, which were also excluded from the fitting procedure of Eq. (\ref{eqrel}). For oxides such $d-$oxides such as Cu$_2$O and ZnO, we observe moderate underestimation in the obtained band gaps, although we observe adequate overestimations for In$_2$O$_3$, NiO, SiO$_2$, SnO$_2$, and TiO$_2$. DD-MGGAC-CAM produces an MAE of $0.36$ eV, much better than the other two variants.

In addition, in Fig.~\ref{band-plot}, we have plotted experimental versus calculated band gaps for all the materials considered in Table~\ref{band-gap}. The linear regression parameters are also shown therein. As shown in Fig.~\ref{band-plot}, for DD-MGGAC-CAM, the experimental and calculated values are aligned more closely to the diagonal line with $\gamma=0.18$ and $\kappa=0.99$ resulting in a uniform distribution of the values. On the other hand, both $r$MGGAC and $\omega r$MGGAC are scattered because of the underestimations in the band gap values of insulators. Furthermore, when we compare the overall error statistics of DD-MGGAC-CAM with other GGA-level screened DDHs such as DD-RSH-CAM~\cite{WeiGiaRigPas2018} (see Table V and Fig 3. in Ref. \cite{WeiGiaRigPas2018}) , the resultant error statistics are close to each other, indicating the requirement of the empirical choice of $\alpha$.

\begin{table}[!ht]
\begin{center}
\caption{\label{d-position} Average positions of the occupied $d-$ band with respect to the Fermi energy (in eV) at $\Gamma-$point. For experimental values, see Table VI of Ref.~\cite{WeiGiaRigPas2018} and references therein.}
\begin{tabular}{lcccccccccccccccccccccccc}
\hline\hline
Solids	&	$r$MGGAC	&	$\omega r$MGGAC	&	DD-MGGAC& Expt.~\cite{WeiGiaRigPas2018}			\\
	&		&		&	-CAM& 			\\
\hline
CdS	&	-7.6	&	-7.9	&	-9.0	&	-9.6	\\
CdSe	&	-7.8	&	-8.2	&	-9.3	&	-10	\\
InP	&	-14.6	&	-15.1	&	-15.5	&	-16.8	\\
GaAs	&	-14.0	&	-14.6	&	-17.0	&	-18.9	\\
GaN	&	-13.5	&	-14.1	&	-18.2	&	-17	\\
GaP	&	-14.7	&	-15.4	&	-16.7	&	-18.7	\\
ZnO	&	-5.6	&	-6.5	&	-7.1	&	-7.5	\\
ZnS	&	-6.0	&	-6.5	&	-8.1	&	-9	\\
ZnSe	&	-6.2	&	-6.8	&	-8.4	&	-9.2	\\
	&		&		&		&		\\
MSE	&	3.0	&	2.4	&	0.8	&		\\
MAE	&	3.0	&	2.4	&	1.1	&		\\
MAPE (\%)&23.88&18.85&8.03\\
\hline\hline
     \end{tabular}
\end{center}
\end{table}
%

\begin{table}
\caption{\label{transition-energy}. Comparison of transition energy gaps (in eV) and percentage error (PE) as calculated using different methods. The experimental values are ZPR corrected, taken from Table IX of Ref.~\cite{WeiGiaRigPas2018} and references therein.}
  \begin{tabular}{lccccccccccccccccccccc}
    \hline\hline
    \multicolumn{1}{l}{Solids} &
      \multicolumn{2}{c}{$r$MGGAC} &&
      \multicolumn{2}{c}{$\omega r$MGGAC} &&
      \multicolumn{2}{c}{DD-MGGAC}&&\multicolumn{2}{c}{Expt.} \\
    \multirow{2}{*}{} &
      \multicolumn{2}{c}{} &&
      \multicolumn{2}{c}{} &&
      \multicolumn{2}{c}{-CAM}&&\multicolumn{2}{c}{} \\
      & {$E_g$} & {PE} && {$E_g$} & {PE} && {$E_g$} & {PE} && {$E_g$} &{Transitions}  \\
      \hline
Si	& 3.07	&-11.2&&	3.38	&-2.4&&	3.26	&-5.7&&	3.46&$\Gamma\to\Gamma$	\\
	&1.23	&-5.9&&	1.45	&10.8&&	1.17	&-11.0&&	1.31&$\Gamma\to X$	\\
	&2.36	&-4.2&&	2.60	&5.6&&	2.55	&3.6&&	2.46&$\Gamma\to L$	\\[0.1 cm]
GaAs	&	2.19	&39.2&&	2.55	&62.4&&	2.81	&78.9&&	1.57&$\Gamma\to\Gamma$	\\
	&2.20	&-1.3&&	2.44	&9.6&&	2.13	&-4.4&&	2.23&$\Gamma\to X$	\\
	&	2.15	&13.5&&	2.45	&29.7&&	2.43	&28.6&&	1.89&$\Gamma\to L$	\\[0.1 cm]
ZnO		&	1.32	&-63.4&&	1.83	&-49.0&&	3.01	&-16.4&&	3.60&	$\Gamma\to\Gamma$	\\[0.1 cm]
C		&	6.22	&-18.9&&	6.76	&-11.7&&	7.15	&-6.8&&	7.67&	$\Gamma\to\Gamma$	\\[0.1 cm]
MgO		&	5.96	&-28.7&&	6.50	&-22.3&&	8.27	&-1.0&&	8.36	&$\Gamma\to\Gamma$	\\
    \hline\hline
  \end{tabular}
\end{table}

\subsubsection{$d-$ band positions and transition energies}
The position of the
semicore $d-$states for $d-$band solids are important indicators that measure the delocalization error for those secies. The DFT+U method is often used beyond the standard semilocal approximation to mitigate the correct $d-$states behavior. It is also shown that hybrid functionals with dielectric dependence can improve this feature~\cite{WeiGiaRigPas2018}. In Table~\ref{d-position}, we assess the performance of different methods for the average positions of the occupied $d-$band to the Fermi energy. $r$MGGAC and $\omega r$MGGAC tend to underestimate the mean position. However, a consistent improvement from the dielectric-dependent hybrid is observed with MAE $\sim$1.1 eV, which is vis-\"{a}-vis similar to DD-RSH-CAM~\cite{WeiGiaRigPas2018}.
\begin{figure}
  \centering
 \includegraphics[width=6 cm,height=8 cm]{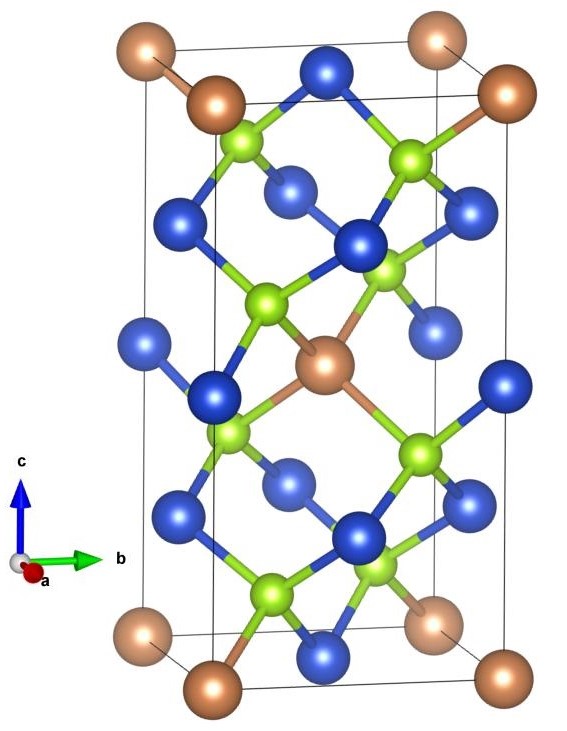}
  \caption{Crystal structure of Cu$_3$(Sb/As)Se$_4$. Violet spheres represent Cu atoms, Maroon spheres represent $Sb/As$ atoms, and LimeGreen spheres represent Se atoms.}
  \label{crystal_structure}
\end{figure}

Regarding the transition energies, earlier studies indicate~\cite{WeiGiaRigPas2018} that the full short-range exchange together with a fraction of long-range exchange $\frac{\alpha[\epsilon_\infty^{-1}]}{ r}$ is important to improve transition energies. Typically, transition energies measure the quality of the orbital energies obtained from gKS calculations. It is measured using energy differences, $\varepsilon_{X}^{occupied}-\varepsilon_{Y}^{un-occupied}$, where $X$ is the occupied band and $Y$ is the unoccupied band. Therefore, one can expect a consistent improvement from semilocal to hybrid because of the lesser de-localization error. We observe this effect in Table~\ref{transition-energy}, where the performance of Si, GaAs, ZnO, C, and MgO are shown. One can note a consistent improvement in transition energies from DD-MGGAC-CAM compared to other methods. However, the percentage error (PE) shows poorer behavior for GaAs where DD-MGGAC-CAM overestimates the experimental band gap. Disregarding this case, the maximum deviation for Si becomes -11\%, -6.8\% for C, -16\% for ZnO, and -1.0\% for MgO, which is comparable with DD-RSH-CAM~\cite{WeiGiaRigPas2018}.

\begin{figure*}
  \centering
 \includegraphics[width=15 cm,height=10 cm]{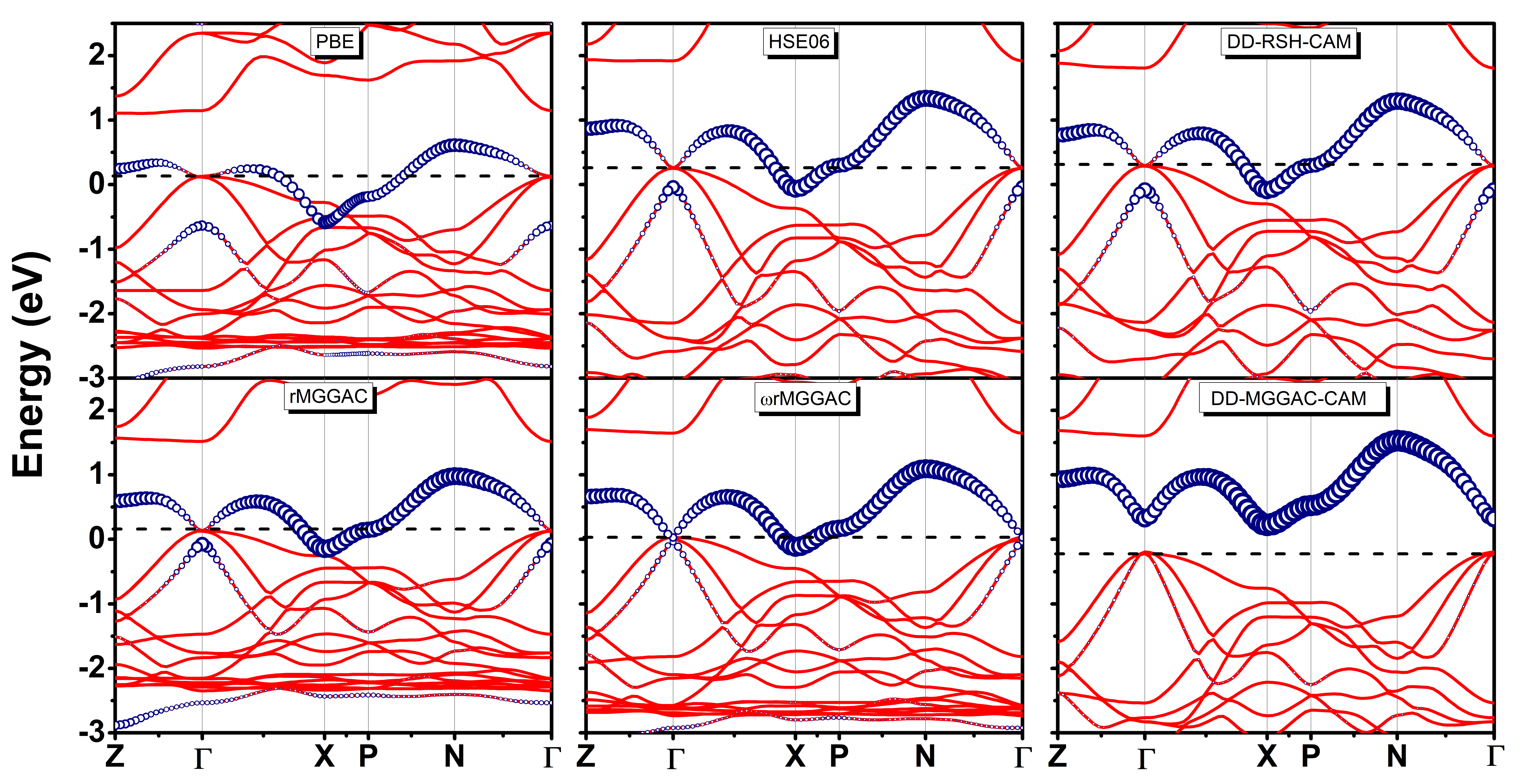}
  \caption{Band structure of Cu$_3$SbSe$_4$ calculated using the different methods. The projection of the band wave functions onto the Sb $s$ states is shown as a circle. The Fermi levels are at zero. The Brillouin zone is sampled with $7\times 7\times 8$ $\Gamma-$centered ${\bf{k}}-$points. 
  The dotted black line corresponds to the valance band maxima.}
  \label{cu3sbse4-band-structure}
\end{figure*}

\begin{figure*}
  \centering
 \includegraphics[width=15 cm,height=10 cm]{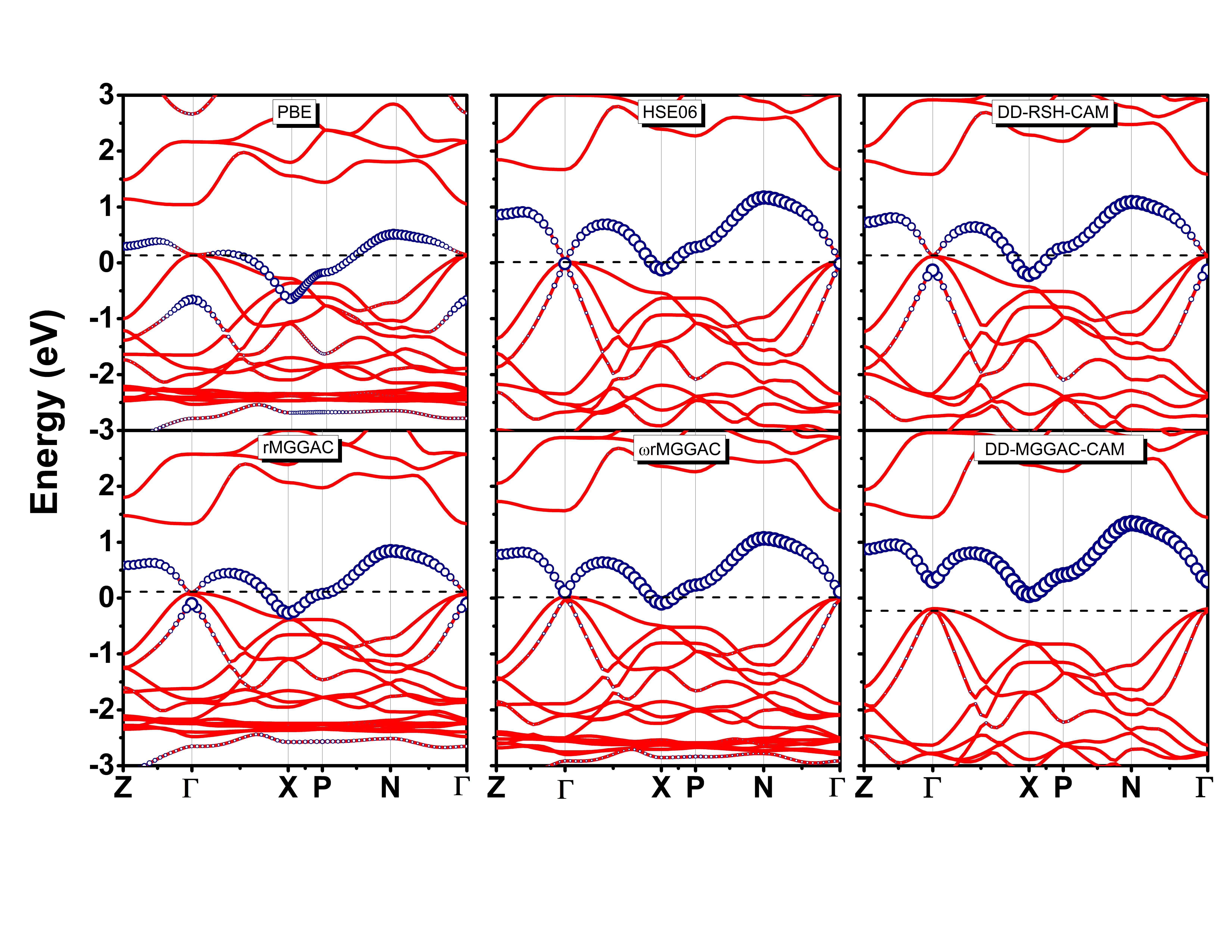}
  \caption{Same as Cu$_3$SbSe$_4$, but the band structure is plotted for Cu$_3$AsSe$_4$. The projection of the band wave functions onto the As $s$ states is shown as a circle.}
  \label{cu3asse4-band-structure}
\end{figure*}

\begin{table}
\begin{center}
\caption{\label{difficult-band-gap} Inversion energies and direct band gaps at $\Gamma-$ point of Cu$_3$SbSe$_4$ and Cu$_3$AsSe$_4$ as obtained from different methods. The direct band gap is calculated
using the difference between the highest occupied state (highest valance band) and
the lowest unoccupied state (lowest
conduction band). All values are in eV.}
\begin{tabular}{lcccccccccccccccccccccccc}
\hline\hline
Methods	&&&&	$\Delta_i$	&&&	$E_g$&&&Topological?			\\
\hline
\multicolumn{11}{c}{Cu$_3$SbSe$_4$}\\
 PBE&&&&-0.80&&&0.00&&&Y\\
 HSE06&&&&-0.64&&&0.00&&&Y\\
 DD-RSH-CAM&&&&-0.25&&&0.00&&&Y\\
 $r$MGGAC&&&&-0.19&&&0.00&&&Y\\
 $\omega r$MGGAC&&&&-0.04&&&0.00&&&Y\\
 DD-MGGAC-CAM&&&&0.49&&&0.49&&&N\\
 Expt.~\cite{Zhang2013Near}&&&& 0.10$-$0.31&&&0.10$-$0.31&&&N\\[0.5 cm]

\multicolumn{11}{c}{Cu$_3$AsSe$_4$}\\
 PBE&&&&-0.75&&&0.00&&&Y\\
 HSE06&&&&0.00&&&0.00&&&N\\
 DD-RSH-CAM&&&&-0.36&&&0.00&&&Y\\
 $r$MGGAC&&&&-0.21&&&0.00&&&Y\\
 $\omega r$MGGAC&&&&0.00&&&0.00&&&N\\
 DD-MGGAC-CAM&&&&0.51&&&0.51&&&N\\
 Expt.~\cite{Zhang2013Near}&&&&0.65$-$0.88&&&0.65$-$0.88&&&N\\[0.5 cm]

\hline\hline
     \end{tabular}
\end{center}
\end{table}

\subsection{Band structure of narrow-gap material Cu$_3$(Sb/As)Se$_4$}

To justify globally such an empirical choice of $\alpha$ in the long-range part, we further apply meta-GGA DDH to assess for narrow gap materials like Cu$_3$SbSe$_4$ and Cu$_3$AsSe$_4$. Note that materials with narrow band gaps have the potential to be used in various fields of applications such as, solar cells and thermoelectric devices ~\cite{Jackson2011New,Walsh2012Kesterite, Kormath2018Database}. Taking into account their potential application,
Cu$_3$(Sb/As)Se$_4$ semiconductors have attracted significant attention due to their promising electronic and optical properties\cite{Goodman1958Prediction,Pamplin1979Spray,Pamplin1965Quinary}. The Cu$_3$SbSe$_4$ have also been theoretically predicted to exhibit unique topological properties. Moreover, Cu$_3$SbSe$_4$ is a direct band gap semiconductor with an absorption onset $\sim 0.20$ eV, indicating its potential as a material for mid-infrared absorption~\cite{Zhang2013Near}.

To fully realize the potential of Cu$-$based materials, a precise understanding of their near-edge electronic properties, including the band gap and band structure topology, is crucial. Here, we investigate two Cu$-$based narrow-gap topological insulators, Cu$_3$SbSe$_4$ and Cu$_3$AsSe$_4$. The Cu$_3$(Sb/As)Se$_4$ compounds are crystallized in the Famatinite structure and belong to the space group I${\bar{4}}$2m, in which the Se atoms are surrounded by three Cu atoms and one Sb/As atom.
The structure is depicted in Fig. \ref{crystal_structure} and characterized by a three-dimensional Cu$-$Se framework composed of distorted [CuSe$_4$] tetrahedra, with a one-dimensional array of inserted [(Sb/As)Se$_4$] tetrahedra. This unique configuration gives rise to two distinct Cu sites, leading to variations in Cu$-$Se bond lengths~\cite{Garcia2018Thermoelectric}. The structural complexity of these multinary semiconductors poses significant challenges to the accurate theoretical prediction of their electronic properties. Conventional applications of density functional theory (DFT) within the local density approximation (LDA) or generalized gradient approximation (GGA) often lead to incorrect predictions of the band gap and band ordering, which can result in false identification of metal~\cite{Vidal2011False}. Further, $G_0W_0$ calculations also encounter difficulties for these narrow-gap systems. Recently, it has been demonstrated that the modified Becke-Johnson potential (mBJ) and mBJ+$U$ can give reasonably accurate band gaps for moderate gaps systems~\cite{Zhang2013Near}. However, it is also shown in ref.~\cite{Zhang2013Near} that applying mBJ only potential to Cu$-$based multinary semiconductors still substantially underestimates their band gaps. For instance, the mBJ method incorrectly predicts the metallic band structure of Cu$_3$(Sb/AS)Se$_4$.

In Fig.~\ref{cu3sbse4-band-structure} and Fig.~\ref{cu3asse4-band-structure}, we
present a comparison of the band structures for Cu$_3$SbSe$_4$ and Cu$_3$AsSe$_4$, calculated using (a) PBE, (b) HSE06, (c) DD-RSH-CAM, (d) $r$MGGAC, (e) $\omega r$MGGAC, and (f) DD-MGGAC-CAM methods. Here, all calculations are performed with PBE-optimized geometries. Key features of the band structures have been discussed below. For Cu$_3$(Sb/As)Se$_4$, the (Sb/As) $5s$ and Se $4p$ states mostly contribute to the conduction band minimum (CBM), whereas the Cu $3d$ and Se $4p$ states largely impact the electrical structure close to the valence band maximum (VBM). We depict the band structure with the projection of (Sb/As) 5s state. Size of the bubbles represent the contribution of (Sb/As) 5s state in the band structure. In the case of PBE calculations, the band structure shows notable overlap between the valence and conduction bands. Specifically, the $s-$like state lies below the degenerate $p(d)-$like states in the valence band at the $\Gamma$ point, which results in negative inversion energy defined as $\Delta_i=\varepsilon_s-\varepsilon_{p,d}$~\cite{Vidal2011False} (-0.80 eV for Cu$_3$SbSe$_4$ and -0.75 eV for Cu$_3$AsSe$_4$ as referred to Table~\ref{difficult-band-gap}). Hence, within PBE-level theory, these materials are identified as topological semimetals with excitation gap $E_g=0$ and $\Delta_i<0$. Here, the actual conduction band minimum (CBM) is identified by projecting the wave function onto the Sb/As $s$ orbitals, as illustrated in Fig.~\ref{cu3sbse4-band-structure} and Fig.~\ref{cu3asse4-band-structure}. The lowest conduction band (highlighted in blue) is particularly important as it influences the electronic and optical properties of the material. Additionally, the lowest conduction band is strongly coupled to the valence bands around the $X$ point. As shown in Table~\ref{difficult-band-gap} and Fig.~\ref{cu3sbse4-band-structure}, calculations based on the HSE06 and DD-RSH-CAM method do not change the tendencies of $E_g=0$ and $\Delta_i<0$. 
For Cu$_3$SbSe$_4$, $\Delta_i$ decreases from -0.80 eV (PBE) to -0.64 eV (HSE06) and -0.25 eV (DD-RSH-CAM). A similar trend is observed in Cu$_3$AsSe$_4$ (Fig.~\ref{cu3asse4-band-structure}), where the $\Delta_i$,  decreases from -0.75 eV (PBE) to -0.64 eV (HSE06) and -0.36 eV (DD-RSH-CAM). The negative inversion energy (within PBE) and the symmetry-protected degeneracy of the $p(d)-$like valence bands suggest Cu$_3$(Sb/As)Se$_4$ as a non-trivial topological semimetal with $\mathbb{Z}_2 = -1$~\cite{Zhang2013Near,Wang2011Topological,Vidal2011False}. This non-trivial topological semimetallic structure remains unchanged within the HSE06 and DD-RSH-CAM methods, as shown in Fig.~\ref{cu3sbse4-band-structure} and Fig.~\ref{cu3asse4-band-structure}. Next, we examine the band structures of the Cu$_3$(Sb/As)Se$_4$ systems using the $r$MGGAC, $\omega r$MGGAC, and DD-MGGAC-CAM methods. The $r$MGGAC and $\omega r$MGGAC methods also predict non-trivial topological semimetallic behavior for both systems (Fig.~\ref{cu3sbse4-band-structure} and Fig.~\ref{cu3asse4-band-structure}). However, in the DD-MGGAC-CAM method, the contribution of Sb/As 5s state in the valance band disappears, providing a correct semiconductor band structure (Fig.~\ref{cu3sbse4-band-structure}), where the band gap is slightly indirect, with a direct gap of 0.49 eV and an indirect gap from $\Gamma$ to $X$ of 0.39 eV. The band gap calculated by DD-MGGAC-CAM agrees with experimental observations ($E_g\sim 0.10-0.31$ eV)~\cite{Zhang2013Near}. As Zhang et al.~\cite{Zhang2013Near} reported, the spin-orbit coupling (SOC) effect in Cu$_3$SbSe$_4$ can reduce the band gap by $\sim 0.09$ eV. Hence, the direct and indirect band gap accounting SOC become 0.40 eV and 0.30 eV, respectively. The band gap with SOC remains within the range of experimentally observed values (see Table I of Zhang et al.~\cite{Zhang2013Near} and references therein). Indirect band gaps are also predicted by the DD-MGGAC-CAM method for Cu$_3$AsSe$_4$ systems (Fig.~\ref{cu3asse4-band-structure}). As shown in Table~\ref{difficult-band-gap}, this system's direct band gap at $\Gamma-$point becomes 0.51, falls within the experimental gap of 0.65$-$0.88 eV~\cite{Zhang2013Near}. The formation of a band gap in these materials leads to the realization of a band insulator with $\Delta_i>0$ 
as predicted by the DD-MGGAC-CAM method. Hence, the false negative band inversion is due to underestimations in band gaps ipenof the DFT methods used in ab-initio calculations, which is corrected by DD-MGGAC-CAM (shown in present work) and mBJ+U (shown in ref.~\cite{Zhang2013Near}).

\begin{figure*}
  \centering
 \includegraphics[scale=0.5]{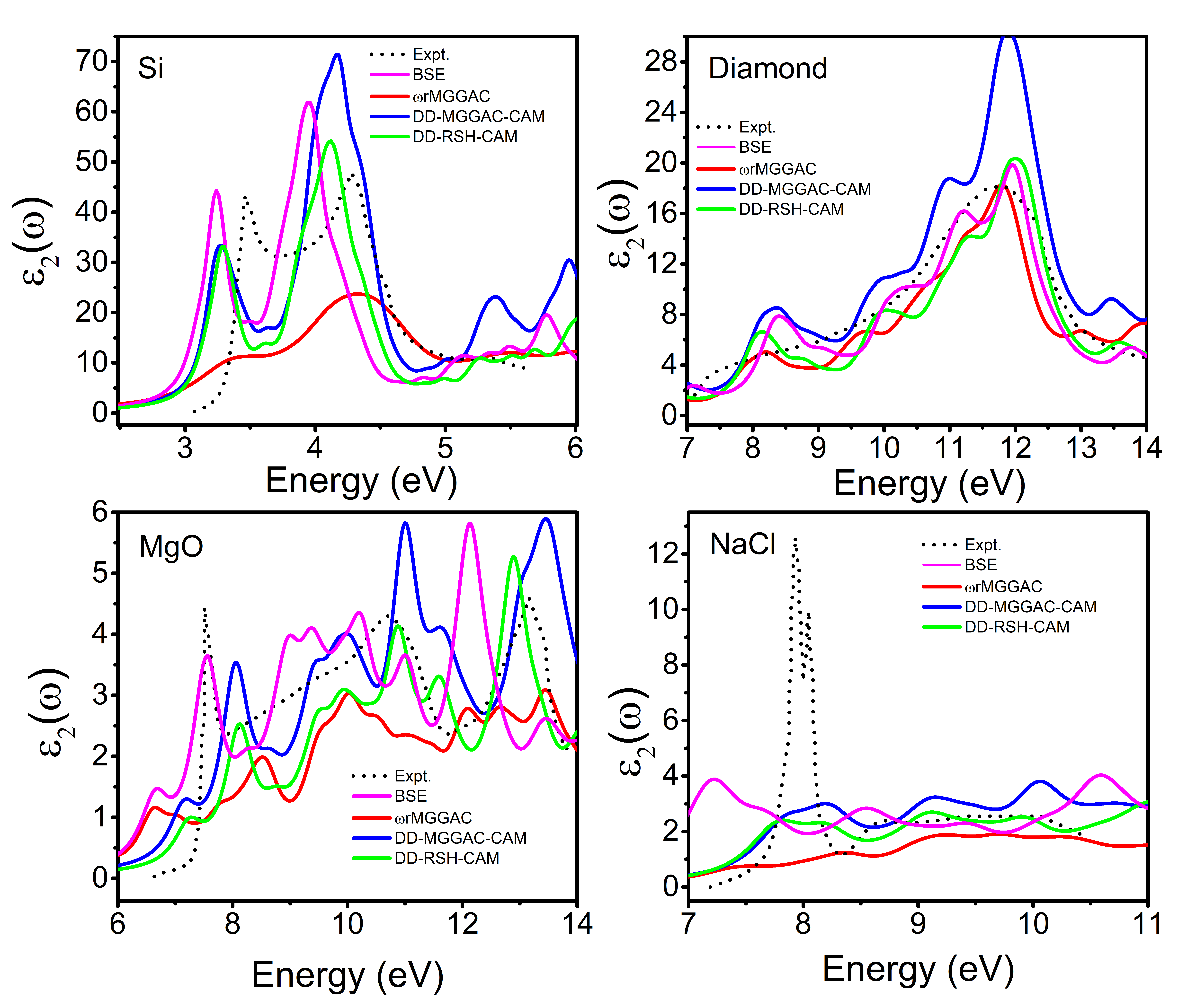}
  \caption{Imaginary part of the frequency-dependent optical absorption spectra, $\epsilon_2(\omega)$ of Si,
C, MgO, and NaCl were calculated using different methods. We use a Gaussian broadening $\sigma$ = 0.1 for Si and  $\sigma$ = 0.3 for C, MgO, and NaCl. Experimental spectra are from Ref.~\cite{BIRKEN1998279} (for Si), Ref.~\cite{LogoLautCard1986} (for C), Ref.~\cite{RoesWalk1968} (for NaCl), Ref.~\cite{BortFrenJone1990} (for MgO).}
  \label{absorption-spectra}
\end{figure*}

\begin{table}
\begin{center}
\caption{\label{exciton} Positions of the first few excitation peaks at BSE and TDDFT (in eV). the experimental values are extracted from references mentioned in Fig.~\ref{absorption-spectra}.}
\begin{tabular}{lcccccccccccccccccccccccc}
\hline\hline
Solids	&	BSE	&	DD-MGGAC-CAM&DD-RSH-CAM&Expt.\\
\hline
Si	&3.2&3.3&3.3&$\sim$3.7 \\
C	&11.9&11.9&12.0&$\sim$11.7\\
MgO&7.5&8.0&8.1&$\sim$7.5\\
NaCl	&7.2&7.8&7.8&$\sim$7.9	\\
&7.7&8.2&8.1&$\sim$8.0\\
\hline\hline
     \end{tabular}
\end{center}
\end{table}

\begin{figure}
  \centering
 \includegraphics[width=8 cm,height=10 cm]{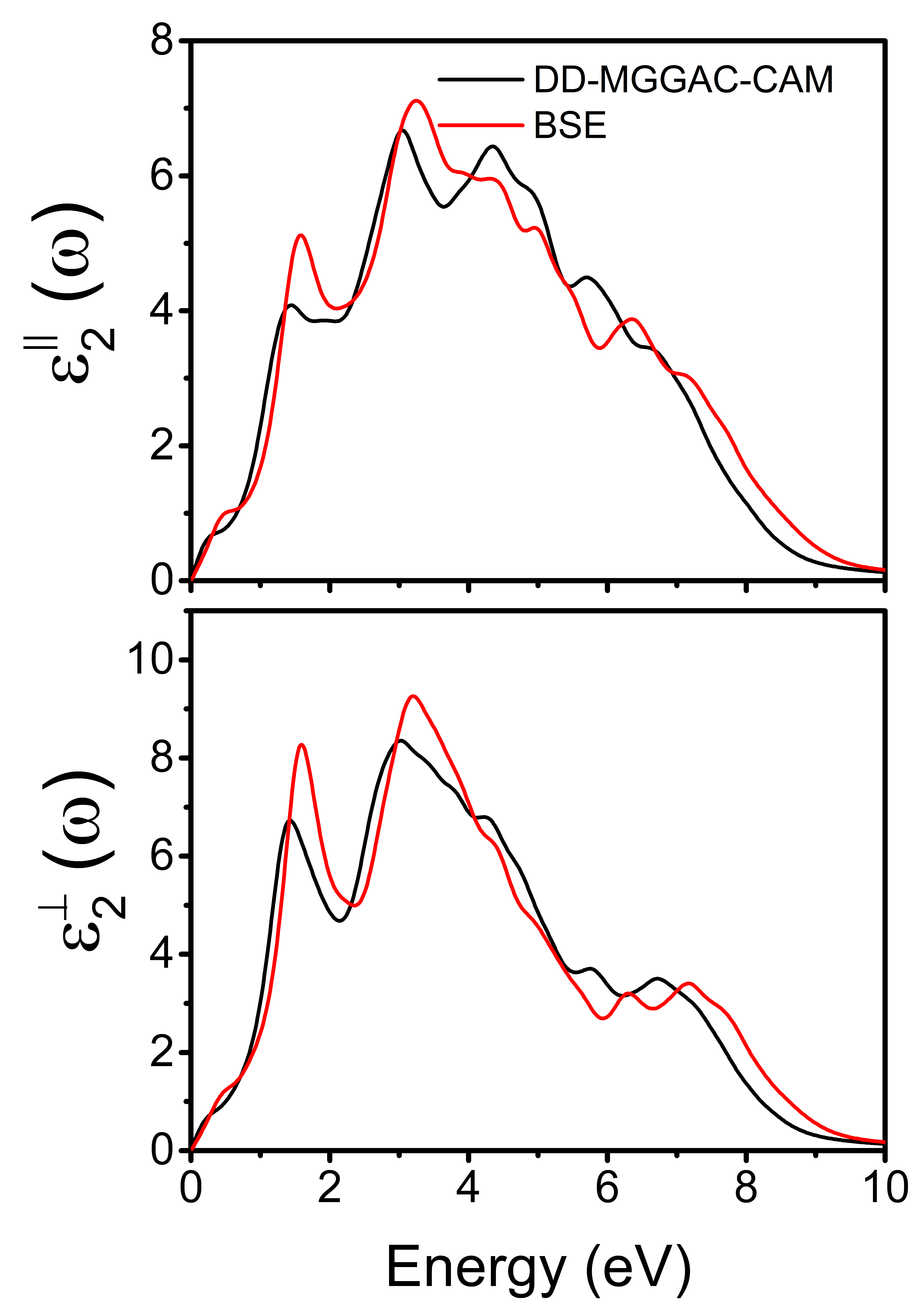}
  \caption{Imaginary part of the frequency-dependent optical absorption spectra for Cu$_3$SbSe$_4$. We use broadening, $\sigma=0.3$. See main text for $\parallel$ and $\perp$ component of $\epsilon_2$.}
  \label{absorption-spectra-cu3asse4}
\end{figure}

\subsection{Optical and excitonic properties}
The optical absorption spectra of solids with linear-response time-dependent DFT (LR-TDDFT)~\cite{Sander2015beyond,Sander2017macroscopic} is mostly calculated by solving the Cassida equation~\cite{cassida-equation}. For hybrid functionals, the gKS orbitals and eigenvalues are first obtained from ground-state calculations and then used to solve the Cassida equation. For optical spectra the BSE@$scGW$ is considered as the state-of-the-art method~\cite{Onida2002electronic}, however, the self-consistent $GW$ (sc$GW$) suffers from computational expenses. As an alternative, solving a ground-state problem with DDH followed by the TDDFT is a useful approach known as the time-dependent DDH~\cite{tal2020accurate}.

Typically, within the Tamm-Dancoff approximation~\cite{cassida-equation,Sander2015beyond,Sander2017macroscopic,tal2020accurate}, the amplitude of the TDDFT spectrum for meta-GGA DDHs is obtained by solving the following matrix elements for excitation and de-excitation~\cite{tal2020accurate},
\begin{eqnarray}
 \Omega_{ij;ab}=\omega_{ia}\delta_{ij;ab}+\langle ib|K_{Hxc}|aj\rangle~,  \label{casida}
\end{eqnarray}
where $\omega_{ia}=\varepsilon_i-\varepsilon_a$ and
\begin{eqnarray}
\langle ib|K_{Hxc}|aj\rangle &=& 2\langle ib|V_H(|{\bf{q+G}}|)|aj\rangle\nonumber\\
&-& \langle ib|f_{xc}^{non-local}|ja\rangle+\langle ib|f_{xc}^{meta-GGA}|aj\rangle\nonumber\\
\end{eqnarray}
with $i,j\in$ unoccupied states, $a,b\in$ occupied states, and $V_H(|{\bf{q+G}}|)$ is the typical Hartree potential. Regarding other terms, those are given by~\cite{tal2020accurate}
\begin{eqnarray}
\langle ib|f_{xc}^{non-local}|ja\rangle &=& \langle ib|\varepsilon^{-1}(|{\bf{q+G}}|)V(|{\bf{q+G}}|)|ja\rangle
\label{kernels}
\end{eqnarray}
and $f_{xc}^{meta-GGA}$ is the adiabatic meta-GGA exchange-correlation kernel, which also includes non-locality or ultra-nonlocality effects~\cite{Nazarov2011Optics}.  One may also note that the exact form of the meta-GGA kernel $f_{xc}^{meta-GGA}({\bf{r}},{\bf{r}'})$ is given by Eq. (5) of Ref.~\cite{Nazarov2011Optics}. Whereas, for GGA $f_{xc}^{GGA}({\bf{r}},{\bf{r}'})$, it is given by Eq.(17) of Ref.~\cite{Singh2019Adiabatic}. However, the investigation of such kernels is beyond the scope of the present paper and has not been implemented in the present version of VASP. Therefore, we consider the derivative involving the Eq.~(\ref{kernels}) with respect to the gradient of density or KS kinetic energy density becomes zero, i.e. within adiabatic local density approximation (ALDA),
\begin{eqnarray}
 f_{xc}^{meta-GGA}\approx f_{xc}^{ALDA}=\frac{\delta^2(E_c^{DFT}-(1-\epsilon_\infty^{-1})E_x^{DFT})}{\delta n({\bf{r}})\delta n({\bf{r}}')}
\end{eqnarray}
This real space representation is valid for DD-MGGAC-CAM (or DD-RSH-CAM~\cite{WeiGiaRigPas2018}), and we restrict ourselves only at ALDA.

Solving Eq.~(\ref{casida}), one obtains energy eigenvalues spectrum with an isolated eigenvalue. This isolated eigenvalue is energetically less than the ionization potential (IP) or direct band gap obtained from DFT calculations. Also noted, both non-local and ALDA kernels are frequency-independent.

Mathematically, for the short-range screened hybrid, $\omega r$MGGAC, the dielectric function $\varepsilon^{-1}(|{\bf{q+G}}|)$ goes to $\beta$ as $q\to 0$, which is a constant and having no screening dependence. However, the correct asymptotic screening-dependent behavior for bulk systems should vary as $\epsilon_\infty^{-1}$ as $q\to 0$. This is respected by DDH, which is the key to improving the optical absorption spectra of the bulk solids~\cite{Ullrichtddft,paier2008dielectric,Yang2015simple,wing2019comparing,stadele1999exact,petersilka1996excitation,kim2002excitonic,Sun2020optical,sun2020lowcost,Kootstra2000application}.

To measure the quality of different methods, we calculate the imaginary part of the frequency-dependent optical absorption spectra  ($\epsilon_2(\omega)$) in the optical limit of the small wave vector of the macroscopic dielectric function $\epsilon^M$ for Si, C, MgO, and NaCl. The plotted quantity, $\epsilon_2=\lim_{q\rightarrow 0}\epsilon^M(q,\omega)$ is given by the Eq. (48) of Ref.~\cite{Sander2017macroscopic}, which is also related to the calculated matrix elements of Eq.~(\ref{casida}). Within the considered meta-GGA methods, $\omega r$MGGAC, and DD-MGGAC-CAM results are reported along with experimental, BSE, DD-RSH-CAM. Unless otherwise stated, the BSE is calculated at the top of the $G_0W_0$@PBE eigen-system. All the results are plotted in Fig.~\ref{absorption-spectra}. In Table~\ref{exciton}, we also calculate the exciton peak positions
with BSE and TDDFT.

Considering the optical absorption spectra of Si as shown in Fig.~\ref{absorption-spectra}, realistic spectra are obtained from various methods. The first peak, which denotes the oscillator strength of optical transition, underestimates for $\omega r$MGGAC. 
Meanwhile, the DD-MGGAC-CAM and DD-RSH-CAM spectra are closer to the BSE and slightly left-shifted than the experimental spectrum.  As shown in Table~\ref{absorption-spectra}, the BSE peak appears at $3.2$ eV, close to the DD-MGGAC-CAM, which appears at 3.3 eV. 
A similar tendency of results is also obtained for another semiconductor structure diamond, where the bright exciton peak of BSE and experimental are significantly closer to DD-MGGAC-CAM. 
Regarding two insulators, MgO and NaCl, the DDH peak amplitude is mostly underestimated. The BSE's first bright peak for MgO appeared at 7.5 eV, considerably closer to the experimental. However, the DD-RSH-CAM and DD-MGGAC-CAM peak positions slightly deviate from the experimental. For NaCl, the BSE shows a slight left shift. The oscillator strength amplitude for BSE and TDDFT spectra shows considerable underestimations compared to experimental spectra. The DD-MGGAC-CAM values are closer to experimental than DD-RSH-CAM. Overall accurate results of DD-MGGAC-CAM conclude that this method can be used for realistic TDDFT calculations.

Further, the usefulness of the DD-MGGAC-CAM is also illustrated from the optical absorption spectra of Cu$_3$SbSe$_4$. As discussed, this system appears metal for all methods except TD-DD-MGGAC-CAM. Hence, because of its band gap opening, only the realistic spectra from DD-MGGAC-CAM can be expected. We calculate the absorption spectra of this system from TD-DD-MGGAC-CAM and BSE (calculated on the top of $G_0W_0$@DD-MGGAC-CAM)  in Fig.~\ref{absorption-spectra-cu3asse4} for the light-polarized perpendicularly to the $c-$axis (denoted by $\epsilon_2^{\perp}(\omega)=\frac{\epsilon_{xx}(\omega)+\epsilon_{yy}(\omega)}{2}$) and along the $c-$axis ($\epsilon_2^{\parallel}(\omega)=\epsilon_{zz}(\omega)$). Both the obtained spectra are realistic and close, especially the BSE results indicate realistic $G_0W_0$ calculation is only possible on top of the DD-MGGAC-CAM for such systems. For both the systems, the self-consistent value of $\epsilon_\infty^{-1}$ becomes
$\sim$0.10 for DD-MGGAC-DDH is significantly larger than the almost zero counterpart obtained from DD-RSH-CAM.

\section{Conclusions}

In this paper, we develop a dielectric and meta-GGA ingredients-dependent range-separated hybrid functional for quantum materials. The construction is based on very simple the MGGAC exchange hole.
In contrary to GGA construction, the meta-GGA functionals allow the inclusion of several exact conditions, which is the main motivation for further incorporating it into the generalized dielectric-dependent framework. We test the functional performance of various quantum material properties, including some challenging cases, such as Cu$-$based quantum materials.

Based on that assessment, we were able to draw some main conclusions regarding meta-RS-DDH functional construction, namely
($i$) overall, we obtained satisfactory performance from the constructed functional for the band gap of solids with MAE $\sim 0.36$ eV; ($ii$) we also obtained good performance for the $d-$band position and transition energies; ($iii$) a reasonably good performance is also obtained for the optical absorption spectra of semiconductors and insulators; ($iv$) the present method also successfully resolves the well-known ``band-gap problem'' of Cu$_3$SbSe$_4$ and Cu$_3$AsSe$_4$ and correctly captures the optical absorption spectra of those materials. The overall accurate results suggest it may be an efficient method for single-shot $G_0W_0$ and BSE calculations for such materials. Based on its overall accuracy, this method can also be considered an all-in-one method for material band gaps, including general-purpose and challenging solids. It can be successfully applied for high-throughput or machine-learning-based material searching for potential candidates for topological insulators.

\section*{Acknowledgements}

S.Ś. thanks to the Polish National Science Center for the partial financial support under
Grant No. 2020/37/B/ST4/02713.

\appendix
\onecolumngrid

\section{Details of the MGGAC and $\omega$MGGAC exchange and potential energy functional}
\label{mggac-details}

In the generalized Kohn-Sham (gKS) framework, the differential operator associated with the meta-GGA potential is given by~\cite{sun2011selfconsistent,Doumont2022Implementation},
\begin{equation}
v_{x}^{MGGAC}({\bf{r}})\to\Big[\frac{\partial (n({\bf{r}})\epsilon_{x}^{MGGAC}({\bf{r}}))}{\partial n({\bf{r}})}-\nabla\frac{\partial (n({\bf{r}})\epsilon_{x}^{MGGAC}({\bf{r}}))}{\partial \nabla n({\bf{r}})}\Big]-\frac{1}{2}\nabla.\{\frac{\partial (n({\bf{r}})\epsilon_{x}^{MGGAC}({\bf{r}}))}{\partial \tau({\bf{r}})} \nabla\}~,
\end{equation}
where $\epsilon_{x}^{MGGAC}({\bf{r}}) = \epsilon_{x}^{LDA}({\bf{r}}) F_x^{MGGAC}[\alpha^{iso}]$ represents the MGGAC exchange energy per particle, with $\epsilon_{x}^{LDA}({\bf{r}})$ being the exchange energy per particle within the local density approximation. Following Ref.~\cite{patra2019relevance}, the form of MGGAC exchange enhancement factor, $F_x^{MGGAC}[\alpha^{iso}]$ is given by~\cite{patra2019relevance}
\begin{eqnarray}
F_x^{MGGAC}[x[\alpha^{iso}]] = \frac{1}{6}\Big(\frac{2\pi^2}{3}\Big)^{1/3}\frac{e^{-2x/3}(-x^2-5x+8e^x-8)}{x(1+x)^{1/3}}~,
\end{eqnarray}
where $x$ satisfies a non-linear equation~\cite{patra2019relevance}
\begin{eqnarray}
\frac{(1+x)^{5/3}}{x-3}e^{-2x/3}=\frac{\beta_1+\beta_2\alpha^{iso}+\beta_3(\alpha^{iso})^2}{1+\beta_4+\beta_5(\alpha^{iso})^2}~,
\end{eqnarray}
with $\beta_1 = 3.712, \beta_2 = 2.0, \beta_4 = 0.1,
\beta_3 = 2.595 + 0.5197\beta_4 + 0.559\beta_2$, and $\beta_5 = -3\beta_3$.

On the other hand, the differential operator associated with the range-separated MGGAC exchange potential is given by,
\begin{equation}
v_{x}^{\omega MGGAC}({\bf{r}})\to\Big[\frac{\partial (n({\bf{r}})\epsilon_{x}^{\omega MGGAC}({\bf{r}}))}{\partial n({\bf{r}})}-\nabla\frac{\partial (n({\bf{r}})\epsilon_{x}^{\omega MGGAC}({\bf{r}}))}{\partial \nabla n({\bf{r}})}\Big]-\frac{1}{2}\nabla.\{\frac{\partial (n({\bf{r}})\epsilon_{x}^{\omega MGGAC}({\bf{r}}))}{\partial \tau({\bf{r}})} \nabla\}~,
\end{equation}
where $\epsilon_{x}^{\omega MGGAC}({\bf{r}}) = \epsilon_{x}^{LDA}({\bf{r}}) F_x^{\omega MGGAC}[\alpha^{iso},\omega,k_F]$, with $F_x^{\omega MGGAC}[\alpha^{iso},\omega,k_F]$ representing the short-range part of the range-separated enhancement factor, given by~\cite{Jana2022Solid}
\begin{eqnarray}
F_x^{\omega MGGAC}(\alpha^{iso},\omega,k_F)&=&\mathcal{A}-\frac{4}{9}\frac{\mathcal{B}}{\lambda}(1-\chi)
-\frac{4}{9}\frac{\mathcal{C}(\alpha^{iso})}{\lambda^2}(1-\frac{3}{2}\chi+\frac{1}{2}\chi^2)-\frac{8}{9}\frac{\mathcal{G}(\alpha^{iso})}
{\lambda^3}(1-\frac{15}{8}\chi+\frac{5}{4}\chi^3-\frac{3}{8}\chi^5)\nonumber\\
&+&2\nu(\sqrt{\zeta+\nu^2}-\sqrt{\eta+\nu^2})+2\zeta\ln(\frac{\nu+\sqrt{\zeta+\nu^2}}{\nu+\sqrt{\lambda+\nu^2}})
-2\eta\ln(\frac{\nu+\sqrt{\eta+\nu^2}}{\nu+\sqrt{\lambda+\nu^2}})~,\nonumber\\
\label{eq20}
\end{eqnarray}
where $\mathcal{A} = 0.757211$, $\mathcal{B} = -0.106364$,
\begin{equation}
\mathcal{C}(\alpha^{iso})=\frac{1}{2}[\frac{1}{5}+a_1\frac{\alpha^{iso}(\alpha^{iso}-1)}{1+(\alpha^{iso})^2}]-\frac{1}{2}\mathcal{H}(\alpha^{iso})-
\frac{9}{8}\mathcal{A}\mathcal{D}^2+\mathcal{B}\mathcal{D}+
\frac{3}{8}\mathcal{A}^3,
\label{eqq3}
\end{equation}
and
\begin{eqnarray}
\mathcal{G}(\alpha^{iso})&=&-\frac{2}{15}\Big(18\sqrt{\mathcal{H}(\alpha^{iso})}\lambda^{7/2}+
6\sqrt{\pi}\lambda^{7/2}+9\mathcal{A}\lambda^3-18\sqrt{\eta}\lambda^{7/2}
+2\mathcal{B}\lambda^2+3\mathcal{C}(\alpha^{iso})\lambda\Big)~.
\label{eqq4}
\end{eqnarray}
Other factors are given by, $\nu=\omega/k_F$, $k_F=(3\pi^2n)^{1/3}$, $\chi=\frac{\nu}{\sqrt{\lambda+\nu^2}}$, $\eta=\mathcal{A}+\mathcal{H}(\alpha^{iso})$, $\lambda=\mathcal{D}+\mathcal{H}(\alpha^{iso})$, $\zeta=\mathcal{H}(\alpha^{iso})$, and
\begin{equation}
\mathcal{H}(\alpha^{iso})=\frac{(\alpha^{iso}-1)^3}{(1+(\alpha^{iso})^3)}\frac{(-c+a_2\alpha^{iso}+c
a_3(\alpha^{iso})^2)}{(1+a_4\alpha^{iso}+a_3(\alpha^{iso})^2)},
\label{eqq5}
\end{equation}
where $\mathcal{D} = 0.609650$, $a_1=2.2$, $c=0.033$, $a_2=0.03999$, $a_3=0.10368$, and $a_4=1.03684$.

\twocolumngrid
\bibliography{reference.bib}

\begin{thebibliography}{123}%
\makeatletter
\providecommand \@ifxundefined [1]{%
 \@ifx{#1\undefined}
}%
\providecommand \@ifnum [1]{%
 \ifnum #1\expandafter \@firstoftwo
 \else \expandafter \@secondoftwo
 \fi
}%
\providecommand \@ifx [1]{%
 \ifx #1\expandafter \@firstoftwo
 \else \expandafter \@secondoftwo
 \fi
}%
\providecommand \natexlab [1]{#1}%
\providecommand \enquote  [1]{``#1''}%
\providecommand \bibnamefont  [1]{#1}%
\providecommand \bibfnamefont [1]{#1}%
\providecommand \citenamefont [1]{#1}%
\providecommand \href@noop [0]{\@secondoftwo}%
\providecommand \href [0]{\begingroup \@sanitize@url \@href}%
\providecommand \@href[1]{\@@startlink{#1}\@@href}%
\providecommand \@@href[1]{\endgroup#1\@@endlink}%
\providecommand \@sanitize@url [0]{\catcode `\\12\catcode `\$12\catcode
  `\&12\catcode `\#12\catcode `\^12\catcode `\_12\catcode `\%12\relax}%
\providecommand \@@startlink[1]{}%
\providecommand \@@endlink[0]{}%
\providecommand \url  [0]{\begingroup\@sanitize@url \@url }%
\providecommand \@url [1]{\endgroup\@href {#1}{\urlprefix }}%
\providecommand \urlprefix  [0]{URL }%
\providecommand \Eprint [0]{\href }%
\providecommand \doibase [0]{https://doi.org/}%
\providecommand \selectlanguage [0]{\@gobble}%
\providecommand \bibinfo  [0]{\@secondoftwo}%
\providecommand \bibfield  [0]{\@secondoftwo}%
\providecommand \translation [1]{[#1]}%
\providecommand \BibitemOpen [0]{}%
\providecommand \bibitemStop [0]{}%
\providecommand \bibitemNoStop [0]{.\EOS\space}%
\providecommand \EOS [0]{\spacefactor3000\relax}%
\providecommand \BibitemShut  [1]{\csname bibitem#1\endcsname}%
\let\auto@bib@innerbib\@empty
\bibitem [{\citenamefont {Kohn}\ and\ \citenamefont
  {Sham}(1965)}]{kohn1965self}%
  \BibitemOpen
  \bibfield  {author} {\bibinfo {author} {\bibfnamefont {W.}~\bibnamefont
  {Kohn}}\ and\ \bibinfo {author} {\bibfnamefont {L.~J.}\ \bibnamefont
  {Sham}},\ }\bibfield  {title} {\bibinfo {title} {Self-consistent equations
  including exchange and correlation effects},\ }\href@noop {} {\bibfield
  {journal} {\bibinfo  {journal} {Phys. Rev.}\ }\textbf {\bibinfo {volume}
  {140}},\ \bibinfo {pages} {A1133} (\bibinfo {year} {1965})}\BibitemShut
  {NoStop}%
\bibitem [{\citenamefont {Hohenberg}\ and\ \citenamefont
  {Kohn}(1964)}]{hohenberg1964inhomogeneous}%
  \BibitemOpen
  \bibfield  {author} {\bibinfo {author} {\bibfnamefont {P.}~\bibnamefont
  {Hohenberg}}\ and\ \bibinfo {author} {\bibfnamefont {W.}~\bibnamefont
  {Kohn}},\ }\bibfield  {title} {\bibinfo {title} {Inhomogeneous electron
  gas},\ }\href@noop {} {\bibfield  {journal} {\bibinfo  {journal} {Phys.
  Rev.}\ }\textbf {\bibinfo {volume} {136}},\ \bibinfo {pages} {B864} (\bibinfo
  {year} {1964})}\BibitemShut {NoStop}%
\bibitem [{\citenamefont {Burke}(2012)}]{burke2012perspective}%
  \BibitemOpen
  \bibfield  {author} {\bibinfo {author} {\bibfnamefont {K.}~\bibnamefont
  {Burke}},\ }\bibfield  {title} {\bibinfo {title} {Perspective on density
  functional theory},\ }\href@noop {} {\bibfield  {journal} {\bibinfo
  {journal} {J. Chem. Phys.}\ }\textbf {\bibinfo {volume} {136}},\ \bibinfo
  {pages} {150901} (\bibinfo {year} {2012})}\BibitemShut {NoStop}%
\bibitem [{\citenamefont {Engel}\ and\ \citenamefont
  {Dreizler}(2013)}]{engel2013density}%
  \BibitemOpen
  \bibfield  {author} {\bibinfo {author} {\bibfnamefont {E.}~\bibnamefont
  {Engel}}\ and\ \bibinfo {author} {\bibfnamefont {R.~M.}\ \bibnamefont
  {Dreizler}},\ }\href@noop {} {\emph {\bibinfo {title} {Density functional
  theory}}}\ (\bibinfo  {publisher} {Springer},\ \bibinfo {year}
  {2013})\BibitemShut {NoStop}%
\bibitem [{\citenamefont {Jones}(2015)}]{Jones2015}%
  \BibitemOpen
  \bibfield  {author} {\bibinfo {author} {\bibfnamefont {R.~O.}\ \bibnamefont
  {Jones}},\ }\bibfield  {title} {\bibinfo {title} {Density functional theory:
  Its origins, rise to prominence, and future},\ }\href@noop {} {\bibfield
  {journal} {\bibinfo  {journal} {Rev. Mod. Phys.}\ }\textbf {\bibinfo {volume}
  {87}},\ \bibinfo {pages} {897} (\bibinfo {year} {2015})}\BibitemShut
  {NoStop}%
\bibitem [{\citenamefont {Cohen}\ \emph {et~al.}(2012)\citenamefont {Cohen},
  \citenamefont {Mori-Sánchez},\ and\ \citenamefont
  {Yang}}]{CoheMoriYang2012}%
  \BibitemOpen
  \bibfield  {author} {\bibinfo {author} {\bibfnamefont {A.~J.}\ \bibnamefont
  {Cohen}}, \bibinfo {author} {\bibfnamefont {P.}~\bibnamefont
  {Mori-Sánchez}},\ and\ \bibinfo {author} {\bibfnamefont {W.}~\bibnamefont
  {Yang}},\ }\bibfield  {title} {\bibinfo {title} {Challenges for density
  functional theory},\ }\href@noop {} {\bibfield  {journal} {\bibinfo
  {journal} {Chemical Reviews}\ }\textbf {\bibinfo {volume} {112}},\ \bibinfo
  {pages} {289} (\bibinfo {year} {2012})}\BibitemShut {NoStop}%
\bibitem [{\citenamefont {Hasnip}\ \emph {et~al.}(2014)\citenamefont {Hasnip},
  \citenamefont {Refson}, \citenamefont {Probert}, \citenamefont {Yates},
  \citenamefont {Clark},\ and\ \citenamefont {Pickard}}]{HasnRefsProb2011}%
  \BibitemOpen
  \bibfield  {author} {\bibinfo {author} {\bibfnamefont {P.~J.}\ \bibnamefont
  {Hasnip}}, \bibinfo {author} {\bibfnamefont {K.}~\bibnamefont {Refson}},
  \bibinfo {author} {\bibfnamefont {M.~I.~J.}\ \bibnamefont {Probert}},
  \bibinfo {author} {\bibfnamefont {J.~R.}\ \bibnamefont {Yates}}, \bibinfo
  {author} {\bibfnamefont {S.~J.}\ \bibnamefont {Clark}},\ and\ \bibinfo
  {author} {\bibfnamefont {C.~J.}\ \bibnamefont {Pickard}},\ }\bibfield
  {title} {\bibinfo {title} {Density functional theory in the solid state},\
  }\href@noop {} {\bibfield  {journal} {\bibinfo  {journal} {Philosophical
  Transactions of the Royal Society A: Mathematical, Physical and Engineering
  Sciences}\ }\textbf {\bibinfo {volume} {372}},\ \bibinfo {pages} {20130270}
  (\bibinfo {year} {2014})}\BibitemShut {NoStop}%
\bibitem [{\citenamefont {K{\"u}mmel}\ and\ \citenamefont
  {Kronik}(2008)}]{kummel2008orbital}%
  \BibitemOpen
  \bibfield  {author} {\bibinfo {author} {\bibfnamefont {S.}~\bibnamefont
  {K{\"u}mmel}}\ and\ \bibinfo {author} {\bibfnamefont {L.}~\bibnamefont
  {Kronik}},\ }\bibfield  {title} {\bibinfo {title} {Orbital-dependent density
  functionals: Theory and applications},\ }\href@noop {} {\bibfield  {journal}
  {\bibinfo  {journal} {Reviews of Modern Physics}\ }\textbf {\bibinfo {volume}
  {80}},\ \bibinfo {pages} {3} (\bibinfo {year} {2008})}\BibitemShut {NoStop}%
\bibitem [{\citenamefont {Teale}\ \emph {et~al.}(2022)\citenamefont {Teale},
  \citenamefont {Helgaker}, \citenamefont {Savin}, \citenamefont {Adamo},
  \citenamefont {Aradi}, \citenamefont {Arbuznikov}, \citenamefont {Ayers},
  \citenamefont {Baerends}, \citenamefont {Barone}, \citenamefont {Calaminici},
  \citenamefont {Cancès}, \citenamefont {Carter}, \citenamefont {Chattaraj},
  \citenamefont {Chermette}, \citenamefont {Ciofini}, \citenamefont {Crawford},
  \citenamefont {De~Proft}, \citenamefont {Dobson}, \citenamefont {Draxl},
  \citenamefont {Frauenheim}, \citenamefont {Fromager}, \citenamefont
  {Fuentealba}, \citenamefont {Gagliardi}, \citenamefont {Galli}, \citenamefont
  {Gao}, \citenamefont {Geerlings}, \citenamefont {Gidopoulos}, \citenamefont
  {Gill}, \citenamefont {Gori-Giorgi}, \citenamefont {Görling}, \citenamefont
  {Gould}, \citenamefont {Grimme}, \citenamefont {Gritsenko}, \citenamefont
  {Jensen}, \citenamefont {Johnson}, \citenamefont {Jones}, \citenamefont
  {Kaupp}, \citenamefont {Köster}, \citenamefont {Kronik}, \citenamefont
  {Krylov}, \citenamefont {Kvaal}, \citenamefont {Laestadius}, \citenamefont
  {Levy}, \citenamefont {Lewin}, \citenamefont {Liu}, \citenamefont {Loos},
  \citenamefont {Maitra}, \citenamefont {Neese}, \citenamefont {Perdew},
  \citenamefont {Pernal}, \citenamefont {Pernot}, \citenamefont {Piecuch},
  \citenamefont {Rebolini}, \citenamefont {Reining}, \citenamefont
  {Romaniello}, \citenamefont {Ruzsinszky}, \citenamefont {Salahub},
  \citenamefont {Scheffler}, \citenamefont {Schwerdtfeger}, \citenamefont
  {Staroverov}, \citenamefont {Sun}, \citenamefont {Tellgren}, \citenamefont
  {Tozer}, \citenamefont {Trickey}, \citenamefont {Ullrich}, \citenamefont
  {Vela}, \citenamefont {Vignale}, \citenamefont {Wesolowski}, \citenamefont
  {Xu},\ and\ \citenamefont {Yang}}]{dftsharing2022}%
  \BibitemOpen
  \bibfield  {author} {\bibinfo {author} {\bibfnamefont {A.~M.}\ \bibnamefont
  {Teale}}, \bibinfo {author} {\bibfnamefont {T.}~\bibnamefont {Helgaker}},
  \bibinfo {author} {\bibfnamefont {A.}~\bibnamefont {Savin}}, \bibinfo
  {author} {\bibfnamefont {C.}~\bibnamefont {Adamo}}, \bibinfo {author}
  {\bibfnamefont {B.}~\bibnamefont {Aradi}}, \bibinfo {author} {\bibfnamefont
  {A.~V.}\ \bibnamefont {Arbuznikov}}, \bibinfo {author} {\bibfnamefont
  {P.~W.}\ \bibnamefont {Ayers}}, \bibinfo {author} {\bibfnamefont {E.~J.}\
  \bibnamefont {Baerends}}, \bibinfo {author} {\bibfnamefont {V.}~\bibnamefont
  {Barone}}, \bibinfo {author} {\bibfnamefont {P.}~\bibnamefont {Calaminici}},
  \bibinfo {author} {\bibfnamefont {E.}~\bibnamefont {Cancès}}, \bibinfo
  {author} {\bibfnamefont {E.~A.}\ \bibnamefont {Carter}}, \bibinfo {author}
  {\bibfnamefont {P.~K.}\ \bibnamefont {Chattaraj}}, \bibinfo {author}
  {\bibfnamefont {H.}~\bibnamefont {Chermette}}, \bibinfo {author}
  {\bibfnamefont {I.}~\bibnamefont {Ciofini}}, \bibinfo {author} {\bibfnamefont
  {T.~D.}\ \bibnamefont {Crawford}}, \bibinfo {author} {\bibfnamefont
  {F.}~\bibnamefont {De~Proft}}, \bibinfo {author} {\bibfnamefont {J.~F.}\
  \bibnamefont {Dobson}}, \bibinfo {author} {\bibfnamefont {C.}~\bibnamefont
  {Draxl}}, \bibinfo {author} {\bibfnamefont {T.}~\bibnamefont {Frauenheim}},
  \bibinfo {author} {\bibfnamefont {E.}~\bibnamefont {Fromager}}, \bibinfo
  {author} {\bibfnamefont {P.}~\bibnamefont {Fuentealba}}, \bibinfo {author}
  {\bibfnamefont {L.}~\bibnamefont {Gagliardi}}, \bibinfo {author}
  {\bibfnamefont {G.}~\bibnamefont {Galli}}, \bibinfo {author} {\bibfnamefont
  {J.}~\bibnamefont {Gao}}, \bibinfo {author} {\bibfnamefont {P.}~\bibnamefont
  {Geerlings}}, \bibinfo {author} {\bibfnamefont {N.}~\bibnamefont
  {Gidopoulos}}, \bibinfo {author} {\bibfnamefont {P.~M.~W.}\ \bibnamefont
  {Gill}}, \bibinfo {author} {\bibfnamefont {P.}~\bibnamefont {Gori-Giorgi}},
  \bibinfo {author} {\bibfnamefont {A.}~\bibnamefont {Görling}}, \bibinfo
  {author} {\bibfnamefont {T.}~\bibnamefont {Gould}}, \bibinfo {author}
  {\bibfnamefont {S.}~\bibnamefont {Grimme}}, \bibinfo {author} {\bibfnamefont
  {O.}~\bibnamefont {Gritsenko}}, \bibinfo {author} {\bibfnamefont {H.~J.~A.}\
  \bibnamefont {Jensen}}, \bibinfo {author} {\bibfnamefont {E.~R.}\
  \bibnamefont {Johnson}}, \bibinfo {author} {\bibfnamefont {R.~O.}\
  \bibnamefont {Jones}}, \bibinfo {author} {\bibfnamefont {M.}~\bibnamefont
  {Kaupp}}, \bibinfo {author} {\bibfnamefont {A.~M.}\ \bibnamefont {Köster}},
  \bibinfo {author} {\bibfnamefont {L.}~\bibnamefont {Kronik}}, \bibinfo
  {author} {\bibfnamefont {A.~I.}\ \bibnamefont {Krylov}}, \bibinfo {author}
  {\bibfnamefont {S.}~\bibnamefont {Kvaal}}, \bibinfo {author} {\bibfnamefont
  {A.}~\bibnamefont {Laestadius}}, \bibinfo {author} {\bibfnamefont
  {M.}~\bibnamefont {Levy}}, \bibinfo {author} {\bibfnamefont {M.}~\bibnamefont
  {Lewin}}, \bibinfo {author} {\bibfnamefont {S.}~\bibnamefont {Liu}}, \bibinfo
  {author} {\bibfnamefont {P.-F.}\ \bibnamefont {Loos}}, \bibinfo {author}
  {\bibfnamefont {N.~T.}\ \bibnamefont {Maitra}}, \bibinfo {author}
  {\bibfnamefont {F.}~\bibnamefont {Neese}}, \bibinfo {author} {\bibfnamefont
  {J.~P.}\ \bibnamefont {Perdew}}, \bibinfo {author} {\bibfnamefont
  {K.}~\bibnamefont {Pernal}}, \bibinfo {author} {\bibfnamefont
  {P.}~\bibnamefont {Pernot}}, \bibinfo {author} {\bibfnamefont
  {P.}~\bibnamefont {Piecuch}}, \bibinfo {author} {\bibfnamefont
  {E.}~\bibnamefont {Rebolini}}, \bibinfo {author} {\bibfnamefont
  {L.}~\bibnamefont {Reining}}, \bibinfo {author} {\bibfnamefont
  {P.}~\bibnamefont {Romaniello}}, \bibinfo {author} {\bibfnamefont
  {A.}~\bibnamefont {Ruzsinszky}}, \bibinfo {author} {\bibfnamefont {D.~R.}\
  \bibnamefont {Salahub}}, \bibinfo {author} {\bibfnamefont {M.}~\bibnamefont
  {Scheffler}}, \bibinfo {author} {\bibfnamefont {P.}~\bibnamefont
  {Schwerdtfeger}}, \bibinfo {author} {\bibfnamefont {V.~N.}\ \bibnamefont
  {Staroverov}}, \bibinfo {author} {\bibfnamefont {J.}~\bibnamefont {Sun}},
  \bibinfo {author} {\bibfnamefont {E.}~\bibnamefont {Tellgren}}, \bibinfo
  {author} {\bibfnamefont {D.~J.}\ \bibnamefont {Tozer}}, \bibinfo {author}
  {\bibfnamefont {S.~B.}\ \bibnamefont {Trickey}}, \bibinfo {author}
  {\bibfnamefont {C.~A.}\ \bibnamefont {Ullrich}}, \bibinfo {author}
  {\bibfnamefont {A.}~\bibnamefont {Vela}}, \bibinfo {author} {\bibfnamefont
  {G.}~\bibnamefont {Vignale}}, \bibinfo {author} {\bibfnamefont {T.~A.}\
  \bibnamefont {Wesolowski}}, \bibinfo {author} {\bibfnamefont
  {X.}~\bibnamefont {Xu}},\ and\ \bibinfo {author} {\bibfnamefont
  {W.}~\bibnamefont {Yang}},\ }\bibfield  {title} {\bibinfo {title} {Dft
  exchange: sharing perspectives on the workhorse of quantum chemistry and
  materials science},\ }\href@noop {} {\bibfield  {journal} {\bibinfo
  {journal} {Phys. Chem. Chem. Phys.}\ }\textbf {\bibinfo {volume} {24}},\
  \bibinfo {pages} {28700} (\bibinfo {year} {2022})}\BibitemShut {NoStop}%
\bibitem [{per(2001)}]{perdew2001jacob}%
  \BibitemOpen
  \bibfield  {title} {\bibinfo {title} {Jacob's ladder of density functional
  approximations for the exchange-correlation energy},\ }\href
  {https://doi.org/10.1063/1.1390175} {\bibfield  {journal} {\bibinfo
  {journal} {AIP Conf. Proc.}\ }\textbf {\bibinfo {volume} {577}},\ \bibinfo
  {pages} {1} (\bibinfo {year} {2001})}\BibitemShut {NoStop}%
\bibitem [{\citenamefont {Scuseria}\ and\ \citenamefont
  {Staroverov}(2005)}]{scuseriaREVIEW05}%
  \BibitemOpen
  \bibfield  {author} {\bibinfo {author} {\bibfnamefont {G.~E.}\ \bibnamefont
  {Scuseria}}\ and\ \bibinfo {author} {\bibfnamefont {V.~N.}\ \bibnamefont
  {Staroverov}},\ }\bibfield  {title} {\bibinfo {title} {Progress in the
  development of exchange-correlation functionals},\ }in\ \href@noop {} {\emph
  {\bibinfo {booktitle} {Theory and Application of Computational Chemistry: The
  First 40 Years}}},\ \bibinfo {editor} {edited by\ \bibinfo {editor}
  {\bibfnamefont {C.~E.}\ \bibnamefont {Dykstra}}, \bibinfo {editor}
  {\bibfnamefont {G.}~\bibnamefont {Frenking}}, \bibinfo {editor}
  {\bibfnamefont {K.~S.}\ \bibnamefont {Kim}},\ and\ \bibinfo {editor}
  {\bibfnamefont {G.~E.}\ \bibnamefont {Scuseria}}}\ (\bibinfo  {publisher}
  {Elsevier: Amsterdam},\ \bibinfo {year} {2005})\ pp.\ \bibinfo {pages}
  {669--724}\BibitemShut {NoStop}%
\bibitem [{\citenamefont {Della~Sala}\ \emph {et~al.}(2016)\citenamefont
  {Della~Sala}, \citenamefont {Fabiano},\ and\ \citenamefont
  {Constantin}}]{della2016kinetic}%
  \BibitemOpen
  \bibfield  {author} {\bibinfo {author} {\bibfnamefont {F.}~\bibnamefont
  {Della~Sala}}, \bibinfo {author} {\bibfnamefont {E.}~\bibnamefont
  {Fabiano}},\ and\ \bibinfo {author} {\bibfnamefont {L.~A.}\ \bibnamefont
  {Constantin}},\ }\bibfield  {title} {\bibinfo {title} {Kinetic-energy-density
  dependent semilocal exchange-correlation functionals},\ }\href@noop {}
  {\bibfield  {journal} {\bibinfo  {journal} {Int. J. Quantum Chem.}\ }\textbf
  {\bibinfo {volume} {22}},\ \bibinfo {pages} {1641} (\bibinfo {year}
  {2016})}\BibitemShut {NoStop}%
\bibitem [{\citenamefont {Perdew}\ \emph {et~al.}(2008)\citenamefont {Perdew},
  \citenamefont {Staroverov}, \citenamefont {Tao},\ and\ \citenamefont
  {Scuseria}}]{perdew2008density}%
  \BibitemOpen
  \bibfield  {author} {\bibinfo {author} {\bibfnamefont {J.~P.}\ \bibnamefont
  {Perdew}}, \bibinfo {author} {\bibfnamefont {V.~N.}\ \bibnamefont
  {Staroverov}}, \bibinfo {author} {\bibfnamefont {J.}~\bibnamefont {Tao}},\
  and\ \bibinfo {author} {\bibfnamefont {G.~E.}\ \bibnamefont {Scuseria}},\
  }\bibfield  {title} {\bibinfo {title} {Density functional with full exact
  exchange, balanced nonlocality of correlation, and constraint satisfaction},\
  }\href@noop {} {\bibfield  {journal} {\bibinfo  {journal} {Phys. Rev. A}\
  }\textbf {\bibinfo {volume} {78}},\ \bibinfo {pages} {052513} (\bibinfo
  {year} {2008})}\BibitemShut {NoStop}%
\bibitem [{\citenamefont {Perdew}\ \emph {et~al.}(2005)\citenamefont {Perdew},
  \citenamefont {Ruzsinszky}, \citenamefont {Tao}, \citenamefont {Staroverov},
  \citenamefont {Scuseria},\ and\ \citenamefont
  {Csonka}}]{perdew2005prescription}%
  \BibitemOpen
  \bibfield  {author} {\bibinfo {author} {\bibfnamefont {J.~P.}\ \bibnamefont
  {Perdew}}, \bibinfo {author} {\bibfnamefont {A.}~\bibnamefont {Ruzsinszky}},
  \bibinfo {author} {\bibfnamefont {J.}~\bibnamefont {Tao}}, \bibinfo {author}
  {\bibfnamefont {V.~N.}\ \bibnamefont {Staroverov}}, \bibinfo {author}
  {\bibfnamefont {G.~E.}\ \bibnamefont {Scuseria}},\ and\ \bibinfo {author}
  {\bibfnamefont {G.~I.}\ \bibnamefont {Csonka}},\ }\bibfield  {title}
  {\bibinfo {title} {Prescription for the design and selection of density
  functional approximations: More constraint satisfaction with fewer fits},\
  }\href@noop {} {\bibfield  {journal} {\bibinfo  {journal} {J. Chem. Phys.}\
  }\textbf {\bibinfo {volume} {123}},\ \bibinfo {pages} {062201} (\bibinfo
  {year} {2005})}\BibitemShut {NoStop}%
\bibitem [{\citenamefont {Lehtola}\ \emph {et~al.}(2018)\citenamefont
  {Lehtola}, \citenamefont {Steigemann}, \citenamefont {Oliveira},\ and\
  \citenamefont {Marques}}]{LehtolaSX18}%
  \BibitemOpen
  \bibfield  {author} {\bibinfo {author} {\bibfnamefont {S.}~\bibnamefont
  {Lehtola}}, \bibinfo {author} {\bibfnamefont {C.}~\bibnamefont {Steigemann}},
  \bibinfo {author} {\bibfnamefont {M.~J.~T.}\ \bibnamefont {Oliveira}},\ and\
  \bibinfo {author} {\bibfnamefont {M.~A.~L.}\ \bibnamefont {Marques}},\ }\href
  {https://doi.org/10.1016/j.softx.2017.11.002} {\bibfield  {journal} {\bibinfo
   {journal} {SoftwareX}\ }\textbf {\bibinfo {volume} {7}},\ \bibinfo {pages}
  {1} (\bibinfo {year} {2018})}\BibitemShut {NoStop}%
\bibitem [{\citenamefont {Perdew}\ \emph {et~al.}(1996)\citenamefont {Perdew},
  \citenamefont {Burke},\ and\ \citenamefont {Ernzerhof}}]{perdewPRL96}%
  \BibitemOpen
  \bibfield  {author} {\bibinfo {author} {\bibfnamefont {J.~P.}\ \bibnamefont
  {Perdew}}, \bibinfo {author} {\bibfnamefont {K.}~\bibnamefont {Burke}},\ and\
  \bibinfo {author} {\bibfnamefont {M.}~\bibnamefont {Ernzerhof}},\ }\bibfield
  {title} {\bibinfo {title} {Generalized gradient approximation made simple},\
  }\href@noop {} {\bibfield  {journal} {\bibinfo  {journal} {Phys. Rev. Lett.}\
  }\textbf {\bibinfo {volume} {77}},\ \bibinfo {pages} {3865} (\bibinfo {year}
  {1996})}\BibitemShut {NoStop}%
\bibitem [{\citenamefont {Heyd}\ \emph {et~al.}(2003)\citenamefont {Heyd},
  \citenamefont {Scuseria},\ and\ \citenamefont {Ernzerhof}}]{heyd2003hybrid}%
  \BibitemOpen
  \bibfield  {author} {\bibinfo {author} {\bibfnamefont {J.}~\bibnamefont
  {Heyd}}, \bibinfo {author} {\bibfnamefont {G.~E.}\ \bibnamefont {Scuseria}},\
  and\ \bibinfo {author} {\bibfnamefont {M.}~\bibnamefont {Ernzerhof}},\
  }\bibfield  {title} {\bibinfo {title} {Hybrid functionals based on a screened
  coulomb potential},\ }\href@noop {} {\bibfield  {journal} {\bibinfo
  {journal} {J. Chem. Phys.}\ }\textbf {\bibinfo {volume} {118}},\ \bibinfo
  {pages} {8207} (\bibinfo {year} {2003})}\BibitemShut {NoStop}%
\bibitem [{\citenamefont {Krukau}\ \emph {et~al.}(2006)\citenamefont {Krukau},
  \citenamefont {Vydrov}, \citenamefont {Izmaylov},\ and\ \citenamefont
  {Scuseria}}]{krukau2006influence}%
  \BibitemOpen
  \bibfield  {author} {\bibinfo {author} {\bibfnamefont {A.~V.}\ \bibnamefont
  {Krukau}}, \bibinfo {author} {\bibfnamefont {O.~A.}\ \bibnamefont {Vydrov}},
  \bibinfo {author} {\bibfnamefont {A.~F.}\ \bibnamefont {Izmaylov}},\ and\
  \bibinfo {author} {\bibfnamefont {G.~E.}\ \bibnamefont {Scuseria}},\
  }\bibfield  {title} {\bibinfo {title} {Influence of the exchange screening
  parameter on the performance of screened hybrid functionals},\ }\href@noop {}
  {\bibfield  {journal} {\bibinfo  {journal} {J. Chem. Phys.}\ }\textbf
  {\bibinfo {volume} {125}},\ \bibinfo {pages} {224106} (\bibinfo {year}
  {2006})}\BibitemShut {NoStop}%
\bibitem [{\citenamefont {Heyd}\ and\ \citenamefont
  {Scuseria}(2004)}]{heyd2004efficient}%
  \BibitemOpen
  \bibfield  {author} {\bibinfo {author} {\bibfnamefont {J.}~\bibnamefont
  {Heyd}}\ and\ \bibinfo {author} {\bibfnamefont {G.~E.}\ \bibnamefont
  {Scuseria}},\ }\bibfield  {title} {\bibinfo {title} {Efficient hybrid density
  functional calculations in solids: Assessment of the
  heyd–scuseria–ernzerhof screened coulomb hybrid functional},\ }\href@noop
  {} {\bibfield  {journal} {\bibinfo  {journal} {J. Chem. Phys.}\ }\textbf
  {\bibinfo {volume} {121}},\ \bibinfo {pages} {1187} (\bibinfo {year}
  {2004})}\BibitemShut {NoStop}%
\bibitem [{\citenamefont {Jana}\ \emph
  {et~al.}(2020{\natexlab{a}})\citenamefont {Jana}, \citenamefont {Patra},
  \citenamefont {Constantin},\ and\ \citenamefont {Samal}}]{jana2020screened}%
  \BibitemOpen
  \bibfield  {author} {\bibinfo {author} {\bibfnamefont {S.}~\bibnamefont
  {Jana}}, \bibinfo {author} {\bibfnamefont {A.}~\bibnamefont {Patra}},
  \bibinfo {author} {\bibfnamefont {L.~A.}\ \bibnamefont {Constantin}},\ and\
  \bibinfo {author} {\bibfnamefont {P.}~\bibnamefont {Samal}},\ }\bibfield
  {title} {\bibinfo {title} {Screened range-separated hybrid by balancing the
  compact and slowly varying density regimes: Satisfaction of local density
  linear response},\ }\href@noop {} {\bibfield  {journal} {\bibinfo  {journal}
  {J. Chem. Phys.}\ }\textbf {\bibinfo {volume} {152}},\ \bibinfo {pages}
  {044111} (\bibinfo {year} {2020}{\natexlab{a}})}\BibitemShut {NoStop}%
\bibitem [{\citenamefont {Jana}\ \emph
  {et~al.}(2020{\natexlab{b}})\citenamefont {Jana}, \citenamefont {Patra},
  \citenamefont {\ifmmode~\acute{S}\else \'{S}\fi{}miga}, \citenamefont
  {Constantin},\ and\ \citenamefont {Samal}}]{jana2020improved}%
  \BibitemOpen
  \bibfield  {author} {\bibinfo {author} {\bibfnamefont {S.}~\bibnamefont
  {Jana}}, \bibinfo {author} {\bibfnamefont {B.}~\bibnamefont {Patra}},
  \bibinfo {author} {\bibfnamefont {S.}~\bibnamefont {\ifmmode~\acute{S}\else
  \'{S}\fi{}miga}}, \bibinfo {author} {\bibfnamefont {L.~A.}\ \bibnamefont
  {Constantin}},\ and\ \bibinfo {author} {\bibfnamefont {P.}~\bibnamefont
  {Samal}},\ }\bibfield  {title} {\bibinfo {title} {Improved solid stability
  from a screened range-separated hybrid functional by satisfying semiclassical
  atom theory and local density linear response},\ }\href@noop {} {\bibfield
  {journal} {\bibinfo  {journal} {Phys. Rev. B}\ }\textbf {\bibinfo {volume}
  {102}},\ \bibinfo {pages} {155107} (\bibinfo {year}
  {2020}{\natexlab{b}})}\BibitemShut {NoStop}%
\bibitem [{\citenamefont {Perdew}\ \emph {et~al.}(2017)\citenamefont {Perdew},
  \citenamefont {Yang}, \citenamefont {Burke}, \citenamefont {Yang},
  \citenamefont {Gross}, \citenamefont {Scheffler}, \citenamefont {Scuseria},
  \citenamefont {Henderson}, \citenamefont {Zhang}, \citenamefont {Ruzsinszky},
  \citenamefont {Peng}, \citenamefont {Sun}, \citenamefont {Trushin},\ and\
  \citenamefont {G{\"o}rling}}]{perdew2017understanding}%
  \BibitemOpen
  \bibfield  {author} {\bibinfo {author} {\bibfnamefont {J.~P.}\ \bibnamefont
  {Perdew}}, \bibinfo {author} {\bibfnamefont {W.}~\bibnamefont {Yang}},
  \bibinfo {author} {\bibfnamefont {K.}~\bibnamefont {Burke}}, \bibinfo
  {author} {\bibfnamefont {Z.}~\bibnamefont {Yang}}, \bibinfo {author}
  {\bibfnamefont {E.~K.~U.}\ \bibnamefont {Gross}}, \bibinfo {author}
  {\bibfnamefont {M.}~\bibnamefont {Scheffler}}, \bibinfo {author}
  {\bibfnamefont {G.~E.}\ \bibnamefont {Scuseria}}, \bibinfo {author}
  {\bibfnamefont {T.~M.}\ \bibnamefont {Henderson}}, \bibinfo {author}
  {\bibfnamefont {I.~Y.}\ \bibnamefont {Zhang}}, \bibinfo {author}
  {\bibfnamefont {A.}~\bibnamefont {Ruzsinszky}}, \bibinfo {author}
  {\bibfnamefont {H.}~\bibnamefont {Peng}}, \bibinfo {author} {\bibfnamefont
  {J.}~\bibnamefont {Sun}}, \bibinfo {author} {\bibfnamefont {E.}~\bibnamefont
  {Trushin}},\ and\ \bibinfo {author} {\bibfnamefont {A.}~\bibnamefont
  {G{\"o}rling}},\ }\bibfield  {title} {\bibinfo {title} {Understanding band
  gaps of solids in generalized kohn--sham theory},\ }\href@noop {} {\bibfield
  {journal} {\bibinfo  {journal} {Proc. Natl. Acad. Sci. U. S. A.}\ }\textbf
  {\bibinfo {volume} {114}},\ \bibinfo {pages} {2801} (\bibinfo {year}
  {2017})}\BibitemShut {NoStop}%
\bibitem [{\citenamefont {Tran}\ \emph {et~al.}(2007)\citenamefont {Tran},
  \citenamefont {Blaha},\ and\ \citenamefont {Schwarz}}]{tran2007band}%
  \BibitemOpen
  \bibfield  {author} {\bibinfo {author} {\bibfnamefont {F.}~\bibnamefont
  {Tran}}, \bibinfo {author} {\bibfnamefont {P.}~\bibnamefont {Blaha}},\ and\
  \bibinfo {author} {\bibfnamefont {K.}~\bibnamefont {Schwarz}},\ }\bibfield
  {title} {\bibinfo {title} {Band gap calculations with becke--johnson exchange
  potential},\ }\href@noop {} {\bibfield  {journal} {\bibinfo  {journal} {J.
  Phys.: Condens. Matter}\ }\textbf {\bibinfo {volume} {19}},\ \bibinfo {pages}
  {196208} (\bibinfo {year} {2007})}\BibitemShut {NoStop}%
\bibitem [{\citenamefont {Borlido}\ \emph {et~al.}(2020)\citenamefont
  {Borlido}, \citenamefont {Schmidt}, \citenamefont {Huran}, \citenamefont
  {Tran}, \citenamefont {Marques},\ and\ \citenamefont {Botti}}]{Borlido2020}%
  \BibitemOpen
  \bibfield  {author} {\bibinfo {author} {\bibfnamefont {P.}~\bibnamefont
  {Borlido}}, \bibinfo {author} {\bibfnamefont {J.}~\bibnamefont {Schmidt}},
  \bibinfo {author} {\bibfnamefont {A.~W.}\ \bibnamefont {Huran}}, \bibinfo
  {author} {\bibfnamefont {F.}~\bibnamefont {Tran}}, \bibinfo {author}
  {\bibfnamefont {M.~A.~L.}\ \bibnamefont {Marques}},\ and\ \bibinfo {author}
  {\bibfnamefont {S.}~\bibnamefont {Botti}},\ }\bibfield  {title} {\bibinfo
  {title} {Exchange-correlation functionals for band gaps of solids: benchmark,
  reparametrization and machine learning},\ }\href@noop {} {\bibfield
  {journal} {\bibinfo  {journal} {npj Computational Materials}\ }\textbf
  {\bibinfo {volume} {6}},\ \bibinfo {pages} {96} (\bibinfo {year}
  {2020})}\BibitemShut {NoStop}%
\bibitem [{\citenamefont {Patra}\ \emph
  {et~al.}(2019{\natexlab{a}})\citenamefont {Patra}, \citenamefont {Jana},
  \citenamefont {Constantin},\ and\ \citenamefont
  {Samal}}]{patra2019efficient}%
  \BibitemOpen
  \bibfield  {author} {\bibinfo {author} {\bibfnamefont {B.}~\bibnamefont
  {Patra}}, \bibinfo {author} {\bibfnamefont {S.}~\bibnamefont {Jana}},
  \bibinfo {author} {\bibfnamefont {L.~A.}\ \bibnamefont {Constantin}},\ and\
  \bibinfo {author} {\bibfnamefont {P.}~\bibnamefont {Samal}},\ }\bibfield
  {title} {\bibinfo {title} {Efficient band gap prediction of semiconductors
  and insulators from a semilocal exchange-correlation functional},\
  }\href@noop {} {\bibfield  {journal} {\bibinfo  {journal} {Phys. Rev. B}\
  }\textbf {\bibinfo {volume} {100}},\ \bibinfo {pages} {045147} (\bibinfo
  {year} {2019}{\natexlab{a}})}\BibitemShut {NoStop}%
\bibitem [{\citenamefont {Tran}\ \emph {et~al.}(2018)\citenamefont {Tran},
  \citenamefont {Ehsan},\ and\ \citenamefont {Blaha}}]{fabien2018assessment}%
  \BibitemOpen
  \bibfield  {author} {\bibinfo {author} {\bibfnamefont {F.}~\bibnamefont
  {Tran}}, \bibinfo {author} {\bibfnamefont {S.}~\bibnamefont {Ehsan}},\ and\
  \bibinfo {author} {\bibfnamefont {P.}~\bibnamefont {Blaha}},\ }\bibfield
  {title} {\bibinfo {title} {Assessment of the gllb-sc potential for
  solid-state properties and attempts for improvement},\ }\href@noop {}
  {\bibfield  {journal} {\bibinfo  {journal} {Phys. Rev. Materials}\ }\textbf
  {\bibinfo {volume} {2}},\ \bibinfo {pages} {023802} (\bibinfo {year}
  {2018})}\BibitemShut {NoStop}%
\bibitem [{\citenamefont {Jana}\ \emph {et~al.}(2018)\citenamefont {Jana},
  \citenamefont {Patra},\ and\ \citenamefont {Samal}}]{jana2018assessing}%
  \BibitemOpen
  \bibfield  {author} {\bibinfo {author} {\bibfnamefont {S.}~\bibnamefont
  {Jana}}, \bibinfo {author} {\bibfnamefont {A.}~\bibnamefont {Patra}},\ and\
  \bibinfo {author} {\bibfnamefont {P.}~\bibnamefont {Samal}},\ }\bibfield
  {title} {\bibinfo {title} {Assessing the performance of the tao-mo semilocal
  density functional in the projector-augmented-wave method},\ }\href@noop {}
  {\bibfield  {journal} {\bibinfo  {journal} {J. Chem. Phys.}\ }\textbf
  {\bibinfo {volume} {149}},\ \bibinfo {pages} {044120} (\bibinfo {year}
  {2018})}\BibitemShut {NoStop}%
\bibitem [{\citenamefont {Tran}\ and\ \citenamefont
  {Blaha}(2017)}]{tran2017importance}%
  \BibitemOpen
  \bibfield  {author} {\bibinfo {author} {\bibfnamefont {F.}~\bibnamefont
  {Tran}}\ and\ \bibinfo {author} {\bibfnamefont {P.}~\bibnamefont {Blaha}},\
  }\bibfield  {title} {\bibinfo {title} {Importance of the kinetic energy
  density for band gap calculations in solids with density functional theory},\
  }\href@noop {} {\bibfield  {journal} {\bibinfo  {journal} {J. Phys. Chem. A}\
  }\textbf {\bibinfo {volume} {121}},\ \bibinfo {pages} {3318} (\bibinfo {year}
  {2017})}\BibitemShut {NoStop}%
\bibitem [{\citenamefont {Tran}\ \emph {et~al.}(2019)\citenamefont {Tran},
  \citenamefont {Doumont}, \citenamefont {Kalantari}, \citenamefont {Huran},
  \citenamefont {Marques},\ and\ \citenamefont {Blaha}}]{fabien2019semilocal}%
  \BibitemOpen
  \bibfield  {author} {\bibinfo {author} {\bibfnamefont {F.}~\bibnamefont
  {Tran}}, \bibinfo {author} {\bibfnamefont {J.}~\bibnamefont {Doumont}},
  \bibinfo {author} {\bibfnamefont {L.}~\bibnamefont {Kalantari}}, \bibinfo
  {author} {\bibfnamefont {A.~W.}\ \bibnamefont {Huran}}, \bibinfo {author}
  {\bibfnamefont {M.~A.~L.}\ \bibnamefont {Marques}},\ and\ \bibinfo {author}
  {\bibfnamefont {P.}~\bibnamefont {Blaha}},\ }\bibfield  {title} {\bibinfo
  {title} {Semilocal exchange-correlation potentials for solid-state
  calculations: Current status and future directions},\ }\href@noop {}
  {\bibfield  {journal} {\bibinfo  {journal} {Journal of Applied Physics}\
  }\textbf {\bibinfo {volume} {126}},\ \bibinfo {pages} {110902} (\bibinfo
  {year} {2019})}\BibitemShut {NoStop}%
\bibitem [{\citenamefont {Paier}\ \emph {et~al.}(2008)\citenamefont {Paier},
  \citenamefont {Marsman},\ and\ \citenamefont {Kresse}}]{paier2008dielectric}%
  \BibitemOpen
  \bibfield  {author} {\bibinfo {author} {\bibfnamefont {J.}~\bibnamefont
  {Paier}}, \bibinfo {author} {\bibfnamefont {M.}~\bibnamefont {Marsman}},\
  and\ \bibinfo {author} {\bibfnamefont {G.}~\bibnamefont {Kresse}},\
  }\bibfield  {title} {\bibinfo {title} {Dielectric properties and excitons for
  extended systems from hybrid functionals},\ }\href@noop {} {\bibfield
  {journal} {\bibinfo  {journal} {Phys. Rev. B}\ }\textbf {\bibinfo {volume}
  {78}},\ \bibinfo {pages} {121201} (\bibinfo {year} {2008})}\BibitemShut
  {NoStop}%
\bibitem [{\citenamefont {Wing}\ \emph
  {et~al.}(2019{\natexlab{a}})\citenamefont {Wing}, \citenamefont {Haber},
  \citenamefont {Noff}, \citenamefont {Barker}, \citenamefont {Egger},
  \citenamefont {Ramasubramaniam}, \citenamefont {Louie}, \citenamefont
  {Neaton},\ and\ \citenamefont {Kronik}}]{wing2019comparing}%
  \BibitemOpen
  \bibfield  {author} {\bibinfo {author} {\bibfnamefont {D.}~\bibnamefont
  {Wing}}, \bibinfo {author} {\bibfnamefont {J.~B.}\ \bibnamefont {Haber}},
  \bibinfo {author} {\bibfnamefont {R.}~\bibnamefont {Noff}}, \bibinfo {author}
  {\bibfnamefont {B.}~\bibnamefont {Barker}}, \bibinfo {author} {\bibfnamefont
  {D.~A.}\ \bibnamefont {Egger}}, \bibinfo {author} {\bibfnamefont
  {A.}~\bibnamefont {Ramasubramaniam}}, \bibinfo {author} {\bibfnamefont
  {S.~G.}\ \bibnamefont {Louie}}, \bibinfo {author} {\bibfnamefont {J.~B.}\
  \bibnamefont {Neaton}},\ and\ \bibinfo {author} {\bibfnamefont
  {L.}~\bibnamefont {Kronik}},\ }\bibfield  {title} {\bibinfo {title}
  {Comparing time-dependent density functional theory with many-body
  perturbation theory for semiconductors: Screened range-separated hybrids and
  the $gw$ plus bethe-salpeter approach},\ }\href@noop {} {\bibfield  {journal}
  {\bibinfo  {journal} {Phys. Rev. Materials}\ }\textbf {\bibinfo {volume}
  {3}},\ \bibinfo {pages} {064603} (\bibinfo {year}
  {2019}{\natexlab{a}})}\BibitemShut {NoStop}%
\bibitem [{\citenamefont {St\"adele}\ \emph {et~al.}(1999)\citenamefont
  {St\"adele}, \citenamefont {Moukara}, \citenamefont {Majewski}, \citenamefont
  {Vogl},\ and\ \citenamefont {G\"orling}}]{stadele1999exact}%
  \BibitemOpen
  \bibfield  {author} {\bibinfo {author} {\bibfnamefont {M.}~\bibnamefont
  {St\"adele}}, \bibinfo {author} {\bibfnamefont {M.}~\bibnamefont {Moukara}},
  \bibinfo {author} {\bibfnamefont {J.~A.}\ \bibnamefont {Majewski}}, \bibinfo
  {author} {\bibfnamefont {P.}~\bibnamefont {Vogl}},\ and\ \bibinfo {author}
  {\bibfnamefont {A.}~\bibnamefont {G\"orling}},\ }\bibfield  {title} {\bibinfo
  {title} {Exact exchange kohn-sham formalism applied to semiconductors},\
  }\href@noop {} {\bibfield  {journal} {\bibinfo  {journal} {Phys. Rev. B}\
  }\textbf {\bibinfo {volume} {59}},\ \bibinfo {pages} {10031} (\bibinfo {year}
  {1999})}\BibitemShut {NoStop}%
\bibitem [{\citenamefont {Petersilka}\ \emph {et~al.}(1996)\citenamefont
  {Petersilka}, \citenamefont {Gossmann},\ and\ \citenamefont
  {Gross}}]{petersilka1996excitation}%
  \BibitemOpen
  \bibfield  {author} {\bibinfo {author} {\bibfnamefont {M.}~\bibnamefont
  {Petersilka}}, \bibinfo {author} {\bibfnamefont {U.~J.}\ \bibnamefont
  {Gossmann}},\ and\ \bibinfo {author} {\bibfnamefont {E.~K.~U.}\ \bibnamefont
  {Gross}},\ }\bibfield  {title} {\bibinfo {title} {Excitation energies from
  time-dependent density-functional theory},\ }\href@noop {} {\bibfield
  {journal} {\bibinfo  {journal} {Phys. Rev. Lett.}\ }\textbf {\bibinfo
  {volume} {76}},\ \bibinfo {pages} {1212} (\bibinfo {year}
  {1996})}\BibitemShut {NoStop}%
\bibitem [{\citenamefont {Kim}\ and\ \citenamefont
  {G{\"o}rling}(2002)}]{kim2002excitonic}%
  \BibitemOpen
  \bibfield  {author} {\bibinfo {author} {\bibfnamefont {Y.-H.}\ \bibnamefont
  {Kim}}\ and\ \bibinfo {author} {\bibfnamefont {A.}~\bibnamefont
  {G{\"o}rling}},\ }\bibfield  {title} {\bibinfo {title} {Excitonic optical
  spectrum of semiconductors obtained by time-dependent density-functional
  theory with the exact-exchange kernel},\ }\href@noop {} {\bibfield  {journal}
  {\bibinfo  {journal} {Phys. Rev. Lett.}\ }\textbf {\bibinfo {volume} {89}},\
  \bibinfo {pages} {096402} (\bibinfo {year} {2002})}\BibitemShut {NoStop}%
\bibitem [{\citenamefont {Terentjev}\ \emph {et~al.}(2018)\citenamefont
  {Terentjev}, \citenamefont {Constantin},\ and\ \citenamefont
  {Pitarke}}]{terentjev2018gradient}%
  \BibitemOpen
  \bibfield  {author} {\bibinfo {author} {\bibfnamefont {A.~V.}\ \bibnamefont
  {Terentjev}}, \bibinfo {author} {\bibfnamefont {L.~A.}\ \bibnamefont
  {Constantin}},\ and\ \bibinfo {author} {\bibfnamefont {J.~M.}\ \bibnamefont
  {Pitarke}},\ }\bibfield  {title} {\bibinfo {title} {Gradient-dependent
  exchange-correlation kernel for materials optical properties},\ }\href@noop
  {} {\bibfield  {journal} {\bibinfo  {journal} {Phys. Rev. B}\ }\textbf
  {\bibinfo {volume} {98}},\ \bibinfo {pages} {085123} (\bibinfo {year}
  {2018})}\BibitemShut {NoStop}%
\bibitem [{\citenamefont {Sharma}\ \emph {et~al.}(2011)\citenamefont {Sharma},
  \citenamefont {Dewhurst}, \citenamefont {Sanna},\ and\ \citenamefont
  {Gross}}]{sharma2011bootstrap}%
  \BibitemOpen
  \bibfield  {author} {\bibinfo {author} {\bibfnamefont {S.}~\bibnamefont
  {Sharma}}, \bibinfo {author} {\bibfnamefont {J.~K.}\ \bibnamefont
  {Dewhurst}}, \bibinfo {author} {\bibfnamefont {A.}~\bibnamefont {Sanna}},\
  and\ \bibinfo {author} {\bibfnamefont {E.~K.~U.}\ \bibnamefont {Gross}},\
  }\bibfield  {title} {\bibinfo {title} {Bootstrap approximation for the
  exchange-correlation kernel of time-dependent density-functional theory},\
  }\href@noop {} {\bibfield  {journal} {\bibinfo  {journal} {Phys. Rev. Lett.}\
  }\textbf {\bibinfo {volume} {107}},\ \bibinfo {pages} {186401} (\bibinfo
  {year} {2011})}\BibitemShut {NoStop}%
\bibitem [{\citenamefont {Rigamonti}\ \emph {et~al.}(2015)\citenamefont
  {Rigamonti}, \citenamefont {Botti}, \citenamefont {Veniard}, \citenamefont
  {Draxl}, \citenamefont {Reining},\ and\ \citenamefont
  {Sottile}}]{rigamonti2015estimating}%
  \BibitemOpen
  \bibfield  {author} {\bibinfo {author} {\bibfnamefont {S.}~\bibnamefont
  {Rigamonti}}, \bibinfo {author} {\bibfnamefont {S.}~\bibnamefont {Botti}},
  \bibinfo {author} {\bibfnamefont {V.}~\bibnamefont {Veniard}}, \bibinfo
  {author} {\bibfnamefont {C.}~\bibnamefont {Draxl}}, \bibinfo {author}
  {\bibfnamefont {L.}~\bibnamefont {Reining}},\ and\ \bibinfo {author}
  {\bibfnamefont {F.}~\bibnamefont {Sottile}},\ }\bibfield  {title} {\bibinfo
  {title} {Estimating excitonic effects in the absorption spectra of solids:
  Problems and insight from a guided iteration scheme},\ }\href@noop {}
  {\bibfield  {journal} {\bibinfo  {journal} {Phys. Rev. Lett.}\ }\textbf
  {\bibinfo {volume} {114}},\ \bibinfo {pages} {146402} (\bibinfo {year}
  {2015})}\BibitemShut {NoStop}%
\bibitem [{\citenamefont {Van~Faassen}\ \emph {et~al.}(2002)\citenamefont
  {Van~Faassen}, \citenamefont {De~Boeij}, \citenamefont {Van~Leeuwen},
  \citenamefont {Berger},\ and\ \citenamefont
  {Snijders}}]{van2002ultranonlocality}%
  \BibitemOpen
  \bibfield  {author} {\bibinfo {author} {\bibfnamefont {M.}~\bibnamefont
  {Van~Faassen}}, \bibinfo {author} {\bibfnamefont {P.}~\bibnamefont
  {De~Boeij}}, \bibinfo {author} {\bibfnamefont {R.}~\bibnamefont
  {Van~Leeuwen}}, \bibinfo {author} {\bibfnamefont {J.}~\bibnamefont
  {Berger}},\ and\ \bibinfo {author} {\bibfnamefont {J.}~\bibnamefont
  {Snijders}},\ }\bibfield  {title} {\bibinfo {title} {Ultranonlocality in
  time-dependent current-density-functional theory: Application to conjugated
  polymers},\ }\href@noop {} {\bibfield  {journal} {\bibinfo  {journal} {Phys.
  Rev. Lett.}\ }\textbf {\bibinfo {volume} {88}},\ \bibinfo {pages} {186401}
  (\bibinfo {year} {2002})}\BibitemShut {NoStop}%
\bibitem [{\citenamefont {Cavo}\ \emph {et~al.}(2020)\citenamefont {Cavo},
  \citenamefont {Berger},\ and\ \citenamefont {Romaniello}}]{cavo2020accurate}%
  \BibitemOpen
  \bibfield  {author} {\bibinfo {author} {\bibfnamefont {S.}~\bibnamefont
  {Cavo}}, \bibinfo {author} {\bibfnamefont {J.}~\bibnamefont {Berger}},\ and\
  \bibinfo {author} {\bibfnamefont {P.}~\bibnamefont {Romaniello}},\ }\bibfield
   {title} {\bibinfo {title} {Accurate optical spectra of solids from pure
  time-dependent density functional theory},\ }\href@noop {} {\bibfield
  {journal} {\bibinfo  {journal} {Phys. Rev. B}\ }\textbf {\bibinfo {volume}
  {101}},\ \bibinfo {pages} {115109} (\bibinfo {year} {2020})}\BibitemShut
  {NoStop}%
\bibitem [{\citenamefont {Ohad}\ \emph {et~al.}(2022)\citenamefont {Ohad},
  \citenamefont {Wing}, \citenamefont {Gant}, \citenamefont {Cohen},
  \citenamefont {Haber}, \citenamefont {Sagredo}, \citenamefont {Filip},
  \citenamefont {Neaton},\ and\ \citenamefont {Kronik}}]{OhadWingGant2022}%
  \BibitemOpen
  \bibfield  {author} {\bibinfo {author} {\bibfnamefont {G.}~\bibnamefont
  {Ohad}}, \bibinfo {author} {\bibfnamefont {D.}~\bibnamefont {Wing}}, \bibinfo
  {author} {\bibfnamefont {S.~E.}\ \bibnamefont {Gant}}, \bibinfo {author}
  {\bibfnamefont {A.~V.}\ \bibnamefont {Cohen}}, \bibinfo {author}
  {\bibfnamefont {J.~B.}\ \bibnamefont {Haber}}, \bibinfo {author}
  {\bibfnamefont {F.}~\bibnamefont {Sagredo}}, \bibinfo {author} {\bibfnamefont
  {M.~R.}\ \bibnamefont {Filip}}, \bibinfo {author} {\bibfnamefont {J.~B.}\
  \bibnamefont {Neaton}},\ and\ \bibinfo {author} {\bibfnamefont
  {L.}~\bibnamefont {Kronik}},\ }\bibfield  {title} {\bibinfo {title} {Band
  gaps of halide perovskites from a wannier-localized optimally tuned screened
  range-separated hybrid functional},\ }\href@noop {} {\bibfield  {journal}
  {\bibinfo  {journal} {Phys. Rev. Mater.}\ }\textbf {\bibinfo {volume} {6}},\
  \bibinfo {pages} {104606} (\bibinfo {year} {2022})}\BibitemShut {NoStop}%
\bibitem [{\citenamefont {Wing}\ \emph
  {et~al.}(2019{\natexlab{b}})\citenamefont {Wing}, \citenamefont {Haber},
  \citenamefont {Noff}, \citenamefont {Barker}, \citenamefont {Egger},
  \citenamefont {Ramasubramaniam}, \citenamefont {Louie}, \citenamefont
  {Neaton},\ and\ \citenamefont {Kronik}}]{WingHabeJonah2019}%
  \BibitemOpen
  \bibfield  {author} {\bibinfo {author} {\bibfnamefont {D.}~\bibnamefont
  {Wing}}, \bibinfo {author} {\bibfnamefont {J.~B.}\ \bibnamefont {Haber}},
  \bibinfo {author} {\bibfnamefont {R.}~\bibnamefont {Noff}}, \bibinfo {author}
  {\bibfnamefont {B.}~\bibnamefont {Barker}}, \bibinfo {author} {\bibfnamefont
  {D.~A.}\ \bibnamefont {Egger}}, \bibinfo {author} {\bibfnamefont
  {A.}~\bibnamefont {Ramasubramaniam}}, \bibinfo {author} {\bibfnamefont
  {S.~G.}\ \bibnamefont {Louie}}, \bibinfo {author} {\bibfnamefont {J.~B.}\
  \bibnamefont {Neaton}},\ and\ \bibinfo {author} {\bibfnamefont
  {L.}~\bibnamefont {Kronik}},\ }\bibfield  {title} {\bibinfo {title}
  {Comparing time-dependent density functional theory with many-body
  perturbation theory for semiconductors: Screened range-separated hybrids and
  the $gw$ plus bethe-salpeter approach},\ }\href@noop {} {\bibfield  {journal}
  {\bibinfo  {journal} {Phys. Rev. Mater.}\ }\textbf {\bibinfo {volume} {3}},\
  \bibinfo {pages} {064603} (\bibinfo {year} {2019}{\natexlab{b}})}\BibitemShut
  {NoStop}%
\bibitem [{\citenamefont {Ramasubramaniam}\ \emph {et~al.}(2019)\citenamefont
  {Ramasubramaniam}, \citenamefont {Wing},\ and\ \citenamefont
  {Kronik}}]{AshwinDahvydLeeor2019}%
  \BibitemOpen
  \bibfield  {author} {\bibinfo {author} {\bibfnamefont {A.}~\bibnamefont
  {Ramasubramaniam}}, \bibinfo {author} {\bibfnamefont {D.}~\bibnamefont
  {Wing}},\ and\ \bibinfo {author} {\bibfnamefont {L.}~\bibnamefont {Kronik}},\
  }\bibfield  {title} {\bibinfo {title} {Transferable screened range-separated
  hybrids for layered materials: The cases of ${\mathrm{mos}}_{2}$ and h-bn},\
  }\href@noop {} {\bibfield  {journal} {\bibinfo  {journal} {Phys. Rev.
  Mater.}\ }\textbf {\bibinfo {volume} {3}},\ \bibinfo {pages} {084007}
  (\bibinfo {year} {2019})}\BibitemShut {NoStop}%
\bibitem [{\citenamefont {Perdew}\ and\ \citenamefont
  {Zunger}(1981)}]{PhysRevB.23.5048}%
  \BibitemOpen
  \bibfield  {author} {\bibinfo {author} {\bibfnamefont {J.~P.}\ \bibnamefont
  {Perdew}}\ and\ \bibinfo {author} {\bibfnamefont {A.}~\bibnamefont
  {Zunger}},\ }\bibfield  {title} {\bibinfo {title} {Self-interaction
  correction to density-functional approximations for many-electron systems},\
  }\href {https://doi.org/10.1103/PhysRevB.23.5048} {\bibfield  {journal}
  {\bibinfo  {journal} {Phys. Rev. B}\ }\textbf {\bibinfo {volume} {23}},\
  \bibinfo {pages} {5048} (\bibinfo {year} {1981})}\BibitemShut {NoStop}%
\bibitem [{\citenamefont {Gr\"uneis}\ \emph {et~al.}(2014)\citenamefont
  {Gr\"uneis}, \citenamefont {Kresse}, \citenamefont {Hinuma},\ and\
  \citenamefont {Oba}}]{AndrKressHinu2014}%
  \BibitemOpen
  \bibfield  {author} {\bibinfo {author} {\bibfnamefont {A.}~\bibnamefont
  {Gr\"uneis}}, \bibinfo {author} {\bibfnamefont {G.}~\bibnamefont {Kresse}},
  \bibinfo {author} {\bibfnamefont {Y.}~\bibnamefont {Hinuma}},\ and\ \bibinfo
  {author} {\bibfnamefont {F.}~\bibnamefont {Oba}},\ }\bibfield  {title}
  {\bibinfo {title} {Ionization potentials of solids: The importance of vertex
  corrections},\ }\href@noop {} {\bibfield  {journal} {\bibinfo  {journal}
  {Phys. Rev. Lett.}\ }\textbf {\bibinfo {volume} {112}},\ \bibinfo {pages}
  {096401} (\bibinfo {year} {2014})}\BibitemShut {NoStop}%
\bibitem [{\citenamefont {Shimazaki}\ and\ \citenamefont
  {Nakajima}(2014)}]{ShimNaka2014}%
  \BibitemOpen
  \bibfield  {author} {\bibinfo {author} {\bibfnamefont {T.}~\bibnamefont
  {Shimazaki}}\ and\ \bibinfo {author} {\bibfnamefont {T.}~\bibnamefont
  {Nakajima}},\ }\bibfield  {title} {\bibinfo {title} {Dielectric-dependent
  screened hartree–fock exchange potential and slater-formula with
  coulomb-hole interaction for energy band structure calculations},\
  }\href@noop {} {\bibfield  {journal} {\bibinfo  {journal} {J. Chem. Phys.}\
  }\textbf {\bibinfo {volume} {141}},\ \bibinfo {pages} {114109} (\bibinfo
  {year} {2014})}\BibitemShut {NoStop}%
\bibitem [{\citenamefont {Skone}\ \emph {et~al.}(2014)\citenamefont {Skone},
  \citenamefont {Govoni},\ and\ \citenamefont {Galli}}]{SkonGovoGall2014}%
  \BibitemOpen
  \bibfield  {author} {\bibinfo {author} {\bibfnamefont {J.~H.}\ \bibnamefont
  {Skone}}, \bibinfo {author} {\bibfnamefont {M.}~\bibnamefont {Govoni}},\ and\
  \bibinfo {author} {\bibfnamefont {G.}~\bibnamefont {Galli}},\ }\bibfield
  {title} {\bibinfo {title} {Self-consistent hybrid functional for condensed
  systems},\ }\href@noop {} {\bibfield  {journal} {\bibinfo  {journal} {Phys.
  Rev. B}\ }\textbf {\bibinfo {volume} {89}},\ \bibinfo {pages} {195112}
  (\bibinfo {year} {2014})}\BibitemShut {NoStop}%
\bibitem [{\citenamefont {Brawand}\ \emph {et~al.}(2016)\citenamefont
  {Brawand}, \citenamefont {V\"or\"os}, \citenamefont {Govoni},\ and\
  \citenamefont {Galli}}]{BrawVorosGovo2016}%
  \BibitemOpen
  \bibfield  {author} {\bibinfo {author} {\bibfnamefont {N.~P.}\ \bibnamefont
  {Brawand}}, \bibinfo {author} {\bibfnamefont {M.}~\bibnamefont {V\"or\"os}},
  \bibinfo {author} {\bibfnamefont {M.}~\bibnamefont {Govoni}},\ and\ \bibinfo
  {author} {\bibfnamefont {G.}~\bibnamefont {Galli}},\ }\bibfield  {title}
  {\bibinfo {title} {Generalization of dielectric-dependent hybrid functionals
  to finite systems},\ }\href@noop {} {\bibfield  {journal} {\bibinfo
  {journal} {Phys. Rev. X}\ }\textbf {\bibinfo {volume} {6}},\ \bibinfo {pages}
  {041002} (\bibinfo {year} {2016})}\BibitemShut {NoStop}%
\bibitem [{\citenamefont {Skone}\ \emph
  {et~al.}(2016{\natexlab{a}})\citenamefont {Skone}, \citenamefont {Govoni},\
  and\ \citenamefont {Galli}}]{SkonGovoGall2016}%
  \BibitemOpen
  \bibfield  {author} {\bibinfo {author} {\bibfnamefont {J.~H.}\ \bibnamefont
  {Skone}}, \bibinfo {author} {\bibfnamefont {M.}~\bibnamefont {Govoni}},\ and\
  \bibinfo {author} {\bibfnamefont {G.}~\bibnamefont {Galli}},\ }\bibfield
  {title} {\bibinfo {title} {Nonempirical range-separated hybrid functionals
  for solids and molecules},\ }\href@noop {} {\bibfield  {journal} {\bibinfo
  {journal} {Phys. Rev. B}\ }\textbf {\bibinfo {volume} {93}},\ \bibinfo
  {pages} {235106} (\bibinfo {year} {2016}{\natexlab{a}})}\BibitemShut
  {NoStop}%
\bibitem [{\citenamefont {Chen}\ \emph
  {et~al.}(2018{\natexlab{a}})\citenamefont {Chen}, \citenamefont {Miceli},
  \citenamefont {Rignanese},\ and\ \citenamefont
  {Pasquarello}}]{WeiGiaRigPas2018}%
  \BibitemOpen
  \bibfield  {author} {\bibinfo {author} {\bibfnamefont {W.}~\bibnamefont
  {Chen}}, \bibinfo {author} {\bibfnamefont {G.}~\bibnamefont {Miceli}},
  \bibinfo {author} {\bibfnamefont {G.-M.}\ \bibnamefont {Rignanese}},\ and\
  \bibinfo {author} {\bibfnamefont {A.}~\bibnamefont {Pasquarello}},\
  }\bibfield  {title} {\bibinfo {title} {Nonempirical dielectric-dependent
  hybrid functional with range separation for semiconductors and insulators},\
  }\href@noop {} {\bibfield  {journal} {\bibinfo  {journal} {Phys. Rev.
  Mater.}\ }\textbf {\bibinfo {volume} {2}},\ \bibinfo {pages} {073803}
  (\bibinfo {year} {2018}{\natexlab{a}})}\BibitemShut {NoStop}%
\bibitem [{\citenamefont {Cui}\ \emph {et~al.}(2018)\citenamefont {Cui},
  \citenamefont {Wang}, \citenamefont {Zhang}, \citenamefont {Xu},\ and\
  \citenamefont {Jiang}}]{CuiWangZhang2018}%
  \BibitemOpen
  \bibfield  {author} {\bibinfo {author} {\bibfnamefont {Z.-H.}\ \bibnamefont
  {Cui}}, \bibinfo {author} {\bibfnamefont {Y.-C.}\ \bibnamefont {Wang}},
  \bibinfo {author} {\bibfnamefont {M.-Y.}\ \bibnamefont {Zhang}}, \bibinfo
  {author} {\bibfnamefont {X.}~\bibnamefont {Xu}},\ and\ \bibinfo {author}
  {\bibfnamefont {H.}~\bibnamefont {Jiang}},\ }\bibfield  {title} {\bibinfo
  {title} {Doubly screened hybrid functional: An accurate first-principles
  approach for both narrow- and wide-gap semiconductors},\ }\href@noop {}
  {\bibfield  {journal} {\bibinfo  {journal} {J. Phys. Chem. Lett.}\ }\textbf
  {\bibinfo {volume} {9}},\ \bibinfo {pages} {2338} (\bibinfo {year}
  {2018})}\BibitemShut {NoStop}%
\bibitem [{\citenamefont {Lorke}\ \emph {et~al.}(2020)\citenamefont {Lorke},
  \citenamefont {De\'ak},\ and\ \citenamefont {Frauenheim}}]{MichPeteThom2020}%
  \BibitemOpen
  \bibfield  {author} {\bibinfo {author} {\bibfnamefont {M.}~\bibnamefont
  {Lorke}}, \bibinfo {author} {\bibfnamefont {P.}~\bibnamefont {De\'ak}},\ and\
  \bibinfo {author} {\bibfnamefont {T.}~\bibnamefont {Frauenheim}},\ }\bibfield
   {title} {\bibinfo {title} {Koopmans-compliant screened exchange potential
  with correct asymptotic behavior for semiconductors},\ }\href@noop {}
  {\bibfield  {journal} {\bibinfo  {journal} {Phys. Rev. B}\ }\textbf {\bibinfo
  {volume} {102}},\ \bibinfo {pages} {235168} (\bibinfo {year}
  {2020})}\BibitemShut {NoStop}%
\bibitem [{\citenamefont {Jana}\ \emph
  {et~al.}(2023{\natexlab{a}})\citenamefont {Jana}, \citenamefont {Ghosh},
  \citenamefont {Constantin},\ and\ \citenamefont {Samal}}]{Jana2023Simple}%
  \BibitemOpen
  \bibfield  {author} {\bibinfo {author} {\bibfnamefont {S.}~\bibnamefont
  {Jana}}, \bibinfo {author} {\bibfnamefont {A.}~\bibnamefont {Ghosh}},
  \bibinfo {author} {\bibfnamefont {L.~A.}\ \bibnamefont {Constantin}},\ and\
  \bibinfo {author} {\bibfnamefont {P.}~\bibnamefont {Samal}},\ }\bibfield
  {title} {\bibinfo {title} {Simple and effective screening parameter for
  range-separated dielectric-dependent hybrids},\ }\href
  {https://doi.org/10.1103/PhysRevB.108.045101} {\bibfield  {journal} {\bibinfo
   {journal} {Phys. Rev. B}\ }\textbf {\bibinfo {volume} {108}},\ \bibinfo
  {pages} {045101} (\bibinfo {year} {2023}{\natexlab{a}})}\BibitemShut
  {NoStop}%
\bibitem [{\citenamefont {Bischoff}\ \emph {et~al.}(2019)\citenamefont
  {Bischoff}, \citenamefont {Wiktor}, \citenamefont {Chen},\ and\ \citenamefont
  {Pasquarello}}]{Bischoff2019Nonempirical}%
  \BibitemOpen
  \bibfield  {author} {\bibinfo {author} {\bibfnamefont {T.}~\bibnamefont
  {Bischoff}}, \bibinfo {author} {\bibfnamefont {J.}~\bibnamefont {Wiktor}},
  \bibinfo {author} {\bibfnamefont {W.}~\bibnamefont {Chen}},\ and\ \bibinfo
  {author} {\bibfnamefont {A.}~\bibnamefont {Pasquarello}},\ }\bibfield
  {title} {\bibinfo {title} {Nonempirical hybrid functionals for band gaps of
  inorganic metal-halide perovskites},\ }\href
  {https://doi.org/10.1103/PhysRevMaterials.3.123802} {\bibfield  {journal}
  {\bibinfo  {journal} {Phys. Rev. Mater.}\ }\textbf {\bibinfo {volume} {3}},\
  \bibinfo {pages} {123802} (\bibinfo {year} {2019})}\BibitemShut {NoStop}%
\bibitem [{\citenamefont {Wang}\ \emph {et~al.}(2022)\citenamefont {Wang},
  \citenamefont {Tal}, \citenamefont {Bischoff}, \citenamefont {Gono},\ and\
  \citenamefont {Pasquarello}}]{Wang2022Accurate}%
  \BibitemOpen
  \bibfield  {author} {\bibinfo {author} {\bibfnamefont {H.}~\bibnamefont
  {Wang}}, \bibinfo {author} {\bibfnamefont {A.}~\bibnamefont {Tal}}, \bibinfo
  {author} {\bibfnamefont {T.}~\bibnamefont {Bischoff}}, \bibinfo {author}
  {\bibfnamefont {P.}~\bibnamefont {Gono}},\ and\ \bibinfo {author}
  {\bibfnamefont {A.}~\bibnamefont {Pasquarello}},\ }\bibfield  {title}
  {\bibinfo {title} {Accurate and efficient band-gap predictions for metal
  halide perovskites at finite temperature},\ }\href
  {https://doi.org/10.1038/s41524-022-00869-6} {\bibfield  {journal} {\bibinfo
  {journal} {npj Computational Materials}\ }\textbf {\bibinfo {volume} {8}},\
  \bibinfo {pages} {237} (\bibinfo {year} {2022})}\BibitemShut {NoStop}%
\bibitem [{\citenamefont {Yang}\ \emph {et~al.}(2023)\citenamefont {Yang},
  \citenamefont {Falletta},\ and\ \citenamefont {Pasquarello}}]{Yang2023Range}%
  \BibitemOpen
  \bibfield  {author} {\bibinfo {author} {\bibfnamefont {J.}~\bibnamefont
  {Yang}}, \bibinfo {author} {\bibfnamefont {S.}~\bibnamefont {Falletta}},\
  and\ \bibinfo {author} {\bibfnamefont {A.}~\bibnamefont {Pasquarello}},\
  }\bibfield  {title} {\bibinfo {title} {Range-separated hybrid functionals for
  accurate prediction of band gaps of extended systems},\ }\href
  {https://doi.org/10.1038/s41524-023-01064-x} {\bibfield  {journal} {\bibinfo
  {journal} {npj Computational Materials}\ }\textbf {\bibinfo {volume} {9}},\
  \bibinfo {pages} {108} (\bibinfo {year} {2023})}\BibitemShut {NoStop}%
\bibitem [{\citenamefont {Miceli}\ \emph {et~al.}(2018)\citenamefont {Miceli},
  \citenamefont {Chen}, \citenamefont {Reshetnyak},\ and\ \citenamefont
  {Pasquarello}}]{Miceli2018Nonempirical}%
  \BibitemOpen
  \bibfield  {author} {\bibinfo {author} {\bibfnamefont {G.}~\bibnamefont
  {Miceli}}, \bibinfo {author} {\bibfnamefont {W.}~\bibnamefont {Chen}},
  \bibinfo {author} {\bibfnamefont {I.}~\bibnamefont {Reshetnyak}},\ and\
  \bibinfo {author} {\bibfnamefont {A.}~\bibnamefont {Pasquarello}},\
  }\bibfield  {title} {\bibinfo {title} {Nonempirical hybrid functionals for
  band gaps and polaronic distortions in solids},\ }\href
  {https://doi.org/10.1103/PhysRevB.97.121112} {\bibfield  {journal} {\bibinfo
  {journal} {Phys. Rev. B}\ }\textbf {\bibinfo {volume} {97}},\ \bibinfo
  {pages} {121112} (\bibinfo {year} {2018})}\BibitemShut {NoStop}%
\bibitem [{\citenamefont {Yang}\ \emph {et~al.}(2022)\citenamefont {Yang},
  \citenamefont {Falletta},\ and\ \citenamefont
  {Pasquarello}}]{Yang2022OneShot}%
  \BibitemOpen
  \bibfield  {author} {\bibinfo {author} {\bibfnamefont {J.}~\bibnamefont
  {Yang}}, \bibinfo {author} {\bibfnamefont {S.}~\bibnamefont {Falletta}},\
  and\ \bibinfo {author} {\bibfnamefont {A.}~\bibnamefont {Pasquarello}},\
  }\bibfield  {title} {\bibinfo {title} {One-shot approach for enforcing
  piecewise linearity on hybrid functionals: Application to band gap
  predictions},\ }\href {https://doi.org/10.1021/acs.jpclett.2c00414}
  {\bibfield  {journal} {\bibinfo  {journal} {J. Phys. Chem. Lett.}\ }\textbf
  {\bibinfo {volume} {13}},\ \bibinfo {pages} {3066} (\bibinfo {year}
  {2022})}\BibitemShut {NoStop}%
\bibitem [{\citenamefont {Tal}\ \emph {et~al.}(2020)\citenamefont {Tal},
  \citenamefont {Liu}, \citenamefont {Kresse},\ and\ \citenamefont
  {Pasquarello}}]{tal2020accurate}%
  \BibitemOpen
  \bibfield  {author} {\bibinfo {author} {\bibfnamefont {A.}~\bibnamefont
  {Tal}}, \bibinfo {author} {\bibfnamefont {P.}~\bibnamefont {Liu}}, \bibinfo
  {author} {\bibfnamefont {G.}~\bibnamefont {Kresse}},\ and\ \bibinfo {author}
  {\bibfnamefont {A.}~\bibnamefont {Pasquarello}},\ }\bibfield  {title}
  {\bibinfo {title} {Accurate optical spectra through time-dependent density
  functional theory based on screening-dependent hybrid functionals},\
  }\href@noop {} {\bibfield  {journal} {\bibinfo  {journal} {Phys. Rev. Res.}\
  }\textbf {\bibinfo {volume} {2}},\ \bibinfo {pages} {032019} (\bibinfo {year}
  {2020})}\BibitemShut {NoStop}%
\bibitem [{\citenamefont {Jin}\ \emph {et~al.}(2024)\citenamefont {Jin},
  \citenamefont {Rusishvili}, \citenamefont {Govoni},\ and\ \citenamefont
  {Galli}}]{Jin2024Self}%
  \BibitemOpen
  \bibfield  {author} {\bibinfo {author} {\bibfnamefont {Y.}~\bibnamefont
  {Jin}}, \bibinfo {author} {\bibfnamefont {M.}~\bibnamefont {Rusishvili}},
  \bibinfo {author} {\bibfnamefont {M.}~\bibnamefont {Govoni}},\ and\ \bibinfo
  {author} {\bibfnamefont {G.}~\bibnamefont {Galli}},\ }\bibfield  {title}
  {\bibinfo {title} {Self-trapped excitons in metal-halide perovskites
  investigated by time-dependent density functional theory},\ }\href
  {https://doi.org/10.1021/acs.jpclett.4c00209} {\bibfield  {journal} {\bibinfo
   {journal} {J. Phys. Chem. Lett.}\ }\textbf {\bibinfo {volume} {15}},\
  \bibinfo {pages} {3229} (\bibinfo {year} {2024})}\BibitemShut {NoStop}%
\bibitem [{\citenamefont {Gaiduk}\ \emph {et~al.}(2016)\citenamefont {Gaiduk},
  \citenamefont {Govoni}, \citenamefont {Seidel}, \citenamefont {Skone},
  \citenamefont {Winter},\ and\ \citenamefont
  {Galli}}]{Gaiduk2016Photoelectron}%
  \BibitemOpen
  \bibfield  {author} {\bibinfo {author} {\bibfnamefont {A.~P.}\ \bibnamefont
  {Gaiduk}}, \bibinfo {author} {\bibfnamefont {M.}~\bibnamefont {Govoni}},
  \bibinfo {author} {\bibfnamefont {R.}~\bibnamefont {Seidel}}, \bibinfo
  {author} {\bibfnamefont {J.~H.}\ \bibnamefont {Skone}}, \bibinfo {author}
  {\bibfnamefont {B.}~\bibnamefont {Winter}},\ and\ \bibinfo {author}
  {\bibfnamefont {G.}~\bibnamefont {Galli}},\ }\bibfield  {title} {\bibinfo
  {title} {Photoelectron spectra of aqueous solutions from first principles},\
  }\href {https://doi.org/10.1021/jacs.6b00225} {\bibfield  {journal} {\bibinfo
   {journal} {Journal of the American Chemical Society}\ }\textbf {\bibinfo
  {volume} {138}},\ \bibinfo {pages} {6912} (\bibinfo {year}
  {2016})}\BibitemShut {NoStop}%
\bibitem [{\citenamefont {Jin}\ \emph {et~al.}(2022)\citenamefont {Jin},
  \citenamefont {Govoni},\ and\ \citenamefont {Galli}}]{Jin2022Vibrationally}%
  \BibitemOpen
  \bibfield  {author} {\bibinfo {author} {\bibfnamefont {Y.}~\bibnamefont
  {Jin}}, \bibinfo {author} {\bibfnamefont {M.}~\bibnamefont {Govoni}},\ and\
  \bibinfo {author} {\bibfnamefont {G.}~\bibnamefont {Galli}},\ }\bibfield
  {title} {\bibinfo {title} {Vibrationally resolved optical excitations of the
  nitrogen-vacancy center in diamond},\ }\href
  {https://doi.org/10.1038/s41524-022-00928-y} {\bibfield  {journal} {\bibinfo
  {journal} {npj Computational Materials}\ }\textbf {\bibinfo {volume} {8}},\
  \bibinfo {pages} {238} (\bibinfo {year} {2022})}\BibitemShut {NoStop}%
\bibitem [{\citenamefont {Zheng}\ \emph {et~al.}(2019)\citenamefont {Zheng},
  \citenamefont {Govoni},\ and\ \citenamefont {Galli}}]{ZhengGovoniGalli2019}%
  \BibitemOpen
  \bibfield  {author} {\bibinfo {author} {\bibfnamefont {H.}~\bibnamefont
  {Zheng}}, \bibinfo {author} {\bibfnamefont {M.}~\bibnamefont {Govoni}},\ and\
  \bibinfo {author} {\bibfnamefont {G.}~\bibnamefont {Galli}},\ }\bibfield
  {title} {\bibinfo {title} {Dielectric-dependent hybrid functionals for
  heterogeneous materials},\ }\href@noop {} {\bibfield  {journal} {\bibinfo
  {journal} {Phys. Rev. Mater.}\ }\textbf {\bibinfo {volume} {3}},\ \bibinfo
  {pages} {073803} (\bibinfo {year} {2019})}\BibitemShut {NoStop}%
\bibitem [{\citenamefont {Sun}\ \emph {et~al.}(2015)\citenamefont {Sun},
  \citenamefont {Ruzsinszky},\ and\ \citenamefont {Perdew}}]{sun2015strongly}%
  \BibitemOpen
  \bibfield  {author} {\bibinfo {author} {\bibfnamefont {J.}~\bibnamefont
  {Sun}}, \bibinfo {author} {\bibfnamefont {A.}~\bibnamefont {Ruzsinszky}},\
  and\ \bibinfo {author} {\bibfnamefont {J.~P.}\ \bibnamefont {Perdew}},\
  }\bibfield  {title} {\bibinfo {title} {Strongly constrained and appropriately
  normed semilocal density functional},\ }\href@noop {} {\bibfield  {journal}
  {\bibinfo  {journal} {Phys. Rev. Lett.}\ }\textbf {\bibinfo {volume} {115}},\
  \bibinfo {pages} {036402} (\bibinfo {year} {2015})}\BibitemShut {NoStop}%
\bibitem [{\citenamefont {Tao}\ and\ \citenamefont
  {Mo}(2016)}]{tao2016accurate}%
  \BibitemOpen
  \bibfield  {author} {\bibinfo {author} {\bibfnamefont {J.}~\bibnamefont
  {Tao}}\ and\ \bibinfo {author} {\bibfnamefont {Y.}~\bibnamefont {Mo}},\
  }\bibfield  {title} {\bibinfo {title} {Accurate semilocal density functional
  for condensed-matter physics and quantum chemistry},\ }\href@noop {}
  {\bibfield  {journal} {\bibinfo  {journal} {Phys. Rev. Lett.}\ }\textbf
  {\bibinfo {volume} {117}},\ \bibinfo {pages} {073001} (\bibinfo {year}
  {2016})}\BibitemShut {NoStop}%
\bibitem [{\citenamefont {Furness}\ \emph {et~al.}(2020)\citenamefont
  {Furness}, \citenamefont {Kaplan}, \citenamefont {Ning}, \citenamefont
  {Perdew},\ and\ \citenamefont {Sun}}]{furness2020accurate}%
  \BibitemOpen
  \bibfield  {author} {\bibinfo {author} {\bibfnamefont {J.~W.}\ \bibnamefont
  {Furness}}, \bibinfo {author} {\bibfnamefont {A.~D.}\ \bibnamefont {Kaplan}},
  \bibinfo {author} {\bibfnamefont {J.}~\bibnamefont {Ning}}, \bibinfo {author}
  {\bibfnamefont {J.~P.}\ \bibnamefont {Perdew}},\ and\ \bibinfo {author}
  {\bibfnamefont {J.}~\bibnamefont {Sun}},\ }\bibfield  {title} {\bibinfo
  {title} {Accurate and numerically efficient r2scan meta-generalized gradient
  approximation},\ }\href@noop {} {\bibfield  {journal} {\bibinfo  {journal}
  {J. Phys. Chem. Lett.}\ }\textbf {\bibinfo {volume} {11}},\ \bibinfo {pages}
  {8208} (\bibinfo {year} {2020})}\BibitemShut {NoStop}%
\bibitem [{\citenamefont {Neupane}\ \emph {et~al.}(2021)\citenamefont
  {Neupane}, \citenamefont {Tang}, \citenamefont {Nepal}, \citenamefont
  {Adhikari},\ and\ \citenamefont {Ruzsinszky}}]{Neupane2021Opening}%
  \BibitemOpen
  \bibfield  {author} {\bibinfo {author} {\bibfnamefont {B.}~\bibnamefont
  {Neupane}}, \bibinfo {author} {\bibfnamefont {H.}~\bibnamefont {Tang}},
  \bibinfo {author} {\bibfnamefont {N.~K.}\ \bibnamefont {Nepal}}, \bibinfo
  {author} {\bibfnamefont {S.}~\bibnamefont {Adhikari}},\ and\ \bibinfo
  {author} {\bibfnamefont {A.}~\bibnamefont {Ruzsinszky}},\ }\bibfield  {title}
  {\bibinfo {title} {Opening band gaps of low-dimensional materials at the
  meta-gga level of density functional approximations},\ }\href
  {https://doi.org/10.1103/PhysRevMaterials.5.063803} {\bibfield  {journal}
  {\bibinfo  {journal} {Phys. Rev. Mater.}\ }\textbf {\bibinfo {volume} {5}},\
  \bibinfo {pages} {063803} (\bibinfo {year} {2021})}\BibitemShut {NoStop}%
\bibitem [{\citenamefont {Patra}\ \emph
  {et~al.}(2019{\natexlab{b}})\citenamefont {Patra}, \citenamefont {Jana},
  \citenamefont {Constantin},\ and\ \citenamefont
  {Samal}}]{patra2019relevance}%
  \BibitemOpen
  \bibfield  {author} {\bibinfo {author} {\bibfnamefont {B.}~\bibnamefont
  {Patra}}, \bibinfo {author} {\bibfnamefont {S.}~\bibnamefont {Jana}},
  \bibinfo {author} {\bibfnamefont {L.~A.}\ \bibnamefont {Constantin}},\ and\
  \bibinfo {author} {\bibfnamefont {P.}~\bibnamefont {Samal}},\ }\bibfield
  {title} {\bibinfo {title} {Relevance of the pauli kinetic energy density for
  semilocal functionals},\ }\href@noop {} {\bibfield  {journal} {\bibinfo
  {journal} {Phys. Rev. B}\ }\textbf {\bibinfo {volume} {100}},\ \bibinfo
  {pages} {155140} (\bibinfo {year} {2019}{\natexlab{b}})}\BibitemShut
  {NoStop}%
\bibitem [{\citenamefont {Jana}\ \emph
  {et~al.}(2023{\natexlab{b}})\citenamefont {Jana}, \citenamefont {\'Smiga},
  \citenamefont {Constantin},\ and\ \citenamefont {Samal}}]{JanaACSC}%
  \BibitemOpen
  \bibfield  {author} {\bibinfo {author} {\bibfnamefont {S.}~\bibnamefont
  {Jana}}, \bibinfo {author} {\bibfnamefont {S.}~\bibnamefont {\'Smiga}},
  \bibinfo {author} {\bibfnamefont {L.~A.}\ \bibnamefont {Constantin}},\ and\
  \bibinfo {author} {\bibfnamefont {P.}~\bibnamefont {Samal}},\ }\bibfield
  {title} {\bibinfo {title} {Semilocal meta-gga exchange–correlation
  approximation from adiabatic connection formalism: Extent and limitations},\
  }\href {https://doi.org/10.1021/acs.jpca.3c03976} {\bibfield  {journal}
  {\bibinfo  {journal} {The Journal of Physical Chemistry A}\ }\textbf
  {\bibinfo {volume} {127}},\ \bibinfo {pages} {8685} (\bibinfo {year}
  {2023}{\natexlab{b}})},\ \bibinfo {note} {pMID: 37811903},\ \Eprint
  {https://arxiv.org/abs/https://doi.org/10.1021/acs.jpca.3c03976}
  {https://doi.org/10.1021/acs.jpca.3c03976} \BibitemShut {NoStop}%
\bibitem [{\citenamefont {Lebeda}\ \emph {et~al.}(2024)\citenamefont {Lebeda},
  \citenamefont {Aschebrock},\ and\ \citenamefont
  {K\"ummel}}]{Lebeda2024Balancing}%
  \BibitemOpen
  \bibfield  {author} {\bibinfo {author} {\bibfnamefont {T.}~\bibnamefont
  {Lebeda}}, \bibinfo {author} {\bibfnamefont {T.}~\bibnamefont {Aschebrock}},\
  and\ \bibinfo {author} {\bibfnamefont {S.}~\bibnamefont {K\"ummel}},\
  }\bibfield  {title} {\bibinfo {title} {Balancing the contributions to the
  gradient expansion: Accurate binding and band gaps with a nonempirical
  meta-gga},\ }\href {https://doi.org/10.1103/PhysRevLett.133.136402}
  {\bibfield  {journal} {\bibinfo  {journal} {Phys. Rev. Lett.}\ }\textbf
  {\bibinfo {volume} {133}},\ \bibinfo {pages} {136402} (\bibinfo {year}
  {2024})}\BibitemShut {NoStop}%
\bibitem [{\citenamefont {Ghosh}\ \emph {et~al.}(2022)\citenamefont {Ghosh},
  \citenamefont {Jana}, \citenamefont {Rauch}, \citenamefont {Tran},
  \citenamefont {Marques}, \citenamefont {Botti}, \citenamefont {Constantin},
  \citenamefont {Niranjan},\ and\ \citenamefont {Samal}}]{GhoshJanaRauch2022}%
  \BibitemOpen
  \bibfield  {author} {\bibinfo {author} {\bibfnamefont {A.}~\bibnamefont
  {Ghosh}}, \bibinfo {author} {\bibfnamefont {S.}~\bibnamefont {Jana}},
  \bibinfo {author} {\bibfnamefont {T.}~\bibnamefont {Rauch}}, \bibinfo
  {author} {\bibfnamefont {F.}~\bibnamefont {Tran}}, \bibinfo {author}
  {\bibfnamefont {M.~A.~L.}\ \bibnamefont {Marques}}, \bibinfo {author}
  {\bibfnamefont {S.}~\bibnamefont {Botti}}, \bibinfo {author} {\bibfnamefont
  {L.~A.}\ \bibnamefont {Constantin}}, \bibinfo {author} {\bibfnamefont
  {M.~K.}\ \bibnamefont {Niranjan}},\ and\ \bibinfo {author} {\bibfnamefont
  {P.}~\bibnamefont {Samal}},\ }\bibfield  {title} {\bibinfo {title} {Efficient
  and improved prediction of the band offsets at semiconductor heterojunctions
  from meta-gga density functionals: A benchmark study},\ }\href@noop {}
  {\bibfield  {journal} {\bibinfo  {journal} {J. Chem. Phys.}\ }\textbf
  {\bibinfo {volume} {157}},\ \bibinfo {pages} {124108} (\bibinfo {year}
  {2022})}\BibitemShut {NoStop}%
\bibitem [{\citenamefont {Jana}\ \emph {et~al.}(2022)\citenamefont {Jana},
  \citenamefont {Constantin}, \citenamefont {Śmiga},\ and\ \citenamefont
  {Samal}}]{Jana2022Solid}%
  \BibitemOpen
  \bibfield  {author} {\bibinfo {author} {\bibfnamefont {S.}~\bibnamefont
  {Jana}}, \bibinfo {author} {\bibfnamefont {L.~A.}\ \bibnamefont
  {Constantin}}, \bibinfo {author} {\bibfnamefont {S.}~\bibnamefont {Śmiga}},\
  and\ \bibinfo {author} {\bibfnamefont {P.}~\bibnamefont {Samal}},\ }\bibfield
   {title} {\bibinfo {title} {{Solid-state performance of a meta-GGA screened
  hybrid density functional constructed from Pauli kinetic enhancement factor
  dependent semilocal exchange hole}},\ }\href@noop {} {\bibfield  {journal}
  {\bibinfo  {journal} {J. Chem. Phys.}\ }\textbf {\bibinfo {volume} {157}},\
  \bibinfo {pages} {024102} (\bibinfo {year} {2022})}\BibitemShut {NoStop}%
\bibitem [{\citenamefont {Yanai}\ \emph {et~al.}(2004)\citenamefont {Yanai},
  \citenamefont {Tew},\ and\ \citenamefont {Handy}}]{YANAI200451}%
  \BibitemOpen
  \bibfield  {author} {\bibinfo {author} {\bibfnamefont {T.}~\bibnamefont
  {Yanai}}, \bibinfo {author} {\bibfnamefont {D.~P.}\ \bibnamefont {Tew}},\
  and\ \bibinfo {author} {\bibfnamefont {N.~C.}\ \bibnamefont {Handy}},\
  }\bibfield  {title} {\bibinfo {title} {A new hybrid exchange–correlation
  functional using the coulomb-attenuating method (cam-b3lyp)},\ }\href
  {https://doi.org/https://doi.org/10.1016/j.cplett.2004.06.011} {\bibfield
  {journal} {\bibinfo  {journal} {Chemical Physics Letters}\ }\textbf {\bibinfo
  {volume} {393}},\ \bibinfo {pages} {51} (\bibinfo {year} {2004})}\BibitemShut
  {NoStop}%
\bibitem [{\citenamefont {Skone}\ \emph
  {et~al.}(2016{\natexlab{b}})\citenamefont {Skone}, \citenamefont {Govoni},\
  and\ \citenamefont {Galli}}]{Skone2016Nonempirical}%
  \BibitemOpen
  \bibfield  {author} {\bibinfo {author} {\bibfnamefont {J.~H.}\ \bibnamefont
  {Skone}}, \bibinfo {author} {\bibfnamefont {M.}~\bibnamefont {Govoni}},\ and\
  \bibinfo {author} {\bibfnamefont {G.}~\bibnamefont {Galli}},\ }\bibfield
  {title} {\bibinfo {title} {Nonempirical range-separated hybrid functionals
  for solids and molecules},\ }\href
  {https://doi.org/10.1103/PhysRevB.93.235106} {\bibfield  {journal} {\bibinfo
  {journal} {Phys. Rev. B}\ }\textbf {\bibinfo {volume} {93}},\ \bibinfo
  {pages} {235106} (\bibinfo {year} {2016}{\natexlab{b}})}\BibitemShut
  {NoStop}%
\bibitem [{\citenamefont {Kronik}\ and\ \citenamefont
  {Kümmel}(2018)}]{kronik2018dielectric}%
  \BibitemOpen
  \bibfield  {author} {\bibinfo {author} {\bibfnamefont {L.}~\bibnamefont
  {Kronik}}\ and\ \bibinfo {author} {\bibfnamefont {S.}~\bibnamefont
  {Kümmel}},\ }\bibfield  {title} {\bibinfo {title} {Dielectric screening
  meets optimally tuned density functionals},\ }\href@noop {} {\bibfield
  {journal} {\bibinfo  {journal} {Advanced Materials}\ }\textbf {\bibinfo
  {volume} {30}},\ \bibinfo {pages} {1706560} (\bibinfo {year}
  {2018})}\BibitemShut {NoStop}%
\bibitem [{\citenamefont {Liu}\ \emph {et~al.}(2019)\citenamefont {Liu},
  \citenamefont {Franchini}, \citenamefont {Marsman},\ and\ \citenamefont
  {Kresse}}]{Liu2019Assessing}%
  \BibitemOpen
  \bibfield  {author} {\bibinfo {author} {\bibfnamefont {P.}~\bibnamefont
  {Liu}}, \bibinfo {author} {\bibfnamefont {C.}~\bibnamefont {Franchini}},
  \bibinfo {author} {\bibfnamefont {M.}~\bibnamefont {Marsman}},\ and\ \bibinfo
  {author} {\bibfnamefont {G.}~\bibnamefont {Kresse}},\ }\bibfield  {title}
  {\bibinfo {title} {Assessing model-dielectric-dependent hybrid functionals on
  the antiferromagnetic transition-metal monoxides mno, feo, coo, and nio},\
  }\href {https://doi.org/10.1088/1361-648X/ab4150} {\bibfield  {journal}
  {\bibinfo  {journal} {Journal of Physics: Condensed Matter}\ }\textbf
  {\bibinfo {volume} {32}},\ \bibinfo {pages} {015502} (\bibinfo {year}
  {2019})}\BibitemShut {NoStop}%
\bibitem [{\citenamefont {Marques}\ \emph {et~al.}(2011)\citenamefont
  {Marques}, \citenamefont {Vidal}, \citenamefont {Oliveira}, \citenamefont
  {Reining},\ and\ \citenamefont {Botti}}]{Marques2011density}%
  \BibitemOpen
  \bibfield  {author} {\bibinfo {author} {\bibfnamefont {M.~A.~L.}\
  \bibnamefont {Marques}}, \bibinfo {author} {\bibfnamefont {J.}~\bibnamefont
  {Vidal}}, \bibinfo {author} {\bibfnamefont {M.~J.~T.}\ \bibnamefont
  {Oliveira}}, \bibinfo {author} {\bibfnamefont {L.}~\bibnamefont {Reining}},\
  and\ \bibinfo {author} {\bibfnamefont {S.}~\bibnamefont {Botti}},\ }\bibfield
   {title} {\bibinfo {title} {Density-based mixing parameter for hybrid
  functionals},\ }\href {https://doi.org/10.1103/PhysRevB.83.035119} {\bibfield
   {journal} {\bibinfo  {journal} {Phys. Rev. B}\ }\textbf {\bibinfo {volume}
  {83}},\ \bibinfo {pages} {035119} (\bibinfo {year} {2011})}\BibitemShut
  {NoStop}%
\bibitem [{\citenamefont {Chen}\ \emph
  {et~al.}(2018{\natexlab{b}})\citenamefont {Chen}, \citenamefont {Miceli},
  \citenamefont {Rignanese},\ and\ \citenamefont
  {Pasquarello}}]{Chen2018Nonempirical}%
  \BibitemOpen
  \bibfield  {author} {\bibinfo {author} {\bibfnamefont {W.}~\bibnamefont
  {Chen}}, \bibinfo {author} {\bibfnamefont {G.}~\bibnamefont {Miceli}},
  \bibinfo {author} {\bibfnamefont {G.-M.}\ \bibnamefont {Rignanese}},\ and\
  \bibinfo {author} {\bibfnamefont {A.}~\bibnamefont {Pasquarello}},\
  }\bibfield  {title} {\bibinfo {title} {Nonempirical dielectric-dependent
  hybrid functional with range separation for semiconductors and insulators},\
  }\href {https://doi.org/10.1103/PhysRevMaterials.2.073803} {\bibfield
  {journal} {\bibinfo  {journal} {Phys. Rev. Mater.}\ }\textbf {\bibinfo
  {volume} {2}},\ \bibinfo {pages} {073803} (\bibinfo {year}
  {2018}{\natexlab{b}})}\BibitemShut {NoStop}%
\bibitem [{\citenamefont {Tao}\ \emph {et~al.}(2017)\citenamefont {Tao},
  \citenamefont {Bulik},\ and\ \citenamefont {Scuseria}}]{Tao2017Semilocal}%
  \BibitemOpen
  \bibfield  {author} {\bibinfo {author} {\bibfnamefont {J.}~\bibnamefont
  {Tao}}, \bibinfo {author} {\bibfnamefont {I.~W.}\ \bibnamefont {Bulik}},\
  and\ \bibinfo {author} {\bibfnamefont {G.~E.}\ \bibnamefont {Scuseria}},\
  }\bibfield  {title} {\bibinfo {title} {Semilocal exchange hole with an
  application to range-separated density functionals},\ }\href
  {https://doi.org/10.1103/PhysRevB.95.125115} {\bibfield  {journal} {\bibinfo
  {journal} {Phys. Rev. B}\ }\textbf {\bibinfo {volume} {95}},\ \bibinfo
  {pages} {125115} (\bibinfo {year} {2017})}\BibitemShut {NoStop}%
\bibitem [{\citenamefont {Jana}\ and\ \citenamefont
  {Samal}(2019)}]{jana2019screened}%
  \BibitemOpen
  \bibfield  {author} {\bibinfo {author} {\bibfnamefont {S.}~\bibnamefont
  {Jana}}\ and\ \bibinfo {author} {\bibfnamefont {P.}~\bibnamefont {Samal}},\
  }\bibfield  {title} {\bibinfo {title} {Screened hybrid meta-gga
  exchange–correlation functionals for extended systems},\ }\href@noop {}
  {\bibfield  {journal} {\bibinfo  {journal} {Phys. Chem. Chem. Phys.}\
  }\textbf {\bibinfo {volume} {21}},\ \bibinfo {pages} {3002} (\bibinfo {year}
  {2019})}\BibitemShut {NoStop}%
\bibitem [{\citenamefont {Jana}\ and\ \citenamefont
  {Samal}(2018)}]{janameta2018}%
  \BibitemOpen
  \bibfield  {author} {\bibinfo {author} {\bibfnamefont {S.}~\bibnamefont
  {Jana}}\ and\ \bibinfo {author} {\bibfnamefont {P.}~\bibnamefont {Samal}},\
  }\bibfield  {title} {\bibinfo {title} {A meta-gga level screened
  range-separated hybrid functional by employing short range hartree-fock with
  a long range semilocal functional},\ }\href@noop {} {\bibfield  {journal}
  {\bibinfo  {journal} {Phys. Chem. Chem. Phys.}\ }\textbf {\bibinfo {volume}
  {20}},\ \bibinfo {pages} {8999} (\bibinfo {year} {2018})}\BibitemShut
  {NoStop}%
\bibitem [{\citenamefont {Jana}\ \emph {et~al.}(2021)\citenamefont {Jana},
  \citenamefont {Behera}, \citenamefont {{\'{S}}miga}, \citenamefont
  {Constantin},\ and\ \citenamefont {Samal}}]{jana2021szs}%
  \BibitemOpen
  \bibfield  {author} {\bibinfo {author} {\bibfnamefont {S.}~\bibnamefont
  {Jana}}, \bibinfo {author} {\bibfnamefont {S.~K.}\ \bibnamefont {Behera}},
  \bibinfo {author} {\bibfnamefont {S.}~\bibnamefont {{\'{S}}miga}}, \bibinfo
  {author} {\bibfnamefont {L.~A.}\ \bibnamefont {Constantin}},\ and\ \bibinfo
  {author} {\bibfnamefont {P.}~\bibnamefont {Samal}},\ }\bibfield  {title}
  {\bibinfo {title} {Improving the applicability of the pauli kinetic energy
  density based semilocal functional for solids},\ }\href
  {https://doi.org/10.1088/1367-2630/abfd4d} {\bibfield  {journal} {\bibinfo
  {journal} {New J. Phys.}\ }\textbf {\bibinfo {volume} {23}},\ \bibinfo
  {pages} {063007} (\bibinfo {year} {2021})}\BibitemShut {NoStop}%
\bibitem [{\citenamefont {Ghosh}\ \emph {et~al.}(2024)\citenamefont {Ghosh},
  \citenamefont {Jana}, \citenamefont {Rani}, \citenamefont {Hossain},
  \citenamefont {Niranjan},\ and\ \citenamefont {Samal}}]{Ghosh2024Accurate}%
  \BibitemOpen
  \bibfield  {author} {\bibinfo {author} {\bibfnamefont {A.}~\bibnamefont
  {Ghosh}}, \bibinfo {author} {\bibfnamefont {S.}~\bibnamefont {Jana}},
  \bibinfo {author} {\bibfnamefont {D.}~\bibnamefont {Rani}}, \bibinfo {author}
  {\bibfnamefont {M.}~\bibnamefont {Hossain}}, \bibinfo {author} {\bibfnamefont
  {M.~K.}\ \bibnamefont {Niranjan}},\ and\ \bibinfo {author} {\bibfnamefont
  {P.}~\bibnamefont {Samal}},\ }\bibfield  {title} {\bibinfo {title} {Accurate
  and efficient prediction of the band gaps and optical spectra of chalcopyrite
  semiconductors from a nonempirical range-separated dielectric-dependent
  hybrid: Comparison with many-body perturbation theory},\ }\href
  {https://doi.org/10.1103/PhysRevB.109.045133} {\bibfield  {journal} {\bibinfo
   {journal} {Phys. Rev. B}\ }\textbf {\bibinfo {volume} {109}},\ \bibinfo
  {pages} {045133} (\bibinfo {year} {2024})}\BibitemShut {NoStop}%
\bibitem [{\citenamefont {Rani}\ \emph {et~al.}(2025)\citenamefont {Rani},
  \citenamefont {Jana}, \citenamefont {Niranjan},\ and\ \citenamefont
  {Samal}}]{Rani2025Thermoelectric}%
  \BibitemOpen
  \bibfield  {author} {\bibinfo {author} {\bibfnamefont {D.}~\bibnamefont
  {Rani}}, \bibinfo {author} {\bibfnamefont {S.}~\bibnamefont {Jana}}, \bibinfo
  {author} {\bibfnamefont {M.~K.}\ \bibnamefont {Niranjan}},\ and\ \bibinfo
  {author} {\bibfnamefont {P.}~\bibnamefont {Samal}},\ }\bibfield  {title}
  {\bibinfo {title} {Thermoelectric characteristics of silver-based
  chalcopyrite semiconductors: An ab initio study based on the nonempirical
  range-separated dielectric-dependent hybrid},\ }\href
  {https://doi.org/10.1021/acs.jpcc.4c07037} {\bibfield  {journal} {\bibinfo
  {journal} {The Journal of Physical Chemistry C}\ }\textbf {\bibinfo {volume}
  {129}},\ \bibinfo {pages} {3784} (\bibinfo {year} {2025})},\ \Eprint
  {https://arxiv.org/abs/https://doi.org/10.1021/acs.jpcc.4c07037}
  {https://doi.org/10.1021/acs.jpcc.4c07037} \BibitemShut {NoStop}%
\bibitem [{\citenamefont {Jana}\ \emph {et~al.}(2025)\citenamefont {Jana},
  \citenamefont {Ghosh}, \citenamefont {Bhattacharjee}, \citenamefont {Rani},
  \citenamefont {Hossain},\ and\ \citenamefont {Samal}}]{Jana2025Nonempirical}%
  \BibitemOpen
  \bibfield  {author} {\bibinfo {author} {\bibfnamefont {S.}~\bibnamefont
  {Jana}}, \bibinfo {author} {\bibfnamefont {A.}~\bibnamefont {Ghosh}},
  \bibinfo {author} {\bibfnamefont {A.}~\bibnamefont {Bhattacharjee}}, \bibinfo
  {author} {\bibfnamefont {D.}~\bibnamefont {Rani}}, \bibinfo {author}
  {\bibfnamefont {M.}~\bibnamefont {Hossain}},\ and\ \bibinfo {author}
  {\bibfnamefont {P.}~\bibnamefont {Samal}},\ }\bibfield  {title} {\bibinfo
  {title} {Nonempirical dielectric dependent hybrid as an accurate starting
  point for the single shot g0w0 calculation of chalcopyrite semiconductors},\
  }\href {https://doi.org/10.1063/5.0240012} {\bibfield  {journal} {\bibinfo
  {journal} {The Journal of Chemical Physics}\ }\textbf {\bibinfo {volume}
  {162}},\ \bibinfo {pages} {064104} (\bibinfo {year} {2025})},\ \Eprint
  {https://arxiv.org/abs/https://pubs.aip.org/aip/jcp/article-pdf/doi/10.1063/5.0240012/20386658/064104\_1\_5.0240012.pdf}
  {https://pubs.aip.org/aip/jcp/article-pdf/doi/10.1063/5.0240012/20386658/064104\_1\_5.0240012.pdf}
  \BibitemShut {NoStop}%
\bibitem [{\citenamefont {Patra}\ \emph {et~al.}(2021)\citenamefont {Patra},
  \citenamefont {Jana}, \citenamefont {Samal}, \citenamefont {Tran},
  \citenamefont {Kalantari}, \citenamefont {Doumont},\ and\ \citenamefont
  {Blaha}}]{patra2021efficient}%
  \BibitemOpen
  \bibfield  {author} {\bibinfo {author} {\bibfnamefont {A.}~\bibnamefont
  {Patra}}, \bibinfo {author} {\bibfnamefont {S.}~\bibnamefont {Jana}},
  \bibinfo {author} {\bibfnamefont {P.}~\bibnamefont {Samal}}, \bibinfo
  {author} {\bibfnamefont {F.}~\bibnamefont {Tran}}, \bibinfo {author}
  {\bibfnamefont {L.}~\bibnamefont {Kalantari}}, \bibinfo {author}
  {\bibfnamefont {J.}~\bibnamefont {Doumont}},\ and\ \bibinfo {author}
  {\bibfnamefont {P.}~\bibnamefont {Blaha}},\ }\bibfield  {title} {\bibinfo
  {title} {Efficient band structure calculation of two-dimensional materials
  from semilocal density functionals},\ }\href@noop {} {\bibfield  {journal}
  {\bibinfo  {journal} {J. Phys. Chem. C}\ }\textbf {\bibinfo {volume} {125}},\
  \bibinfo {pages} {11206} (\bibinfo {year} {2021})}\BibitemShut {NoStop}%
\bibitem [{\citenamefont {Tran}\ \emph {et~al.}(2021)\citenamefont {Tran},
  \citenamefont {Doumont}, \citenamefont {Kalantari}, \citenamefont {Blaha},
  \citenamefont {Rauch}, \citenamefont {Borlido}, \citenamefont {Botti},
  \citenamefont {Marques}, \citenamefont {Patra}, \citenamefont {Jana},\ and\
  \citenamefont {Samal}}]{Tran2021bandgap}%
  \BibitemOpen
  \bibfield  {author} {\bibinfo {author} {\bibfnamefont {F.}~\bibnamefont
  {Tran}}, \bibinfo {author} {\bibfnamefont {J.}~\bibnamefont {Doumont}},
  \bibinfo {author} {\bibfnamefont {L.}~\bibnamefont {Kalantari}}, \bibinfo
  {author} {\bibfnamefont {P.}~\bibnamefont {Blaha}}, \bibinfo {author}
  {\bibfnamefont {T.}~\bibnamefont {Rauch}}, \bibinfo {author} {\bibfnamefont
  {P.}~\bibnamefont {Borlido}}, \bibinfo {author} {\bibfnamefont
  {S.}~\bibnamefont {Botti}}, \bibinfo {author} {\bibfnamefont {M.~A.~L.}\
  \bibnamefont {Marques}}, \bibinfo {author} {\bibfnamefont {A.}~\bibnamefont
  {Patra}}, \bibinfo {author} {\bibfnamefont {S.}~\bibnamefont {Jana}},\ and\
  \bibinfo {author} {\bibfnamefont {P.}~\bibnamefont {Samal}},\ }\bibfield
  {title} {\bibinfo {title} {Bandgap of two-dimensional materials: Thorough
  assessment of modern exchange–correlation functionals},\ }\href@noop {}
  {\bibfield  {journal} {\bibinfo  {journal} {J. Chem. Phys.}\ }\textbf
  {\bibinfo {volume} {155}},\ \bibinfo {pages} {104103} (\bibinfo {year}
  {2021})}\BibitemShut {NoStop}%
\bibitem [{\citenamefont {Aschebrock}\ and\ \citenamefont
  {K\"ummel}(2019)}]{Aschebrock2019Ultranonlocality}%
  \BibitemOpen
  \bibfield  {author} {\bibinfo {author} {\bibfnamefont {T.}~\bibnamefont
  {Aschebrock}}\ and\ \bibinfo {author} {\bibfnamefont {S.}~\bibnamefont
  {K\"ummel}},\ }\bibfield  {title} {\bibinfo {title} {Ultranonlocality and
  accurate band gaps from a meta-generalized gradient approximation},\ }\href
  {https://doi.org/10.1103/PhysRevResearch.1.033082} {\bibfield  {journal}
  {\bibinfo  {journal} {Phys. Rev. Res.}\ }\textbf {\bibinfo {volume} {1}},\
  \bibinfo {pages} {033082} (\bibinfo {year} {2019})}\BibitemShut {NoStop}%
\bibitem [{\citenamefont {Lebeda}\ \emph {et~al.}(2022)\citenamefont {Lebeda},
  \citenamefont {Aschebrock},\ and\ \citenamefont
  {K\"ummel}}]{Lebeda2022First}%
  \BibitemOpen
  \bibfield  {author} {\bibinfo {author} {\bibfnamefont {T.}~\bibnamefont
  {Lebeda}}, \bibinfo {author} {\bibfnamefont {T.}~\bibnamefont {Aschebrock}},\
  and\ \bibinfo {author} {\bibfnamefont {S.}~\bibnamefont {K\"ummel}},\
  }\bibfield  {title} {\bibinfo {title} {First steps towards achieving both
  ultranonlocality and a reliable description of electronic binding in a
  meta-generalized gradient approximation},\ }\href
  {https://doi.org/10.1103/PhysRevResearch.4.023061} {\bibfield  {journal}
  {\bibinfo  {journal} {Phys. Rev. Res.}\ }\textbf {\bibinfo {volume} {4}},\
  \bibinfo {pages} {023061} (\bibinfo {year} {2022})}\BibitemShut {NoStop}%
\bibitem [{\citenamefont {Lebeda}\ \emph {et~al.}(2023)\citenamefont {Lebeda},
  \citenamefont {Aschebrock}, \citenamefont {Sun}, \citenamefont {Leppert},\
  and\ \citenamefont {K\"ummel}}]{Lebeda2023Right}%
  \BibitemOpen
  \bibfield  {author} {\bibinfo {author} {\bibfnamefont {T.}~\bibnamefont
  {Lebeda}}, \bibinfo {author} {\bibfnamefont {T.}~\bibnamefont {Aschebrock}},
  \bibinfo {author} {\bibfnamefont {J.}~\bibnamefont {Sun}}, \bibinfo {author}
  {\bibfnamefont {L.}~\bibnamefont {Leppert}},\ and\ \bibinfo {author}
  {\bibfnamefont {S.}~\bibnamefont {K\"ummel}},\ }\bibfield  {title} {\bibinfo
  {title} {Right band gaps for the right reason at low computational cost with
  a meta-gga},\ }\href {https://doi.org/10.1103/PhysRevMaterials.7.093803}
  {\bibfield  {journal} {\bibinfo  {journal} {Phys. Rev. Mater.}\ }\textbf
  {\bibinfo {volume} {7}},\ \bibinfo {pages} {093803} (\bibinfo {year}
  {2023})}\BibitemShut {NoStop}%
\bibitem [{\citenamefont {Aschebrock}\ \emph {et~al.}(2023)\citenamefont
  {Aschebrock}, \citenamefont {Lebeda}, \citenamefont {Brütting},
  \citenamefont {Richter}, \citenamefont {Schelter},\ and\ \citenamefont
  {Kümmel}}]{Aschebrock2023Exact}%
  \BibitemOpen
  \bibfield  {author} {\bibinfo {author} {\bibfnamefont {T.}~\bibnamefont
  {Aschebrock}}, \bibinfo {author} {\bibfnamefont {T.}~\bibnamefont {Lebeda}},
  \bibinfo {author} {\bibfnamefont {M.}~\bibnamefont {Brütting}}, \bibinfo
  {author} {\bibfnamefont {R.}~\bibnamefont {Richter}}, \bibinfo {author}
  {\bibfnamefont {I.}~\bibnamefont {Schelter}},\ and\ \bibinfo {author}
  {\bibfnamefont {S.}~\bibnamefont {Kümmel}},\ }\bibfield  {title} {\bibinfo
  {title} {Exact exchange-like electric response from a meta-generalized
  gradient approximation: A semilocal realization of ultranonlocality},\ }\href
  {https://doi.org/10.1063/5.0173776} {\bibfield  {journal} {\bibinfo
  {journal} {J. Chem. Phys.}\ }\textbf {\bibinfo {volume} {159}},\ \bibinfo
  {pages} {234107} (\bibinfo {year} {2023})}\BibitemShut {NoStop}%
\bibitem [{\citenamefont {Kresse}\ and\ \citenamefont {Hafner}(1993)}]{vasp1}%
  \BibitemOpen
  \bibfield  {author} {\bibinfo {author} {\bibfnamefont {G.}~\bibnamefont
  {Kresse}}\ and\ \bibinfo {author} {\bibfnamefont {J.}~\bibnamefont
  {Hafner}},\ }\bibfield  {title} {\bibinfo {title} {Ab initio molecular
  dynamics for liquid metals},\ }\href@noop {} {\bibfield  {journal} {\bibinfo
  {journal} {Phys. Rev. B}\ }\textbf {\bibinfo {volume} {47}},\ \bibinfo
  {pages} {558} (\bibinfo {year} {1993})}\BibitemShut {NoStop}%
\bibitem [{\citenamefont {Kresse}\ and\ \citenamefont
  {Furthm\"uller}(1996)}]{vasp2}%
  \BibitemOpen
  \bibfield  {author} {\bibinfo {author} {\bibfnamefont {G.}~\bibnamefont
  {Kresse}}\ and\ \bibinfo {author} {\bibfnamefont {J.}~\bibnamefont
  {Furthm\"uller}},\ }\bibfield  {title} {\bibinfo {title} {Efficient iterative
  schemes for ab initio total-energy calculations using a plane-wave basis
  set},\ }\href@noop {} {\bibfield  {journal} {\bibinfo  {journal} {Phys. Rev.
  B}\ }\textbf {\bibinfo {volume} {54}},\ \bibinfo {pages} {11169} (\bibinfo
  {year} {1996})}\BibitemShut {NoStop}%
\bibitem [{\citenamefont {Kresse}\ and\ \citenamefont {Joubert}(1999)}]{vasp3}%
  \BibitemOpen
  \bibfield  {author} {\bibinfo {author} {\bibfnamefont {G.}~\bibnamefont
  {Kresse}}\ and\ \bibinfo {author} {\bibfnamefont {D.}~\bibnamefont
  {Joubert}},\ }\bibfield  {title} {\bibinfo {title} {From ultrasoft
  pseudopotentials to the projector augmented-wave method},\ }\href@noop {}
  {\bibfield  {journal} {\bibinfo  {journal} {Phys. Rev. B}\ }\textbf {\bibinfo
  {volume} {59}},\ \bibinfo {pages} {1758} (\bibinfo {year}
  {1999})}\BibitemShut {NoStop}%
\bibitem [{\citenamefont {Kresse}\ and\ \citenamefont
  {Furthmüller}(1996)}]{vasp4}%
  \BibitemOpen
  \bibfield  {author} {\bibinfo {author} {\bibfnamefont {G.}~\bibnamefont
  {Kresse}}\ and\ \bibinfo {author} {\bibfnamefont {J.}~\bibnamefont
  {Furthmüller}},\ }\bibfield  {title} {\bibinfo {title} {Efficiency of
  ab-initio total energy calculations for metals and semiconductors using a
  plane-wave basis set},\ }\href@noop {} {\bibfield  {journal} {\bibinfo
  {journal} {Comput. Mater. Sci.}\ }\textbf {\bibinfo {volume} {6}},\ \bibinfo
  {pages} {15 } (\bibinfo {year} {1996})}\BibitemShut {NoStop}%
\bibitem [{\citenamefont {Baroni}\ \emph {et~al.}(2001)\citenamefont {Baroni},
  \citenamefont {de~Gironcoli}, \citenamefont {Dal~Corso},\ and\ \citenamefont
  {Giannozzi}}]{Baroni2001phonons}%
  \BibitemOpen
  \bibfield  {author} {\bibinfo {author} {\bibfnamefont {S.}~\bibnamefont
  {Baroni}}, \bibinfo {author} {\bibfnamefont {S.}~\bibnamefont
  {de~Gironcoli}}, \bibinfo {author} {\bibfnamefont {A.}~\bibnamefont
  {Dal~Corso}},\ and\ \bibinfo {author} {\bibfnamefont {P.}~\bibnamefont
  {Giannozzi}},\ }\bibfield  {title} {\bibinfo {title} {Phonons and related
  crystal properties from density-functional perturbation theory},\ }\href
  {https://doi.org/10.1103/RevModPhys.73.515} {\bibfield  {journal} {\bibinfo
  {journal} {Rev. Mod. Phys.}\ }\textbf {\bibinfo {volume} {73}},\ \bibinfo
  {pages} {515} (\bibinfo {year} {2001})}\BibitemShut {NoStop}%
\bibitem [{\citenamefont {Gajdo\ifmmode~\check{s}\else \v{s}\fi{}}\ \emph
  {et~al.}(2006)\citenamefont {Gajdo\ifmmode~\check{s}\else \v{s}\fi{}},
  \citenamefont {Hummer}, \citenamefont {Kresse}, \citenamefont
  {Furthm\"uller},\ and\ \citenamefont {Bechstedt}}]{Gajdo2006linear}%
  \BibitemOpen
  \bibfield  {author} {\bibinfo {author} {\bibfnamefont {M.}~\bibnamefont
  {Gajdo\ifmmode~\check{s}\else \v{s}\fi{}}}, \bibinfo {author} {\bibfnamefont
  {K.}~\bibnamefont {Hummer}}, \bibinfo {author} {\bibfnamefont
  {G.}~\bibnamefont {Kresse}}, \bibinfo {author} {\bibfnamefont
  {J.}~\bibnamefont {Furthm\"uller}},\ and\ \bibinfo {author} {\bibfnamefont
  {F.}~\bibnamefont {Bechstedt}},\ }\bibfield  {title} {\bibinfo {title}
  {Linear optical properties in the projector-augmented wave methodology},\
  }\href {https://doi.org/10.1103/PhysRevB.73.045112} {\bibfield  {journal}
  {\bibinfo  {journal} {Phys. Rev. B}\ }\textbf {\bibinfo {volume} {73}},\
  \bibinfo {pages} {045112} (\bibinfo {year} {2006})}\BibitemShut {NoStop}%
\bibitem [{\citenamefont {Zhang}\ \emph {et~al.}(2013)\citenamefont {Zhang},
  \citenamefont {Zhang}, \citenamefont {Gao}, \citenamefont {Abtew},
  \citenamefont {Wang}, \citenamefont {Zhang},\ and\ \citenamefont
  {Zhang}}]{Zhang2013Near}%
  \BibitemOpen
  \bibfield  {author} {\bibinfo {author} {\bibfnamefont {Y.}~\bibnamefont
  {Zhang}}, \bibinfo {author} {\bibfnamefont {J.}~\bibnamefont {Zhang}},
  \bibinfo {author} {\bibfnamefont {W.}~\bibnamefont {Gao}}, \bibinfo {author}
  {\bibfnamefont {T.~A.}\ \bibnamefont {Abtew}}, \bibinfo {author}
  {\bibfnamefont {Y.}~\bibnamefont {Wang}}, \bibinfo {author} {\bibfnamefont
  {P.}~\bibnamefont {Zhang}},\ and\ \bibinfo {author} {\bibfnamefont
  {W.}~\bibnamefont {Zhang}},\ }\bibfield  {title} {\bibinfo {title} {Near-edge
  band structures and band gaps of cu-based semiconductors predicted by the
  modified becke-johnson potential plus an on-site coulomb u},\ }\href
  {https://doi.org/10.1063/1.4828864} {\bibfield  {journal} {\bibinfo
  {journal} {J. Chem. Phys.}\ }\textbf {\bibinfo {volume} {139}},\ \bibinfo
  {pages} {184706} (\bibinfo {year} {2013})}\BibitemShut {NoStop}%
\bibitem [{\citenamefont {Jackson}\ \emph {et~al.}(2011)\citenamefont
  {Jackson}, \citenamefont {Hariskos}, \citenamefont {Lotter}, \citenamefont
  {Paetel}, \citenamefont {Wuerz}, \citenamefont {Menner}, \citenamefont
  {Wischmann},\ and\ \citenamefont {Powalla}}]{Jackson2011New}%
  \BibitemOpen
  \bibfield  {author} {\bibinfo {author} {\bibfnamefont {P.}~\bibnamefont
  {Jackson}}, \bibinfo {author} {\bibfnamefont {D.}~\bibnamefont {Hariskos}},
  \bibinfo {author} {\bibfnamefont {E.}~\bibnamefont {Lotter}}, \bibinfo
  {author} {\bibfnamefont {S.}~\bibnamefont {Paetel}}, \bibinfo {author}
  {\bibfnamefont {R.}~\bibnamefont {Wuerz}}, \bibinfo {author} {\bibfnamefont
  {R.}~\bibnamefont {Menner}}, \bibinfo {author} {\bibfnamefont
  {W.}~\bibnamefont {Wischmann}},\ and\ \bibinfo {author} {\bibfnamefont
  {M.}~\bibnamefont {Powalla}},\ }\bibfield  {title} {\bibinfo {title} {New
  world record efficiency for cu(in,ga)se2 thin-film solar cells beyond 20\%},\
  }\href {https://doi.org/https://doi.org/10.1002/pip.1078} {\bibfield
  {journal} {\bibinfo  {journal} {Progress in Photovoltaics: Research and
  Applications}\ }\textbf {\bibinfo {volume} {19}},\ \bibinfo {pages} {894}
  (\bibinfo {year} {2011})},\ \Eprint
  {https://arxiv.org/abs/https://onlinelibrary.wiley.com/doi/pdf/10.1002/pip.1078}
  {https://onlinelibrary.wiley.com/doi/pdf/10.1002/pip.1078} \BibitemShut
  {NoStop}%
\bibitem [{\citenamefont {Walsh}\ \emph {et~al.}(2012)\citenamefont {Walsh},
  \citenamefont {Chen}, \citenamefont {Wei},\ and\ \citenamefont
  {Gong}}]{Walsh2012Kesterite}%
  \BibitemOpen
  \bibfield  {author} {\bibinfo {author} {\bibfnamefont {A.}~\bibnamefont
  {Walsh}}, \bibinfo {author} {\bibfnamefont {S.}~\bibnamefont {Chen}},
  \bibinfo {author} {\bibfnamefont {S.-H.}\ \bibnamefont {Wei}},\ and\ \bibinfo
  {author} {\bibfnamefont {X.-G.}\ \bibnamefont {Gong}},\ }\bibfield  {title}
  {\bibinfo {title} {Kesterite thin-film solar cells: Advances in materials
  modelling of cu2znsns4},\ }\href
  {https://doi.org/https://doi.org/10.1002/aenm.201100630} {\bibfield
  {journal} {\bibinfo  {journal} {Advanced Energy Materials}\ }\textbf
  {\bibinfo {volume} {2}},\ \bibinfo {pages} {400} (\bibinfo {year}
  {2012})}\BibitemShut {NoStop}%
\bibitem [{\citenamefont {Kormath Madam~Raghupathy}\ \emph
  {et~al.}(2018)\citenamefont {Kormath Madam~Raghupathy}, \citenamefont
  {Wiebeler}, \citenamefont {K{\"u}hne}, \citenamefont {Felser},\ and\
  \citenamefont {Mirhosseini}}]{Kormath2018Database}%
  \BibitemOpen
  \bibfield  {author} {\bibinfo {author} {\bibfnamefont {R.}~\bibnamefont
  {Kormath Madam~Raghupathy}}, \bibinfo {author} {\bibfnamefont
  {H.}~\bibnamefont {Wiebeler}}, \bibinfo {author} {\bibfnamefont {T.~D.}\
  \bibnamefont {K{\"u}hne}}, \bibinfo {author} {\bibfnamefont {C.}~\bibnamefont
  {Felser}},\ and\ \bibinfo {author} {\bibfnamefont {H.}~\bibnamefont
  {Mirhosseini}},\ }\bibfield  {title} {\bibinfo {title} {Database screening of
  ternary chalcogenides for p-type transparent conductors},\ }\href
  {https://doi.org/10.1021/acs.chemmater.8b02719} {\bibfield  {journal}
  {\bibinfo  {journal} {Chemistry of Materials}\ }\textbf {\bibinfo {volume}
  {30}},\ \bibinfo {pages} {6794} (\bibinfo {year} {2018})}\BibitemShut
  {NoStop}%
\bibitem [{\citenamefont {Goodman}(1958)}]{Goodman1958Prediction}%
  \BibitemOpen
  \bibfield  {author} {\bibinfo {author} {\bibfnamefont {C.}~\bibnamefont
  {Goodman}},\ }\bibfield  {title} {\bibinfo {title} {The prediction of
  semiconducting properties in inorganic compounds},\ }\href
  {https://doi.org/https://doi.org/10.1016/0022-3697(58)90050-7} {\bibfield
  {journal} {\bibinfo  {journal} {Journal of Physics and Chemistry of Solids}\
  }\textbf {\bibinfo {volume} {6}},\ \bibinfo {pages} {305} (\bibinfo {year}
  {1958})}\BibitemShut {NoStop}%
\bibitem [{\citenamefont {Pamplin}(1979)}]{Pamplin1979Spray}%
  \BibitemOpen
  \bibfield  {author} {\bibinfo {author} {\bibfnamefont {B.}~\bibnamefont
  {Pamplin}},\ }\bibfield  {title} {\bibinfo {title} {Spray pyrolysis of
  ternary and quaternary solar cell materials},\ }\href
  {https://doi.org/https://doi.org/10.1016/0146-3535(79)90005-4} {\bibfield
  {journal} {\bibinfo  {journal} {Progress in Crystal Growth and
  Characterization}\ }\textbf {\bibinfo {volume} {1}},\ \bibinfo {pages} {395}
  (\bibinfo {year} {1979})}\BibitemShut {NoStop}%
\bibitem [{\citenamefont {PAMPLIN}\ and\ \citenamefont
  {SHAH}(1965)}]{Pamplin1965Quinary}%
  \BibitemOpen
  \bibfield  {author} {\bibinfo {author} {\bibfnamefont {B.~R.}\ \bibnamefont
  {PAMPLIN}}\ and\ \bibinfo {author} {\bibfnamefont {J.~S.}\ \bibnamefont
  {SHAH}},\ }\bibfield  {title} {\bibinfo {title} {Quinary adamantine
  semiconductors},\ }\href {https://doi.org/10.1038/207180a0} {\bibfield
  {journal} {\bibinfo  {journal} {Nature}\ }\textbf {\bibinfo {volume} {207}},\
  \bibinfo {pages} {180} (\bibinfo {year} {1965})}\BibitemShut {NoStop}%
\bibitem [{\citenamefont {García}\ \emph {et~al.}(2018)\citenamefont
  {García}, \citenamefont {Palacios}, \citenamefont {Cabot},\ and\
  \citenamefont {Wahnón}}]{Garcia2018Thermoelectric}%
  \BibitemOpen
  \bibfield  {author} {\bibinfo {author} {\bibfnamefont {G.}~\bibnamefont
  {García}}, \bibinfo {author} {\bibfnamefont {P.}~\bibnamefont {Palacios}},
  \bibinfo {author} {\bibfnamefont {A.}~\bibnamefont {Cabot}},\ and\ \bibinfo
  {author} {\bibfnamefont {P.}~\bibnamefont {Wahnón}},\ }\bibfield  {title}
  {\bibinfo {title} {Thermoelectric properties of doped-cu3sbse4 compounds: A
  first-principles insight},\ }\href
  {https://doi.org/10.1021/acs.inorgchem.8b00980} {\bibfield  {journal}
  {\bibinfo  {journal} {Inorganic Chemistry}\ }\textbf {\bibinfo {volume}
  {57}},\ \bibinfo {pages} {7321} (\bibinfo {year} {2018})}\BibitemShut
  {NoStop}%
\bibitem [{\citenamefont {Vidal}\ \emph {et~al.}(2011)\citenamefont {Vidal},
  \citenamefont {Zhang}, \citenamefont {Yu}, \citenamefont {Luo},\ and\
  \citenamefont {Zunger}}]{Vidal2011False}%
  \BibitemOpen
  \bibfield  {author} {\bibinfo {author} {\bibfnamefont {J.}~\bibnamefont
  {Vidal}}, \bibinfo {author} {\bibfnamefont {X.}~\bibnamefont {Zhang}},
  \bibinfo {author} {\bibfnamefont {L.}~\bibnamefont {Yu}}, \bibinfo {author}
  {\bibfnamefont {J.-W.}\ \bibnamefont {Luo}},\ and\ \bibinfo {author}
  {\bibfnamefont {A.}~\bibnamefont {Zunger}},\ }\bibfield  {title} {\bibinfo
  {title} {False-positive and false-negative assignments of topological
  insulators in density functional theory and hybrids},\ }\href
  {https://doi.org/10.1103/PhysRevB.84.041109} {\bibfield  {journal} {\bibinfo
  {journal} {Phys. Rev. B}\ }\textbf {\bibinfo {volume} {84}},\ \bibinfo
  {pages} {041109} (\bibinfo {year} {2011})}\BibitemShut {NoStop}%
\bibitem [{\citenamefont {Wang}\ \emph {et~al.}(2011)\citenamefont {Wang},
  \citenamefont {Lin}, \citenamefont {Das}, \citenamefont {Hasan},\ and\
  \citenamefont {Bansil}}]{Wang2011Topological}%
  \BibitemOpen
  \bibfield  {author} {\bibinfo {author} {\bibfnamefont {Y.~J.}\ \bibnamefont
  {Wang}}, \bibinfo {author} {\bibfnamefont {H.}~\bibnamefont {Lin}}, \bibinfo
  {author} {\bibfnamefont {T.}~\bibnamefont {Das}}, \bibinfo {author}
  {\bibfnamefont {M.~Z.}\ \bibnamefont {Hasan}},\ and\ \bibinfo {author}
  {\bibfnamefont {A.}~\bibnamefont {Bansil}},\ }\bibfield  {title} {\bibinfo
  {title} {Topological insulators in the quaternary chalcogenide compounds and
  ternary famatinite compounds},\ }\href
  {https://doi.org/10.1088/1367-2630/13/8/085017} {\bibfield  {journal}
  {\bibinfo  {journal} {New Journal of Physics}\ }\textbf {\bibinfo {volume}
  {13}},\ \bibinfo {pages} {085017} (\bibinfo {year} {2011})}\BibitemShut
  {NoStop}%
\bibitem [{\citenamefont {BIRKEN}\ \emph {et~al.}(1998)\citenamefont {BIRKEN},
  \citenamefont {BLESSING},\ and\ \citenamefont {KUNZ}}]{BIRKEN1998279}%
  \BibitemOpen
  \bibfield  {author} {\bibinfo {author} {\bibfnamefont {H.-G.}\ \bibnamefont
  {BIRKEN}}, \bibinfo {author} {\bibfnamefont {C.}~\bibnamefont {BLESSING}},\
  and\ \bibinfo {author} {\bibfnamefont {C.}~\bibnamefont {KUNZ}},\ }\bibfield
  {title} {\bibinfo {title} {Chapter 12 - determination of optical constants
  from angular-dependent, photoelectric-yield measurements},\ }in\ \href
  {https://doi.org/https://doi.org/10.1016/B978-0-08-055630-7.50015-8} {\emph
  {\bibinfo {booktitle} {Handbook of Optical Constants of Solids}}},\ \bibinfo
  {editor} {edited by\ \bibinfo {editor} {\bibfnamefont {E.~D.}\ \bibnamefont
  {PALIK}}}\ (\bibinfo  {publisher} {Academic Press},\ \bibinfo {address}
  {Boston},\ \bibinfo {year} {1998})\ pp.\ \bibinfo {pages}
  {279--292}\BibitemShut {NoStop}%
\bibitem [{\citenamefont {Logothetidis}\ \emph {et~al.}(1986)\citenamefont
  {Logothetidis}, \citenamefont {Lautenschlager},\ and\ \citenamefont
  {Cardona}}]{LogoLautCard1986}%
  \BibitemOpen
  \bibfield  {author} {\bibinfo {author} {\bibfnamefont {S.}~\bibnamefont
  {Logothetidis}}, \bibinfo {author} {\bibfnamefont {P.}~\bibnamefont
  {Lautenschlager}},\ and\ \bibinfo {author} {\bibfnamefont {M.}~\bibnamefont
  {Cardona}},\ }\bibfield  {title} {\bibinfo {title} {Temperature dependence of
  the dielectric function and the interband critical points in orthorhombic
  ges},\ }\href@noop {} {\bibfield  {journal} {\bibinfo  {journal} {Phys. Rev.
  B}\ }\textbf {\bibinfo {volume} {33}},\ \bibinfo {pages} {1110} (\bibinfo
  {year} {1986})}\BibitemShut {NoStop}%
\bibitem [{\citenamefont {Roessler}\ and\ \citenamefont
  {Walker}(1968)}]{RoesWalk1968}%
  \BibitemOpen
  \bibfield  {author} {\bibinfo {author} {\bibfnamefont {D.~M.}\ \bibnamefont
  {Roessler}}\ and\ \bibinfo {author} {\bibfnamefont {W.~C.}\ \bibnamefont
  {Walker}},\ }\bibfield  {title} {\bibinfo {title} {Electronic spectra of
  crystalline nacl and kcl},\ }\href@noop {} {\bibfield  {journal} {\bibinfo
  {journal} {Phys. Rev.}\ }\textbf {\bibinfo {volume} {166}},\ \bibinfo {pages}
  {599} (\bibinfo {year} {1968})}\BibitemShut {NoStop}%
\bibitem [{\citenamefont {Bortz}\ \emph {et~al.}(1990)\citenamefont {Bortz},
  \citenamefont {French}, \citenamefont {Jones}, \citenamefont {Kasowski},\
  and\ \citenamefont {Ohuchi}}]{BortFrenJone1990}%
  \BibitemOpen
  \bibfield  {author} {\bibinfo {author} {\bibfnamefont {M.~L.}\ \bibnamefont
  {Bortz}}, \bibinfo {author} {\bibfnamefont {R.~H.}\ \bibnamefont {French}},
  \bibinfo {author} {\bibfnamefont {D.~J.}\ \bibnamefont {Jones}}, \bibinfo
  {author} {\bibfnamefont {R.~V.}\ \bibnamefont {Kasowski}},\ and\ \bibinfo
  {author} {\bibfnamefont {F.~S.}\ \bibnamefont {Ohuchi}},\ }\bibfield  {title}
  {\bibinfo {title} {Temperature dependence of the electronic structure of
  oxides: Mgo, mgal2o4 and al2o3},\ }\href
  {https://doi.org/10.1088/0031-8949/41/4/036} {\bibfield  {journal} {\bibinfo
  {journal} {Physica Scripta}\ }\textbf {\bibinfo {volume} {41}},\ \bibinfo
  {pages} {537} (\bibinfo {year} {1990})}\BibitemShut {NoStop}%
\bibitem [{\citenamefont {Sander}\ \emph {et~al.}(2015)\citenamefont {Sander},
  \citenamefont {Maggio},\ and\ \citenamefont {Kresse}}]{Sander2015beyond}%
  \BibitemOpen
  \bibfield  {author} {\bibinfo {author} {\bibfnamefont {T.}~\bibnamefont
  {Sander}}, \bibinfo {author} {\bibfnamefont {E.}~\bibnamefont {Maggio}},\
  and\ \bibinfo {author} {\bibfnamefont {G.}~\bibnamefont {Kresse}},\
  }\bibfield  {title} {\bibinfo {title} {Beyond the tamm-dancoff approximation
  for extended systems using exact diagonalization},\ }\href@noop {} {\bibfield
   {journal} {\bibinfo  {journal} {Phys. Rev. B}\ }\textbf {\bibinfo {volume}
  {92}},\ \bibinfo {pages} {045209} (\bibinfo {year} {2015})}\BibitemShut
  {NoStop}%
\bibitem [{\citenamefont {Sander}\ and\ \citenamefont
  {Kresse}(2017)}]{Sander2017macroscopic}%
  \BibitemOpen
  \bibfield  {author} {\bibinfo {author} {\bibfnamefont {T.}~\bibnamefont
  {Sander}}\ and\ \bibinfo {author} {\bibfnamefont {G.}~\bibnamefont
  {Kresse}},\ }\bibfield  {title} {\bibinfo {title} {{Macroscopic dielectric
  function within time-dependent density functional theory—Real time
  evolution versus the Casida approach}},\ }\href@noop {} {\bibfield  {journal}
  {\bibinfo  {journal} {J. Chem. Phys.}\ }\textbf {\bibinfo {volume} {146}},\
  \bibinfo {pages} {064110} (\bibinfo {year} {2017})}\BibitemShut {NoStop}%
\bibitem [{\citenamefont {CASIDA}()}]{cassida-equation}%
  \BibitemOpen
  \bibfield  {author} {\bibinfo {author} {\bibfnamefont {M.~E.}\ \bibnamefont
  {CASIDA}},\ }\bibinfo {title} {Time-dependent density functional response
  theory for molecules},\ in\ \href@noop {} {\emph {\bibinfo {booktitle}
  {Recent Advances in Density Functional Methods}}},\ pp.\ \bibinfo {pages}
  {155--192}\BibitemShut {NoStop}%
\bibitem [{\citenamefont {Onida}\ \emph {et~al.}(2002)\citenamefont {Onida},
  \citenamefont {Reining},\ and\ \citenamefont {Rubio}}]{Onida2002electronic}%
  \BibitemOpen
  \bibfield  {author} {\bibinfo {author} {\bibfnamefont {G.}~\bibnamefont
  {Onida}}, \bibinfo {author} {\bibfnamefont {L.}~\bibnamefont {Reining}},\
  and\ \bibinfo {author} {\bibfnamefont {A.}~\bibnamefont {Rubio}},\ }\bibfield
   {title} {\bibinfo {title} {Electronic excitations: density-functional versus
  many-body green's-function approaches},\ }\href
  {https://doi.org/10.1103/RevModPhys.74.601} {\bibfield  {journal} {\bibinfo
  {journal} {Rev. Mod. Phys.}\ }\textbf {\bibinfo {volume} {74}},\ \bibinfo
  {pages} {601} (\bibinfo {year} {2002})}\BibitemShut {NoStop}%
\bibitem [{\citenamefont {Nazarov}\ and\ \citenamefont
  {Vignale}(2011)}]{Nazarov2011Optics}%
  \BibitemOpen
  \bibfield  {author} {\bibinfo {author} {\bibfnamefont {V.~U.}\ \bibnamefont
  {Nazarov}}\ and\ \bibinfo {author} {\bibfnamefont {G.}~\bibnamefont
  {Vignale}},\ }\bibfield  {title} {\bibinfo {title} {Optics of semiconductors
  from meta-generalized-gradient-approximation-based time-dependent
  density-functional theory},\ }\href
  {https://doi.org/10.1103/PhysRevLett.107.216402} {\bibfield  {journal}
  {\bibinfo  {journal} {Phys. Rev. Lett.}\ }\textbf {\bibinfo {volume} {107}},\
  \bibinfo {pages} {216402} (\bibinfo {year} {2011})}\BibitemShut {NoStop}%
\bibitem [{\citenamefont {Singh}\ \emph {et~al.}(2019)\citenamefont {Singh},
  \citenamefont {Elliott}, \citenamefont {Nautiyal}, \citenamefont {Dewhurst},\
  and\ \citenamefont {Sharma}}]{Singh2019Adiabatic}%
  \BibitemOpen
  \bibfield  {author} {\bibinfo {author} {\bibfnamefont {N.}~\bibnamefont
  {Singh}}, \bibinfo {author} {\bibfnamefont {P.}~\bibnamefont {Elliott}},
  \bibinfo {author} {\bibfnamefont {T.}~\bibnamefont {Nautiyal}}, \bibinfo
  {author} {\bibfnamefont {J.~K.}\ \bibnamefont {Dewhurst}},\ and\ \bibinfo
  {author} {\bibfnamefont {S.}~\bibnamefont {Sharma}},\ }\bibfield  {title}
  {\bibinfo {title} {Adiabatic generalized gradient approximation kernel in
  time-dependent density functional theory},\ }\href
  {https://doi.org/10.1103/PhysRevB.99.035151} {\bibfield  {journal} {\bibinfo
  {journal} {Phys. Rev. B}\ }\textbf {\bibinfo {volume} {99}},\ \bibinfo
  {pages} {035151} (\bibinfo {year} {2019})}\BibitemShut {NoStop}%
\bibitem [{\citenamefont {Ullrich}(2011)}]{Ullrichtddft}%
  \BibitemOpen
  \bibfield  {author} {\bibinfo {author} {\bibfnamefont {C.~A.}\ \bibnamefont
  {Ullrich}},\ }\href@noop {} {\emph {\bibinfo {title} {{Time-Dependent
  Density-Functional Theory: Concepts and Applications}}}}\ (\bibinfo
  {publisher} {Oxford University Press},\ \bibinfo {year} {2011})\BibitemShut
  {NoStop}%
\bibitem [{\citenamefont {Yang}\ \emph {et~al.}(2015)\citenamefont {Yang},
  \citenamefont {Sottile},\ and\ \citenamefont {Ullrich}}]{Yang2015simple}%
  \BibitemOpen
  \bibfield  {author} {\bibinfo {author} {\bibfnamefont {Z.-h.}\ \bibnamefont
  {Yang}}, \bibinfo {author} {\bibfnamefont {F.}~\bibnamefont {Sottile}},\ and\
  \bibinfo {author} {\bibfnamefont {C.~A.}\ \bibnamefont {Ullrich}},\
  }\bibfield  {title} {\bibinfo {title} {Simple screened exact-exchange
  approach for excitonic properties in solids},\ }\href@noop {} {\bibfield
  {journal} {\bibinfo  {journal} {Phys. Rev. B}\ }\textbf {\bibinfo {volume}
  {92}},\ \bibinfo {pages} {035202} (\bibinfo {year} {2015})}\BibitemShut
  {NoStop}%
\bibitem [{\citenamefont {Sun}\ and\ \citenamefont
  {Ullrich}(2020)}]{Sun2020optical}%
  \BibitemOpen
  \bibfield  {author} {\bibinfo {author} {\bibfnamefont {J.}~\bibnamefont
  {Sun}}\ and\ \bibinfo {author} {\bibfnamefont {C.~A.}\ \bibnamefont
  {Ullrich}},\ }\bibfield  {title} {\bibinfo {title} {Optical properties of
  $\mathrm{Cs}{\mathrm{cu}}_{2}{X}_{3}$ $(x=\mathrm{Cl}, \mathrm{Br},
  \mathrm{and} \mathrm{I})$: A comparative study between hybrid time-dependent
  density-functional theory and the bethe-salpeter equation},\ }\href
  {https://doi.org/10.1103/PhysRevMaterials.4.095402} {\bibfield  {journal}
  {\bibinfo  {journal} {Phys. Rev. Mater.}\ }\textbf {\bibinfo {volume} {4}},\
  \bibinfo {pages} {095402} (\bibinfo {year} {2020})}\BibitemShut {NoStop}%
\bibitem [{\citenamefont {Sun}\ \emph {et~al.}(2020)\citenamefont {Sun},
  \citenamefont {Yang},\ and\ \citenamefont {Ullrich}}]{sun2020lowcost}%
  \BibitemOpen
  \bibfield  {author} {\bibinfo {author} {\bibfnamefont {J.}~\bibnamefont
  {Sun}}, \bibinfo {author} {\bibfnamefont {J.}~\bibnamefont {Yang}},\ and\
  \bibinfo {author} {\bibfnamefont {C.~A.}\ \bibnamefont {Ullrich}},\
  }\bibfield  {title} {\bibinfo {title} {Low-cost alternatives to the
  bethe-salpeter equation: Towards simple hybrid functionals for excitonic
  effects in solids},\ }\href
  {https://doi.org/10.1103/PhysRevResearch.2.013091} {\bibfield  {journal}
  {\bibinfo  {journal} {Phys. Rev. Res.}\ }\textbf {\bibinfo {volume} {2}},\
  \bibinfo {pages} {013091} (\bibinfo {year} {2020})}\BibitemShut {NoStop}%
\bibitem [{\citenamefont {Kootstra}\ \emph {et~al.}(2000)\citenamefont
  {Kootstra}, \citenamefont {de~Boeij},\ and\ \citenamefont
  {Snijders}}]{Kootstra2000application}%
  \BibitemOpen
  \bibfield  {author} {\bibinfo {author} {\bibfnamefont {F.}~\bibnamefont
  {Kootstra}}, \bibinfo {author} {\bibfnamefont {P.~L.}\ \bibnamefont
  {de~Boeij}},\ and\ \bibinfo {author} {\bibfnamefont {J.~G.}\ \bibnamefont
  {Snijders}},\ }\bibfield  {title} {\bibinfo {title} {Application of
  time-dependent density-functional theory to the dielectric function of
  various nonmetallic crystals},\ }\href
  {https://doi.org/10.1103/PhysRevB.62.7071} {\bibfield  {journal} {\bibinfo
  {journal} {Phys. Rev. B}\ }\textbf {\bibinfo {volume} {62}},\ \bibinfo
  {pages} {7071} (\bibinfo {year} {2000})}\BibitemShut {NoStop}%
\bibitem [{\citenamefont {Sun}\ \emph {et~al.}(2011)\citenamefont {Sun},
  \citenamefont {Marsman}, \citenamefont {Csonka}, \citenamefont {Ruzsinszky},
  \citenamefont {Hao}, \citenamefont {Kim}, \citenamefont {Kresse},\ and\
  \citenamefont {Perdew}}]{sun2011selfconsistent}%
  \BibitemOpen
  \bibfield  {author} {\bibinfo {author} {\bibfnamefont {J.}~\bibnamefont
  {Sun}}, \bibinfo {author} {\bibfnamefont {M.}~\bibnamefont {Marsman}},
  \bibinfo {author} {\bibfnamefont {G.~I.}\ \bibnamefont {Csonka}}, \bibinfo
  {author} {\bibfnamefont {A.}~\bibnamefont {Ruzsinszky}}, \bibinfo {author}
  {\bibfnamefont {P.}~\bibnamefont {Hao}}, \bibinfo {author} {\bibfnamefont
  {Y.-S.}\ \bibnamefont {Kim}}, \bibinfo {author} {\bibfnamefont
  {G.}~\bibnamefont {Kresse}},\ and\ \bibinfo {author} {\bibfnamefont {J.~P.}\
  \bibnamefont {Perdew}},\ }\bibfield  {title} {\bibinfo {title}
  {Self-consistent meta-generalized gradient approximation within the
  projector-augmented-wave method},\ }\href
  {https://doi.org/10.1103/PhysRevB.84.035117} {\bibfield  {journal} {\bibinfo
  {journal} {Phys. Rev. B}\ }\textbf {\bibinfo {volume} {84}},\ \bibinfo
  {pages} {035117} (\bibinfo {year} {2011})}\BibitemShut {NoStop}%
\bibitem [{\citenamefont {Doumont}\ \emph {et~al.}(2022)\citenamefont
  {Doumont}, \citenamefont {Tran},\ and\ \citenamefont
  {Blaha}}]{Doumont2022Implementation}%
  \BibitemOpen
  \bibfield  {author} {\bibinfo {author} {\bibfnamefont {J.}~\bibnamefont
  {Doumont}}, \bibinfo {author} {\bibfnamefont {F.}~\bibnamefont {Tran}},\ and\
  \bibinfo {author} {\bibfnamefont {P.}~\bibnamefont {Blaha}},\ }\bibfield
  {title} {\bibinfo {title} {Implementation of self-consistent mgga functionals
  in augmented plane wave based methods},\ }\href
  {https://doi.org/10.1103/PhysRevB.105.195138} {\bibfield  {journal} {\bibinfo
   {journal} {Phys. Rev. B}\ }\textbf {\bibinfo {volume} {105}},\ \bibinfo
  {pages} {195138} (\bibinfo {year} {2022})}\BibitemShut {NoStop}%
\end{thebibliography}%

\end{document}